\documentclass[english,british,usenatbib]{mn2e}
\usepackage[T1]{fontenc}
\usepackage[latin9]{inputenc}
\usepackage{refstyle}
\usepackage{units}
\usepackage{url}
\usepackage{amssymb}
\usepackage{graphicx}
\usepackage[authoryear]{natbib}

\makeatletter

\AtBeginDocument{\providecommand\secref[1]{\ref{sec:#1}}}
\AtBeginDocument{\providecommand\tabref[1]{\ref{tab:#1}}}
\AtBeginDocument{\providecommand\figref[1]{\ref{fig:#1}}}
\AtBeginDocument{\providecommand\fnref[1]{\ref{fn:#1}}}
\AtBeginDocument{\providecommand\enuref[1]{\ref{enu:#1}}}
\AtBeginDocument{\providecommand\eqref[1]{\ref{eq:#1}}}
\newcommand{\noun}[1]{\textsc{#1}}

\RS@ifundefined{subref}
  {\def\RSsubtxt{section~}\newref{sub}{name = \RSsubtxt}}
  {}
\RS@ifundefined{thmref}
  {\def\RSthmtxt{theorem~}\newref{thm}{name = \RSthmtxt}}
  {}
\RS@ifundefined{lemref}
  {\def\RSlemtxt{lemma~}\newref{lem}{name = \RSlemtxt}}
  {}

\makeatletter{}

\usepackage{siunitx}
\let\jnl@style=\rm
\def\ref@jnl#1{{\jnl@style#1}}

\def\aj{\ref@jnl{AJ}}                   
\def\actaa{\ref@jnl{Acta Astron.}}      
\def\araa{\ref@jnl{ARA\&A}}             
\def\apj{\ref@jnl{ApJ}}                 
\def\apjl{\ref@jnl{ApJ}}                
\def\apjs{\ref@jnl{ApJS}}               
\def\ao{\ref@jnl{Appl.~Opt.}}           
\def\apss{\ref@jnl{Ap\&SS}}             
\def\aap{\ref@jnl{A\&A}}                
\def\aapr{\ref@jnl{A\&A~Rev.}}          
\def\aaps{\ref@jnl{A\&AS}}              
\def\azh{\ref@jnl{AZh}}                 
\def\baas{\ref@jnl{BAAS}}               
\def\bac{\ref@jnl{Bull. astr. Inst. Czechosl.}}
\def\caa{\ref@jnl{Chinese Astron. Astrophys.}}
\def\cjaa{\ref@jnl{Chinese J. Astron. Astrophys.}}
\def\icarus{\ref@jnl{Icarus}}           
\def\jcap{\ref@jnl{J. Cosmology Astropart. Phys.}}
\def\jrasc{\ref@jnl{JRASC}}             
\def\memras{\ref@jnl{MmRAS}}            
\def\mnras{\ref@jnl{MNRAS}}             
\def\na{\ref@jnl{New A}}                
\def\nar{\ref@jnl{New A Rev.}}          
\def\pra{\ref@jnl{Phys.~Rev.~A}}        
\def\prb{\ref@jnl{Phys.~Rev.~B}}        
\def\prc{\ref@jnl{Phys.~Rev.~C}}        
\def\prd{\ref@jnl{Phys.~Rev.~D}}        
\def\pre{\ref@jnl{Phys.~Rev.~E}}        
\def\prl{\ref@jnl{Phys.~Rev.~Lett.}}    
\def\pasa{\ref@jnl{PASA}}               
\def\pasp{\ref@jnl{PASP}}               
\def\pasj{\ref@jnl{PASJ}}               
\def\rmxaa{\ref@jnl{Rev. Mexicana Astron. Astrofis.}}%
\def\qjras{\ref@jnl{QJRAS}}             
\def\skytel{\ref@jnl{S\&T}}             
\def\solphys{\ref@jnl{Sol.~Phys.}}      
\def\sovast{\ref@jnl{Soviet~Ast.}}      
\def\ssr{\ref@jnl{Space~Sci.~Rev.}}     
\def\zap{\ref@jnl{ZAp}}                 
\def\nat{\ref@jnl{Nature}}              
\def\iaucirc{\ref@jnl{IAU~Circ.}}       
\def\aplett{\ref@jnl{Astrophys.~Lett.}} 
\def\apspr{\ref@jnl{Astrophys.~Space~Phys.~Res.}}
\def\bain{\ref@jnl{Bull.~Astron.~Inst.~Netherlands}} 
\def\fcp{\ref@jnl{Fund.~Cosmic~Phys.}}  
\def\gca{\ref@jnl{Geochim.~Cosmochim.~Acta}}   
\def\grl{\ref@jnl{Geophys.~Res.~Lett.}} 
\def\jcp{\ref@jnl{J.~Chem.~Phys.}}      
\def\jgr{\ref@jnl{J.~Geophys.~Res.}}    
\def\jqsrt{\ref@jnl{J.~Quant.~Spec.~Radiat.~Transf.}}
\def\memsai{\ref@jnl{Mem.~Soc.~Astron.~Italiana}}
\def\nphysa{\ref@jnl{Nucl.~Phys.~A}}   
\def\physrep{\ref@jnl{Phys.~Rep.}}   
\def\physscr{\ref@jnl{Phys.~Scr}}   
\def\planss{\ref@jnl{Planet.~Space~Sci.}}   
\def\procspie{\ref@jnl{Proc.~SPIE}}   

\usepackage{multirow}
\usepackage{url}
\usepackage{booktabs}
\usepackage{datetime}
\usepackage{bm}

\newcommand\colfigspacer{\hspace{1 cm}}

\makeatletter{}

\newcommand{\mean}[1]{\ensuremath{\overline{#1}}}
\newcommand{\standarddeviation}[1]{\ensuremath{\sigma_{#1}}}
\newcommand{\standarderroron}[1]{\ensuremath{\alpha_{#1}}}

\newcommand{\magnitude}{\ensuremath{M}} 
\newcommand{\luminosity}{\ensuremath{L}} 
\newcommand{\colour}{\ensuremath{C}}

\newcommand{\REXCESS}{REXCESS}
\newcommand{\rexcess}{\REXCESS} 
\newcommand{\XMM}{\emph{XMM-Newton}}
\newcommand{\Chandra}{\emph{Chandra}}
\newcommand{\Suzaku}{\emph{Suzaku}}
\newcommand{\ROSAT}{\emph{ROSAT}}



\newcommand{\chisqr}{\ensuremath{\chi^2}}
\newcommand{\redchi}{\ensuremath{\chisqr_r}}

\newcommand{\band}[1]{\ensuremath{#1}}
\newcommand{\Rband}{\band{R}}

\newcommand{\Vband}{\band{V}}
\newcommand{\Bband}{\band{B}}
\newcommand{\Rluminosity}{\ensuremath{L}}

\newcommand{\WFIABR}{\ensuremath{R_\mathrm{AB}}} 

\newcommand{\JCstring}{\ensuremath{\mathrm{J}}}
\newcommand{\kcorrstring}{\ensuremath{\mathrm{K}}}
\newcommand{\rawmagstring}{\ensuremath{\mathrm{raw}}}
\newcommand{\absmagstring}{\ensuremath{\mathrm{abs}}}
\newcommand{\ABmagstring}{\ensuremath{\mathrm{AB}}}

\newcommand{\JC}[1]{\ensuremath{#1_{\JCstring}}} 
\newcommand{\JCk}[1]{\ensuremath{#1_{\JCstring \, \kcorrstring}}} 
\newcommand{\WFIraw}[1]{\ensuremath{#1_{\rawmagstring}}}
\newcommand{\WFIAB}[1]{\ensuremath{#1_{\ABmagstring}}}
\newcommand{\WFIABabsk}[1]{\ensuremath{#1_{\ABmagstring \, \absmagstring \, \kcorrstring}}}
\newcommand{\absmag}[1]{\ensuremath{#1_{\absmagstring}}}

\newcommand{\VJband}{\JC{\Vband}}
\newcommand{\BJband}{\JC{\Bband}}

\newcommand{\Vabsband}{\absmag{\Vband}}

\newcommand{\zeropoint}[1]{\ensuremath{#1{}_{\mathrm{zero}}}}

\newcommand{\Bzero}{\zeropoint{\Bband}}
\newcommand{\Vzero}{\zeropoint{\Vband}}
\newcommand{\Rzero}{\zeropoint{\Rband}}

\newcommand{\gscmagnitude}{\ensuremath{\magnitude_{\mathrm{GSC}}}}

\newcommand{\colourterm}[1][\magnitude]{\ensuremath{c_#1}}
\newcommand{\extinctionterm}[1][\magnitude]{\ensuremath{e_#1}}
\newcommand{\aboffset}[1][\magnitude]{\ensuremath{O_#1}}
\newcommand{\airmass}[1][\magnitude]{\ensuremath{Z_#1}}

\newcommand{\colourtermR}{\colourterm[\Rband]}
\newcommand{\colourtermV}{\colourterm[\Vband]}
\newcommand{\colourtermB}{\colourterm[\Bband]}

\newcommand{\cataloguecontamination}{\ensuremath{K}}

\newcommand{\colouroffset}[1][\colour]{\ensuremath{\kappa_{#1}}}

\newcommand{\MSol}{\ensuremath{\mathrm{M_{\odot}}}} 

\newcommand{\semimajoraxis}{\ensuremath{a}}

\newcommand{\akron}{\ensuremath{\semimajoraxis_{\mathrm{Kron}}}} 

\newcommand{\rfl}[1]{\ensuremath{r_{\mathrm{#1}}}}
\newcommand{\rfh}{\rfl{500}}
\newcommand{\rth}{\rfl{200}}

\newcommand{\stellarity}{\ensuremath{s}}

\newcommand{\mass}{\ensuremath{M}}
\newcommand{\massl}[1]{\ensuremath{\mass_{#1}}}

\newcommand{\luml}[1]{\ensuremath{\luminosity_{#1}}}

\newcommand{\regionweight}{\ensuremath{w}}

\newcommand{\RA}{RA}
\newcommand{\Dec}{Dec}

\newcommand{\schechter}[1]{\ensuremath{{#1}_*}}
\newcommand{\mschechter}{\schechter} 

\newcommand{\galaxydensityfactor}[1][]{\ensuremath{f_{\galaxysymbol #1}}}
\newcommand{\combinedgalaxycountfunction}{\ensuremath{\Xi}}
\newcommand{\metcalfegalaxycountfunction}{\ensuremath{\xi_{\mathrm{Metcalfe}}}}
\newcommand{\fallofffunction}{\ensuremath{\xi_{\mathrm{falloff}}}}
\newcommand{\falloffmagnitude}[1][\magnitude]{\ensuremath{#1 {}_{\mathrm{falloff}}}}
\newcommand{\falloffwidth}{\ensuremath{W_{\mathrm{falloff}}}}

\newcommand{\schechterfunction}{\ensuremath{\Phi}}
\newcommand{\schechternormalisation}{\ensuremath{\phi}}
\newcommand{\schechterslope}{\ensuremath{\alpha}}
\newcommand{\schechterluminosity}{\ensuremath{\schechter{\luminosity}}}

\newcommand{\galaxysymbol}{\ensuremath{\mathrm{g}}}

\newcommand{\totalgalaxycount}{\ensuremath{n_{\galaxysymbol}}}
\newcommand{\totalgalaxycountredblue}{\ensuremath{n_{\galaxysymbol,\mathrm{r+b}}}}

\newcommand{\redshift}{\ensuremath{z}}
\newcommand{\halooverdensity}{\ensuremath{\Delta}}
\newcommand{\concentration}{\ensuremath{c}}
\newcommand{\concentrationmatter}[1]{\ensuremath{\concentration_{#1,m}}}
\newcommand{\concentrationcritical}[1]{\ensuremath{\concentration_{#1}}}
\newcommand{\NFWscalelength}{\ensuremath{r_s}}
\newcommand{\densityparameter}{\Omega}

\newcommand{\mlnorm}{\ensuremath{\eta}}
\newcommand{\mlindex}{\ensuremath{\epsilon}}
\newcommand{\mlpivotmass}{\ensuremath{\mass_{\mathrm{pivot}}}}

\newcommand{\XrayLuminosity}{\ensuremath{L_\Xraysym}}
\newcommand{\XrayY}{\ensuremath{Y_\Xraysym}}

\newcommand{\rccen}{\ensuremath{r}} 

\newcommand{\areafractionincentralbin}{\ensuremath{A_c}} 

\newcommand{\brightcut}[1][\detectionband]{\ensuremath{\mschechter{#1} + 2.5}}
\newcommand{\faintcut}[1][\detectionband]{\ensuremath{\mschechter{#1} + 5}}

\newcommand{\brightcutsymbol}[1][\detectionband]{\ensuremath{{#1}{}_{*+2.5}}}
\newcommand{\faintcutsymbol}[1][\detectionband]{\ensuremath{{#1}{}_{*+5}}}

\newcommand{\bright}[1][\detectionband]{\ensuremath{#1 < \brightcutsymbol[#1]}}

\newcommand{\brightfaint}[1][\detectionband]{\ensuremath{#1 < \faintcutsymbol[#1]}}

\newcommand{\cmrgradient}{\ensuremath{g}}
\newcommand{\rsgradient}{\cmrgradient}
\newcommand{\cmrintercept}{\ensuremath{k}}
\newcommand{\rsintercept}{\ensuremath{k}} 
\newcommand{\rs}{\ensuremath{w_{\mathrm{rs}}}} 
\newcommand{\rso}[1] {\ensuremath{\left| \rs \right| < {#1}}} 
\newcommand{\rsul}[1] {\ensuremath{\rs < {#1}}} 

\newcommand{\cmrmodel}{\ensuremath{\colour_{\mathrm{model}}}}
\newcommand{\cmrresidual}{\ensuremath{\colour_{\mathrm{residual}}}}
\newcommand{\cmrpivot}{\ensuremath{Z}}

\newcommand{\redseqwidth}{\ensuremath{\sigma_r}}
\newcommand{\bluecloudwidth}{\ensuremath{\sigma_b}}
\newcommand{\redseqoffset}{\ensuremath{o_r}}
\newcommand{\bluecloudoffset}{\ensuremath{o_b}}

\newcommand{\scattermodel}{\Psi}
\newcommand{\scattermodelrednorm}{\psi_r}
\newcommand{\scattermodelbluenorm}{\psi_b}

\newcommand{\Xraysym}{\text{X}}

\newcommand{\offtargetarea}[1]{\ensuremath{A_{r>\rfl{#1}}}}

\newcommand{\disturbed}{disturbed} 
\newcommand{\regular}{regular} 
\newcommand{\coolcore}{cool-core} 
\newcommand{\coolcores}{\coolcore s} 
\newcommand{\noncoolcore}{non-cool-core} 
\newcommand{\massive}{massive} %
\newcommand{\lowmass}{low mass}
\newcommand{\Allclusters}{All clusters} 
\newcommand{\Disturbed}{Disturbed} 
\newcommand{\Regular}{Regular} 
\newcommand{\Coolcore}{Cool-core} 
\newcommand{\Noncoolcore}{Non-cool-core} 
\newcommand{\Lowmass}{Low mass}
\newcommand{\Massive}{Massive}
\newcommand{\disturbedsym}{\ensuremath{\mathrm{D}}} 
\newcommand{\coolcoresym}{\ensuremath{\mathrm{CC}}} 
\newcommand{\massivesym}{\ensuremath{\mathrm{M}}} 

\usepackage{color}
\newcommand{\surfacedensity}{\ensuremath{{S}}} 
\newcommand{\volumedensity}{\ensuremath{{\rho}}} 

\newcommand{\electronsymbol}{\ensuremath{\mathrm{e}}}

\newcommand{\electronsurfacedensity}{\ensuremath{\surfacedensity_{\electronsymbol}}}
\newcommand{\masssurfacedensity}{\ensuremath{\surfacedensity_{\mass}}}

\newcommand{\background}[1]{\ensuremath{#1{}_{\mathrm{bg}}}}

\newcommand{\normsymbol}{\ensuremath{\mathrm{norm}}}

\newcommand{\galaxydensitynorm}[1]{\ensuremath{#1_{\normsymbol}}} 
\newcommand{\electrondensitynorm}[1]{\ensuremath{#1_{\electronsymbol,\normsymbol}}}
\newcommand{\massdensitynorm}[1]{\ensuremath{#1_{\mass,\normsymbol}}}

\newcommand{\annulusarea}{\ensuremath{a}}

\newcommand{\simplebackground}{\ensuremath{\background{\surfacedensity}{}_{\mathrm{simple}}}}
\newcommand{\sectorbackground}{\ensuremath{\background{\surfacedensity}{}_{\mathrm{sector}}}}

\newcommand{\simplebackgroundalpha}{\standarderroron{\simplebackground}}
\newcommand{\sectorbackgroundalpha}{\standarderroron{\sectorbackground}}

\newcommand{\NFWareaconstant}{\ensuremath{\background{\surfacedensity}{}_{\mathrm{NFW}}}}

\newcommand{\rcutoff}{\ensuremath{r_{\mathrm{cutoff}}}}

\newcommand{\betagal}{\ensuremath{\beta_{\mathrm{g}}}}
\newcommand{\betael}{\ensuremath{\beta_{\mathrm{e}}}}

\usepackage{siunitx}

\DeclareSIQualifier{\solar}{\ensuremath{\odot}}

\DeclareSIQualifier{\fh}{500}
\DeclareSIQualifier{\th}{200}
\DeclareSIQualifier{\seventy}{70}

\DeclareSIUnit\SIMass{M}
\DeclareSIUnit\SILuminosity{L}

\DeclareSIUnit\solarmass{\SIMass\solar}
\DeclareSIUnit\solarluminosity{\SILuminosity\solar}
\DeclareSIUnit\solMass{\solarmass}
\DeclareSIUnit\solLum{\solarluminosity}
\DeclareSIUnit\Msun{\solarmass}
\DeclareSIUnit\Lsun{\solarluminosity}

\DeclareSIUnit\pixel{pixel}
\DeclareSIUnit\pix{\pixel}
\DeclareSIUnit\arcminuteword{arcmin}
\DeclareSIUnit\arcsecondword{arcsec}
\DeclareSIUnit\degreeword{degree}
\DeclareSIUnit\parsec{pc} 
\DeclareSIUnit\lightyear{ly}
\DeclareSIUnit\count{count}
\DeclareSIUnit\ct{\count}
\DeclareSIUnit\radial{\textit{r}}
\DeclareSIUnit\rfhsi{\radial\fh}
\DeclareSIUnit\rthsi{\radial\th}

\DeclareSIUnit\rfh{\radial\fh}
\DeclareSIUnit\rth{\radial\th}

\DeclareSIUnit\year{yr}
\DeclareSIUnit\mag{mag}
\DeclareSIUnit\erg{erg}
\DeclareSIUnit\hparam{\textit{h}}
\DeclareSIUnit\hst{\hparam\seventy}
\DeclareSIUnit\pixel{pixel}
\DeclareSIUnit\kilopixel{kilopixel}
\DeclareSIUnit\dex{dex}

\newcommand{\softwarepackage}[1]{\textsc{#1}} 
\newcommand{\softwarevariable}[1]{\texttt{#1}} 
\newcommand{\softwarescript}[1]{\texttt{#1}} 

\newcommand{\sextractor}{\softwarepackage{sextractor}}

\newcommand{\magtoflux}{\softwarepackage{mag2flux}}
\newcommand{\theli}{\softwarepackage{theli}}
\newcommand{\iraf}{\softwarepackage{iraf}}
\newcommand{\esomvm}{\softwarepackage{eso/mvm (alambic)}}
\newcommand{\automag}{\softwarevariable{auto\textunderscore mag}}
\newcommand{\automask}{\softwarescript{automask.sh}}
\newcommand{\psfmatch}{\softwarescript{psfmatch}}
\newcommand{\wregister}{\softwarescript{wregister}}

\newcommand{\astropy}{\softwarepackage{astropy}}
\newcommand{\python}{\softwarepackage{python}}
\newcommand{\scipy}{\softwarepackage{scipy}}
\newcommand{\matplotlib}{\softwarepackage{matplotlib}}

\makeatletter
\@ifpackagelater{siunitx}{2012/01/01}
{}
{
\DeclareSIUnit\keV{\kilo\electronvolt}
\DeclareSIUnit\cm{\centi\metre}
}
 
\makeatletter{}

\makeatletter
\title[REXCESS optical and X-ray galaxy cluster profiles]{Optical and X-ray profiles in the REXCESS sample of galaxy clusters\thanks{Based on observations collected at the European Southern Observatory (La Silla, Chile).}} \let\Title\@title
\author[Holland et al.]{John G. Holland$^1$, Hans B\"ohringer$^1$, Gayoung Chon$^1$ and Daniele Pierini$^{2,3,4}$\\
$^1$Max-Planck-Institut f\"ur extraterrestrische Physik, D 85748 Garching, Germany\\
$^2$Institut de Recherche en Astrophysique et Plan\'etologie (IRAP), CNRS - Universit\'e Paul Sabatier, 9 avenue du Colonel Roche, \\BP 44346 - 311028 Toulouse Cedex 4, France \\ $^3$Universit\'e Toulouse III - Paul Sabatier, 118 route de Narbonne, 31062 Toulouse Cedex 9, France \\ $^4$Aix-Marseille Universit\'e, Jardin du Pharo, 58 bd Charles Livon, 13284 Marseille Cedex 7, France }\let\Author\@author
\date{Accepted 2015 January 15. Received 2015 January 15; in original form 2014 April 30}\let\Date\@date

\pubyear{2015}

\makeatother
 
\makeatletter

\@ifundefined{showcaptionsetup}{}{%
 \PassOptionsToPackage{caption=false}{subfig}}
\usepackage{subfig}
\makeatother

\usepackage{babel}
\begin{document}
\maketitle

\makeatletter{}
\begin{abstract}
Galaxy clusters' structure, dominated by dark matter, is traced by
member galaxies in the optical and hot intra-cluster medium (ICM)
in X-rays. We compare the radial distribution of these components
and determine the mass-to-light ratio vs. system mass relation. 

We use 14 clusters from the \REXCESS~sample which is representative
of clusters detected in X-ray surveys. Photometric observations with
the Wide Field Imager on the 2.2m MPG/ESO telescope are used to determine
the number density profiles of the galaxy distribution out to $\rth$.
These are compared to electron density profiles of the ICM obtained
using \XMM\unskip, and dark matter profiles inferred from scaling
relations and an NFW model.

While red sequence galaxies trace the total matter profile, the blue
galaxy distribution is much shallower. We see a deficit of faint galaxies
in the central regions of massive and regular clusters, and strong
suppression of bright and faint blue galaxies in the centres of \coolcore\ clusters,
attributable to ram pressure stripping of gas from blue galaxies in
high density regions of ICM and disruption of faint galaxies due to
galaxy interactions. We find a mass-to-light ratio vs. mass relation
within $\rth$ of \makeatletter{}
\ensuremath{\SI{3.0(4)e+02}{\hparam\solarmass\per\solarluminosity}}
 at
$\SI{1e15}{\solarmass}$ with slope \makeatletter{}
\ensuremath{0.16 \pm 0.14}
 \unskip,
consistent with most previous results. 
\end{abstract}


\begin{keywords}
galaxies: clusters: general -- 
X-rays: galaxies: clusters -- 
intergalactic medium --
dark matter
\end{keywords}

\makeatletter{}

\section{Introduction}

\label{sec:Introduction}

Galaxy clusters are the most massive gravitationally bound and virialised
objects in the observable Universe, with masses up to a few $\times\SI{1e15}{\solMass}$
(e.g. \citet{2014:Kohlinger.F;Schmidt.R:Strong-lensing-in-RX-J1347.5-1145-revisited:article}
measure the projected mass within $\SI{200}{\kilo\parsec}$ of RXC~J1347.5--1145
-- one of the most X-ray luminous clusters found -- to be in the range
\SIrange[fixed-exponent=15,scientific-notation=fixed]{2.19e15}{2.47e15}{\solMass}).
Superclusters of galaxies can be more massive and bound, but are unvirialised,
e.g. \citet*{2013:Chon.G;Bohringer.H;Nowak.N:The-extended-ROSAT-ESO-Flux-Limited-X-ray-Galaxy-Cluster-Survey-REFLEX-II---III.-Construction:article}.
Galaxy clusters are interesting as their contents are reasonably representative
of the contents of the universe as a whole and they contain a population
of coeval galaxies whose appearance is affected by a whole range of
astrophysical processes and the interactions between the processes.

The dominant baryonic component of galaxy clusters is the intracluster
medium, (ICM), a hot ($\SI{>1e7}{\kelvin}$), X-ray emitting plasma
which contains most of the baryonic mass \citep[e.g.][]{2003:Lin.Y;Mohr.J;Stanford.S:Near-Infrared-Properties-of-Galaxy-Clusters:-Luminosity-as-a-Binding-Mass-Predictor-and-the-State-of-Cluster:article}.
The gas becomes X-ray luminous after being heated by adiabatic compression
and shocks during cluster collapse \citep[e.g.][]{1972:Gunn.J;Gott.J:On-the-Infall-of-Matter-Into-Clusters-of-Galaxies-and-Some-Effects-on-Their-Evolution:article,2012:Kravtsov.A;Borgani.S:Formation-of-Galaxy-Clusters:article}.
A fraction of the clusters have a centre where the density is high
and where entropies can be low enough that cooling should take place
on the order of the Hubble time \citep{1994:Fabian.A:Cooling-Flows-in-Clusters-of-Galaxies:article},
but feedback mechanisms quench cooling flows (\citealp{2006:Bower.R;Benson.A;Malbon.R:Breaking-the-hierarchy-of-galaxy-formation:article};
\citealp{2012:Fabian.A:Observational-Evidence-of-Active-Galactic-Nuclei-Feedback:article}
is a recent review), adding additional energy to the initially gravitationally
heated gas and causing it to be more broadly distributed than the
dark matter potential. Cluster mass profile estimates from X-ray data,
including ICM density measurements, are limited to the region in which
robust temperature measurements can be made. For \XMM~and \Chandra,
this is typically $\lesssim\rfh$; \Suzaku~and \ROSAT~can reach
$\rth$ with substantially lower spatial resolution \citep{2013:Reiprich.T;Basu.K;Ettori.S:Outskirts-of-Galaxy-Clusters:article}
($\halooverdensity$ is the `halo overdensity' and $\rfl{\halooverdensity}$
refers to the radius of the volume in which the mean density is $\halooverdensity$
times the critical density of the universe). 

Cluster galaxies can be used to probe the cluster environment to greater
cluster-centric distances than gas and can be treated as approximately
collisionless particles moving in the dark matter potential well of
the cluster. The well known morphology-density relation -- higher
fractions of elliptical galaxies in high density environments like
galaxy clusters, compared to low fractions of elliptical galaxies
in the lower density field environment \citep{1984:Dressler.A:The-Evolution-of-Galaxies-in-Clusters:article}
-- is caused by the ICM and the presence of other galaxies, but a
precise description of the way the different components involved interact
is not yet available. The critical processes are: ram-pressure stripping,
where weakly bound gas is stripped away from galaxies by interaction
with the ICM \citep{1972:Gunn.J;Gott.J:On-the-Infall-of-Matter-Into-Clusters-of-Galaxies-and-Some-Effects-on-Their-Evolution:article};
strangulation, where galaxies are starved of cool gas in their haloes
-- gravitational heating combined with the active galactic nucleus
(AGN)/wind/supernova feedback mechanisms already mentioned lead to
ICM entropy which is too high for effective cooling and replenishment
of the cool gas \citep{1980:Larson.R;Tinsley.B;Caldwell.C:The-evolution-of-disk-galaxies-and-the-origin-of-S0-galaxies:article};
harassment -- gravitational interactions between galaxies which increase
internal energy and lead to morphological change (\citealp{1981:Farouki.R;Shapiro.S:Computer-simulations-of-environmental-influences-on-galaxy-evolution-in-dense-clusters.-II---Rapid-tidal:article};
\citealp{1996:Moore.B;Katz.N;Lake.G:Galaxy-harassment-and-the-evolution-of-clusters-of-galaxies:article};
\citealp*{1998:Moore.B;Lake.G;Katz.N:Morphological-Transformation-from-Galaxy-Harassment:article});
and galactic cannibalism, where dynamical friction reduces the velocity
of satellites relative to the central galaxy below the velocity dispersion
of the satellite, allowing it to be accreted on to the central galaxy
\citep{1977:Ostriker.J;Hausman.M:Cannibalism-among-the-galaxies---Dynamically-produced-evolution-of-cluster-luminosity-functions:article}.
Galaxies at low cluster-centric distances are preferentially harassed
and starved due to the high ICM density and more frequent encounters
with other cluster galaxies, including cD galaxies which are often
the brightest cluster galaxy (BCG). Faint and dwarf galaxies with
the weakest dark matter haloes and the weakest hold on their gas reservoirs
fare the worst in this environment such that the population of cluster
galaxies seen at $z>0.4$ which includes a large fraction of star
forming galaxies changes by $z=0$ into a population dominated by
galaxies with very low star formation rates \citep{1984:Butcher.H;Oemler.A:The-evolution-of-galaxies-in-clusters.-V---A-study-of-populations-since-Z-approximately-equal-to-0.5:article,1994:Dressler.A;Oemler.A;Butcher.H:The-morphology-of-distant-cluster-galaxies.-1:-HST-observations-of-CL-09394713:article,1997:Oemler.A;Dressler.A;Butcher.H:The-Morphology-of-Distant-Cluster-Galaxies.-II.-HST-Observations-of-Four-Rich-Clusters-at-Z-0.4:article,2006:Popesso.P;Biviano.A;Bohringer.H:RASS-SDSS-Galaxy-cluster-survey.-IV.-A-ubiquitous-dwarf-galaxy-population-in-clusters:article,2006:Boselli.A;Gavazzi.G:Environmental-Effects-on-Late-Type-Galaxies-in-Nearby-Clusters:article}. 

It should be noted that an X-ray selected sample like the Representative
$\XMM$ Cluster Structure Survey (\REXCESS, \citealp{2007:Bohringer.H;Schuecker.P;Pratt.G:The-representative-XMM-Newton-cluster-structure-survey-REXCESS-of-an-X-ray-luminosity-selected-galaxy:article}),
used in the present study, preferentially represents objects with
higher ICM densities, so we could expect that any ICM-dominated effects
would be stronger in our results than in other, optically selected
samples \citep[e.g.][]{1997:Carlberg.R;Yee.H;Ellingson.E:The-Average-Mass-and-Light-Profiles-of-Galaxy-Clusters:article}
or partially optically selected samples \citep[e.g.][]{2004:Popesso.P;Bohringer.H;Brinkmann.J:RASS-SDSS-Galaxy-clusters-survey.-I.-The-catalog-and-the-correlation-of-X-ray-and-optical-properties:article}.
They also preferentially include clusters with deep potential wells
and thus more time available for galactic evolution. \citet{2004:Bohringer.H;Matsushita.K;Churazov.E:Implications-of-the-central-metal-abundance-peak-in-cooling-core-clusters-of-galaxies:article}
show that cool-cores in clusters must be preserved on very long time-scales
and we could expect these to cause distinctive features in the population
of galaxies. Conversely, if we assume clusters which show disturbances
in their X-ray morphology are relatively recent mergers, we would
expect their galactic populations to be less evolved than in other
types of cluster and might also show some trace of disturbance in
the galaxy distribution.

In this work we study the relationship between the optical density
profiles of the galaxy distribution and the density profiles of X-ray
emitting gas in X-ray selected galaxy clusters. We explore the extent
to which galaxies and gas trace one-another and the underlying dark
matter. We also investigate how the red and blue galaxy populations
are distributed. We investigate the total mass to optical light ratio
of galaxy clusters, and measure how this varies with respect to total
cluster mass and morphology (presence/absence of cool-cores, regular/disturbed
ICM). By taking into account the morphology of our sample, we show
clear differences in the distribution of different types of galaxies
in the centres of clusters which have had relatively stable morphology
for a long period of time -- massive clusters or those which have
cool-cores -- when compared to clusters with signs of more radical
recent evolution -- disturbed and non cool-core clusters.

The paper is structured as follows. In \secref{Sample-Description}
the sample characteristics, X-ray data and optical data are described.
In \secref{Analysis} we describe all stages of the analysis including
object detection, classification, optical data calibration, red sequence
fitting, radial profile generation and luminosity measurements, along
with the results generated at each stage. The results are discussed
in \secref{Discussion}. The conclusions are summarised in \secref{Conclusions}. 

Throughout this paper, radial distances are measured in units of $\rfh$.
The influence of the cluster may extend further than this, so we typically
use the region outside $\SI{1.5}{\rfhsi}\sim\rth$ as the off-target
region. ($\rth=1.51\rfh$ for a concentration $\concentrationcritical{500}=3.2$,
\citep*{2007:Arnaud.M;Pointecouteau.E;Pratt.G:Calibration-of-the-galaxy-cluster-M00hBcY-relation-with-XMM-Newton:article}).
Magnitudes $M$ in a given broad band filter $\left(R,V,B\right)$
 are signified by a subscript suffix, AB magnitudes by $_{\ABmagstring}$,
Johnson magnitudes by $_{\JCstring}$ and K-correction by $_{\kcorrstring}$.
We adopt a flat cosmology where $\si{\hparam}=0.7$ (and $\si{\hparam}=\SI{0.7}{\hst}$),
$\densityparameter_{m}=0.3$, $\densityparameter_{\Lambda}=0.7$ and
$H_{0}=\SI{100}{\hparam\kilo\metre\per\second\per\mega\parsec}$.


\makeatletter{}

\section{Sample description}

\label{sec:Sample-Description}

\subsection{REXCESS sample}

The \REXCESS\ sample has been compiled as a galaxy cluster sample,
representative of clusters detected by their X-ray luminosity and
independent of their morphology. The sample selection is described
in \citet{2007:Bohringer.H;Schuecker.P;Pratt.G:The-representative-XMM-Newton-cluster-structure-survey-REXCESS-of-an-X-ray-luminosity-selected-galaxy:article}.
The clusters have redshifts between $\redshift=0.055$ and $\redshift=0.183$
and luminosities above $\SI{0.4e44}{\per\hst\squared\erg\per\second}$
in the \SIrange{0.1}{2.4}{\keV} band. The $\rfl{500}$ region, where
the mean density is $500\times$ the critical density, plus a region
outside where the background can be estimated are within the \XMM\ field-of-view
$(\sim\SI{30}{\arcmin})$. The mass range of the clusters is $\massl{200}=$
\SIrange[fixed-exponent=14,scientific-notation=fixed]{1.36e14}{10.8e14}{\solarmass}.
They represent a relatively homogeneous population in X-ray luminosity,
$\XrayLuminosity$.

The 14 objects comprising the subset of the \REXCESS\ sample used
in this work are tabulated with their key parameters in \tabref{Overview-of-the-clusters}
and comprise approximately half of the complete \REXCESS\ sample.
The objects were selected by right ascension for ease of follow-up
observation scheduling, and were observed first. 

\begin{table*}
\protect\caption{\label{tab:Overview-of-the-clusters}Overview of the \REXCESS\ clusters
analysed in this paper. \coolcoresym\ = \coolcore, \disturbedsym\ =
\disturbed, \massivesym\ = \massive. Abell Name, RA, Dec, $\redshift$
and $\XrayLuminosity$ are from table 3 in \citet{2007:Bohringer.H;Schuecker.P;Pratt.G:The-representative-XMM-Newton-cluster-structure-survey-REXCESS-of-an-X-ray-luminosity-selected-galaxy:article}.
\rfh, \coolcoresym\ and \disturbedsym\ are from table 1 in \citet{2009:Pratt.G;Croston.J;Arnaud.M:Galaxy-cluster-X-ray-luminosity-scaling-relations-from-a-representative-local-sample:article};
$\massl{500}$ was derived from $\rfl{500}$ and $\redshift$ using
the fiducial cosmology. Massive objects are those with $\massl{500}$
above the median of the entire \REXCESS\ population, \protect\makeatletter{}
\ensuremath{\SI{2.95e+14}{\Msun}}
 \unskip.
ID is used in Figures in \secref{Mass_to_light_ratio} to distinguish
between the clusters in the sample.}
\makeatletter{}
\begin{tabular}{lllllSSSllll}
\hline
Object & Abell Name & \RA & \Dec & \redshift & \XrayLuminosity & \rfh & \massl{500} & \disturbedsym & \coolcoresym & \massivesym & ID \\
 &  &  &  &  & \ensuremath{\SI{1e+37}{\watt}} & \ensuremath{\si{\kilo\parsec}} & \ensuremath{\SI{1e+14}{\Msun}} &  &  &  &  \\
\hline
RXC\,J0006.0--3443 & A2721 & $00^\mathrm{h}06^\mathrm{m}03.0^\mathrm{s}$ & $-34^\circ43{}^\prime27.0{}^{\prime\prime}$ & 0.1147 & 1.875 & 1059.3 & \num{4.21} & \disturbedsym &  & \massivesym & 1 \\
RXC\,J0049.4--2931 & S0084 & $00^\mathrm{h}49^\mathrm{m}24.0^\mathrm{s}$ & $-29^\circ31{}^\prime28.0{}^{\prime\prime}$ & 0.1084 & 1.503 & 807.8 & \num{1.84} &  &  &  & 2 \\
RXC\,J0345.7--4112 & S0384 & $03^\mathrm{h}45^\mathrm{m}45.7^\mathrm{s}$ & $-41^\circ12{}^\prime27.0{}^{\prime\prime}$ & 0.0603 & 0.495 & 688.4 & \num{1.04} &  & \coolcoresym &  & 3 \\
RXC\,J0547.6--3152 & A3364 & $05^\mathrm{h}47^\mathrm{m}38.2^\mathrm{s}$ & $-31^\circ52{}^\prime31.0{}^{\prime\prime}$ & 0.1483 & 4.667 & 1133.7 & \num{5.53} &  &  & \massivesym & 4 \\
RXC\,J0605.8--3518 & A3378 & $06^\mathrm{h}05^\mathrm{m}52.8^\mathrm{s}$ & $-35^\circ18{}^\prime02.0{}^{\prime\prime}$ & 0.1392 & 4.478 & 1045.9 & \num{4.26} &  & \coolcoresym & \massivesym & 5 \\
RXC\,J0616.8--4748 &  & $06^\mathrm{h}16^\mathrm{m}53.6^\mathrm{s}$ & $-47^\circ48{}^\prime18.0{}^{\prime\prime}$ & 0.1164 & 1.597 & 939.2 & \num{2.95} & \disturbedsym &  &  & 6 \\
RXC\,J0645.4--5413 & A3404 & $06^\mathrm{h}45^\mathrm{m}29.3^\mathrm{s}$ & $-54^\circ13{}^\prime08.0{}^{\prime\prime}$ & 0.1644 & 7.139 & 1280.0 & \num{8.23} &  &  & \massivesym & 7 \\
RXC\,J0821.8+0112 & A0653 & $08^\mathrm{h}21^\mathrm{m}51.7^\mathrm{s}$ & $+01^\circ12{}^\prime42.0{}^{\prime\prime}$ & 0.0822 & 0.673 & 755.9 & \num{1.44} &  &  &  & 8 \\
RXC\,J2023.0--2056 & S0868 & $20^\mathrm{h}23^\mathrm{m}01.6^\mathrm{s}$ & $-20^\circ56{}^\prime55.0{}^{\prime\prime}$ & 0.0564 & 0.411 & 739.5 & \num{1.28} & \disturbedsym &  &  & 9 \\
RXC\,J2048.1--1750 & A2328 & $20^\mathrm{h}48^\mathrm{m}10.6^\mathrm{s}$ & $-17^\circ50{}^\prime38.0{}^{\prime\prime}$ & 0.1475 & 3.215 & 1078.0 & \num{4.75} & \disturbedsym &  & \massivesym & A \\
RXC\,J2129.8--5048 & A3771 & $21^\mathrm{h}29^\mathrm{m}51.0^\mathrm{s}$ & $-50^\circ48{}^\prime04.0{}^{\prime\prime}$ & 0.0796 & 0.767 & 900.6 & \num{2.42} & \disturbedsym &  &  & B \\
RXC\,J2218.6--3853 & A3856 & $22^\mathrm{h}18^\mathrm{m}40.2^\mathrm{s}$ & $-38^\circ53{}^\prime51.0{}^{\prime\prime}$ & 0.1411 & 3.516 & 1130.1 & \num{5.40} & \disturbedsym &  & \massivesym & C \\
RXC\,J2234.5--3744 & A3888 & $22^\mathrm{h}34^\mathrm{m}31.0^\mathrm{s}$ & $-37^\circ44{}^\prime06.0{}^{\prime\prime}$ & 0.1510 & 6.363 & 1283.2 & \num{8.06} &  &  & \massivesym & D \\
RXC\,J2319.6--7313 & A3992 & $23^\mathrm{h}19^\mathrm{m}41.8^\mathrm{s}$ & $-73^\circ13{}^\prime51.0{}^{\prime\prime}$ & 0.0984 & 0.937 & 788.7 & \num{1.68} & \disturbedsym & \coolcoresym &  & E \\
\hline
\end{tabular}
 
\end{table*}

\subsection{X-ray data}

\label{sec:X-ray_sample_description}

The X-ray observations are described in \citet{2007:Bohringer.H;Schuecker.P;Pratt.G:The-representative-XMM-Newton-cluster-structure-survey-REXCESS-of-an-X-ray-luminosity-selected-galaxy:article}.
Each cluster was observed using all three detectors (MOS1, MOS2 and
PN), and the mean final exposure after cleaning was $\SI{1.4\pm0.7e4}{\second}$
for PN and $\SI{2.1\pm0.9e4}{\second}$ for each of the MOS detectors.
All exposures were cleaned from times of high background due to Solar
flares and the PN data were corrected for out-of-time events. The
mean fraction of exposure time lost to Solar flares was $\sim0.35$
for PN and $\sim0.25$ for MOS1/2.

Cluster centres were set by finding the density peak of the X-ray
image on a scale of \makeatletter{}
\ensuremath{\SI{8.2}{\arcsecond}}
 (corresponding
to $2\times$ the PN pixel width), and all radial distances, $r$,
were measured from these centres. $\rfl{500}$ values are from \citet{2009:Pratt.G;Croston.J;Arnaud.M:Galaxy-cluster-X-ray-luminosity-scaling-relations-from-a-representative-local-sample:article}
and were found through iteration of the $\rfl{500}-\XrayY$ relation
for morphologically relaxed clusters (\citealp[eq. 1 in][]{2009:Pratt.G;Croston.J;Arnaud.M:Galaxy-cluster-X-ray-luminosity-scaling-relations-from-a-representative-local-sample:article},
\citealp{2007:Arnaud.M;Pointecouteau.E;Pratt.G:Calibration-of-the-galaxy-cluster-M00hBcY-relation-with-XMM-Newton:article}).
Electron density profiles are from \citet{2008:Croston.J;Pratt.G;Bohringer.H:Galaxy-cluster-gas-density-distributions-of-the-representative-XMM-Newton-cluster-structure-survey-REXCESS:article}.
They were derived from surface brightness profiles using the non-parametric
method of \citet{2006:Croston.J;Arnaud.M;Pointecouteau.E:An-improved-deprojection-and-PSF-deconvolution-technique-for-galaxy-cluster-X-ray-surface-brightness-profiles:article}
which performs a direct deprojection based on the assumption of spherical
symmetry and a regularisation procedure, and involves a point spread
function deconvolution, rather than fitting of a pre-determined gas
density distribution to the surface brightness profile.

\subsection{Optical data}

\label{sec:Optical_data}

We used the Wide Field Imager on the MPG/ESO $\SI{2.2}{\metre}$ Telescope
at La Silla, which is well suited to \XMM\ follow up as it has a
similar field of view $(\SI{34x33}{\arcmin})$. Each set of optical
data cover the cluster and a region outside $\rfl{200}$ which we
use for the background assessment. The nominal resolution is $\SI{0.238}{\arcsec\per\pixel}$
and the detector is a $\num{4x2}$ array of $\SI{2x4}{\kilopixel}$
CCDs.

Dithered observations with total exposure times listed in \tabref{Details-of-observations}
were taken in $\Bband,\,\Vband$ and $\Rband$ bands (ESO filters
B/123, V/89, and Rc/162). The raw frames were reduced and co-added
using \esomvm. These were aligned (shifted and rotated), then cropped
to exclude regions where any band was missing using the \iraf\ task
\wregister.\noun{ }\sextractor\ \citep[version 2.8.6;][]{1996:Bertin.E;Arnouts.S:SExtractor:-Software-for-source-extraction:article}
was run on each image and the seeing measured by taking the median
full width at half maximum (FWHM) of objects in the unsaturated part
of the stellar locus in the FWHM-Magnitude diagram. Bright but unsaturated
isolated point sources, with\noun{ }\sextractor-measured stellarity
$>0.965$ in all three bands, were confirmed as point sources by eye
and used to calculate convolution kernels with which to degrade each
set of images to a common seeing for photometry measurements using
the \iraf\ task \psfmatch. The final seeing was equal to the worst
seeing in each set of three images. Star-galaxy separation was performed
on the original, non-PSF-matched images.

\begin{table*}
\protect\caption{\label{tab:Details-of-observations}Details of optical observations.
Seeing was measured from the images, and the observation date refers
to the date when the observations were started. $\offtargetarea{500}$
and $\offtargetarea{200}$ are the areas of the regions outside $\rfl{500}$
and $\rfl{200}$ which can be used for background estimation. }
\makeatletter{}
\begin{tabular}{llllllllllll}
\hline
Object & ${R}_{\mathrm{exp}}$ & ${V}_{\mathrm{exp}}$ & ${B}_{\mathrm{exp}}$ & ${R}_{\mathrm{see}}$ & ${V}_{\mathrm{see}}$ & ${B}_{\mathrm{see}}$ & ${R}_{\mathrm{date}}$ & ${V}_{\mathrm{date}}$ & ${B}_{\mathrm{date}}$ & $\offtargetarea{500}$ & $\offtargetarea{200}$ \\
 & \ensuremath{\si{\hour}} & \ensuremath{\si{\hour}} & \ensuremath{\si{\hour}} & \ensuremath{\si{\arcsecondword}} & \ensuremath{\si{\arcsecondword}} & \ensuremath{\si{\arcsecondword}} &  &  &  & \ensuremath{\si{\arcminuteword\tothe{2}}} & \ensuremath{\si{\arcminuteword\tothe{2}}} \\
\hline
RXC\,J0006.0--3443 & 0.50 & 0.50 & 0.75 & 0.99 & 0.96 & 1.05 & 2008--09--24 & 2008--09--24 & 2008--09--24 & 647.6 & 416.7 \\
RXC\,J0049.4--2931 & 0.50 & 0.61 & 1.25 & 0.99 & 0.97 & 1.03 & 2008--09--23 & 2008--09--23 & 2008--09--23 & 711.4 & 561.5 \\
RXC\,J0345.7--4112 & 0.50 & 0.50 & 0.75 & 0.88 & 0.97 & 1.03 & 2007--11--19 & 2007--11--19 & 2007--11--19 & 602.5 & 268.0 \\
RXC\,J0547.6--3152 & 0.50 & 1.00 & 0.50 & 0.76 & 1.04 & 0.95 & 2000--01--19 & 2000--01--18 & 2007--11--14 & 454.7 & 315.9 \\
RXC\,J0605.8--3518 & 0.50 & 0.50 & 1.37 & 0.76 & 0.78 & 0.92 & 2007--11--27 & 2007--11--27 & 2007--11--15 & 730.6 & 586.3 \\
RXC\,J0616.8--4748 & 0.50 & 0.50 & 0.75 & 0.93 & 1.02 & 1.08 & 2007--11--15 & 2007--11--15 & 2007--11--15 & 740.9 & 572.4 \\
RXC\,J0645.4--5413 & 0.50 & 0.50 & 0.75 & 1.06 & 1.08 & 1.19 & 2000--01--03 & 2000--01--03 & 2000--01--03 & 765.0 & 606.2 \\
RXC\,J0821.8+0112 & 0.83 & 0.58 & 0.62 & 0.95 & 0.98 & 1.12 & 2007--11--15 & 2007--11--16 & 2007--11--16 & 633.1 & 414.1 \\
RXC\,J2023.0--2056 & 0.56 & 0.67 & 1.00 & 0.90 & 1.01 & 1.19 & 2008--07--01 & 2008--07--01 & 2008--07--01 & 462.3 & 85.8 \\
RXC\,J2048.1--1750 & 0.50 & 0.50 & 0.75 & 0.78 & 0.90 & 0.99 & 2008--07--01 & 2008--07--01 & 2008--07--03 & 667.7 & 526.6 \\
RXC\,J2129.8--5048 & 0.50 & 0.50 & 0.75 & 1.21 & 1.25 & 1.40 & 2008--06--30 & 2008--06--30 & 2008--06--30 & 574.0 & 249.2 \\
RXC\,J2218.6--3853 & 0.50 & 0.50 & 0.87 & 1.18 & 1.15 & 1.04 & 2008--09--20 & 2008--09--20 & 2008--09--20 & 744.9 & 573.4 \\
RXC\,J2234.5--3744 & 0.50 & 0.50 & 0.75 & 1.11 & 1.22 & 1.41 & 2008--06--30 & 2008--06--30 & 2008--06--30 & 674.8 & 477.6 \\
RXC\,J2319.6--7313 & 0.44 & 0.56 & 0.75 & 1.07 & 1.06 & 1.15 & 2008--09--21 & 2008--09--21 & 2008--09--21 & 733.2 & 562.3 \\
\hline
\end{tabular}
 
\end{table*}

Examples of stacked, flat-fielded images where stars have been excised
are shown in Figures \ref{fig:RXC-J0006.0-3443-in-R}, \ref{fig:RXC-J0616.8-4748-in-R},
\ref{fig:RXC-J2234.5-3744-in-R} and \ref{fig:RXC-J0006.0-3443-in-R-Detail}
(online material).

\subsection{Subsamples}

\label{sec:Subsamples}

We use subsamples of the 14 objects in our catalogue, based on their
X-ray parameters as given in \tabref{Overview-of-the-clusters}. We
use the morphological classifications of \citet[§2.3]{2009:Pratt.G;Croston.J;Arnaud.M:Galaxy-cluster-X-ray-luminosity-scaling-relations-from-a-representative-local-sample:article}.

`\Massive' objects are those with $\massl{500}$ above the median
of the entire \REXCESS\ population \protect\makeatletter{}
\ensuremath{\SI{2.95e+14}{\Msun}}
 \unskip. 

`\Coolcore' objects have central electron density $h(z)^{-2}\, n_{e,0}>\SI{4e2}{\cm}$
and have central cooling times $\SI{<1e9}{\year}$ \citep[§2.3.1]{2009:Pratt.G;Croston.J;Arnaud.M:Galaxy-cluster-X-ray-luminosity-scaling-relations-from-a-representative-local-sample:article}. 

`\Disturbed' objects are classified based on their X-ray centroid
shifts $w_{i}$; $w_{i}=\nicefrac{d_{i}}{r_{500}}$ for $d_{i}$ the
projected separation between the X-ray peak and centroid in apertures
with radii in the range \SIrange{0.1}{1}{\rfhsi}. If the standard
deviation $\left\langle w_{i}\right\rangle $ is above the threshold
value $\SI{0.01}{\rfhsi}$, the object is classified as disturbed.
A detailed description of the determination of this morphological
parameter is given in \citet[§2.4]{2010:Bohringer.H;Pratt.G;Arnaud.M:Substructure-of-the-galaxy-clusters-in-the-REXCESS-sample:-observed-statistics-and-comparison-to-numerical-simulations:article}.


\makeatletter{}

\section{Analysis}

\label{sec:Analysis}

\makeatletter{}

\subsection{Source detection and classification}

Source detection and measurement was carried out on PSF-matched images
(see \secref{Optical_data}) using \sextractor\ in double image mode,
with $\Rband$ as the detection band. Groups of $\geq5$ pixels with
values $\geq2\sigma$ above the median filtered background per pixel
were treated as objects. We used\texttt{ \automag\ }in \sextractor\ for
magnitude measurements.

\begin{figure}
\includegraphics[width=8cm]{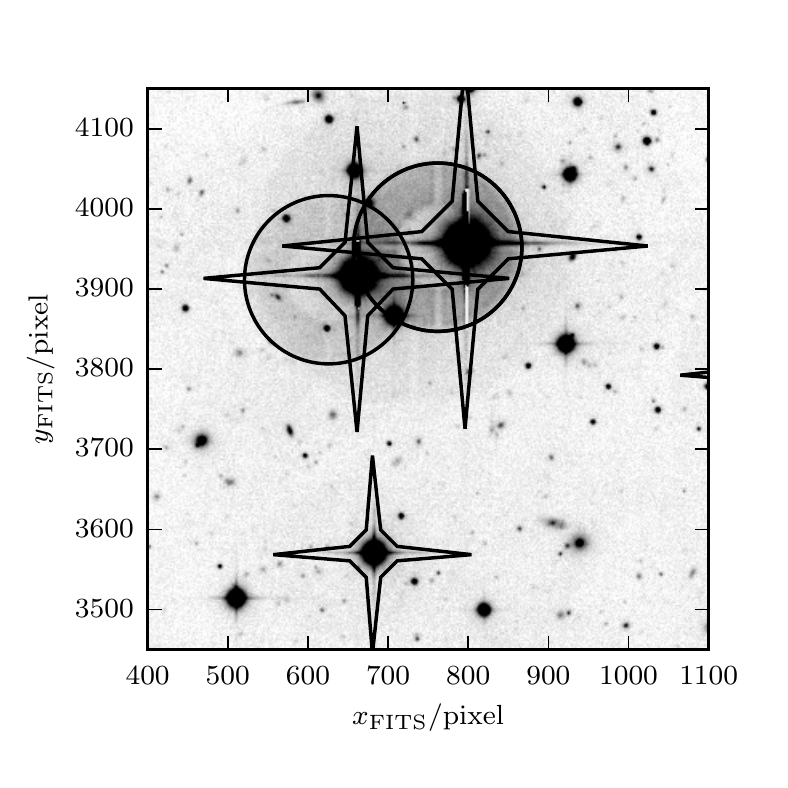}\protect\caption{\label{fig:Example-star-masks}Example masks produced by hand for
bright objects. The circles have radius $\SI{25}{\arcsec}$.}
\end{figure}

Masks were placed on stars:
\begin{itemize}
\item magnitude $\gscmagnitude<14$ in the Guide Star Catalogue 1.2 \citep{2001:Morrison.J;Roser.S;McLean.B:The-Guide-Star-Catalog-Version-1.2:-An-Astrometric-Recalibration-and-Other-Refinements:article},
\item and/or with prominent diffraction or blooming spikes,
\item and/or with prominent secondary or higher order reflections.
\end{itemize}
Initial placement of the masks was performed using a modified version
of\texttt{ \automask\ }from \theli\ \citep{2005:Erben.T;Schirmer.M;Dietrich.J:GaBoDS:-The-Garching-Bonn-Deep-Survey.-IV.-Methods-for-the-image-reduction-of-multi-chip-cameras-demonstrated:article}
and then finely positioned and scaled by hand. Examples are shown
in \figref{Example-star-masks}. Masking introduces sharp edges which
produce spurious detections not present when using unmasked images,
so objects which only appeared in masked images were ignored. When
a bright object overlaid the edge of a mask, that object was masked
individually to avoid blending it with nearby extended objects. Objects
from partially exposed regions at the edges of the images -- a result
of the telescope dithering and frame stacking -- were filtered out
of the catalogues.

\begin{figure}
\subfloat[\label{fig:Stellarity-vs-mag}Stellarity vs. detection band magnitude.]{\includegraphics[width=8cm]{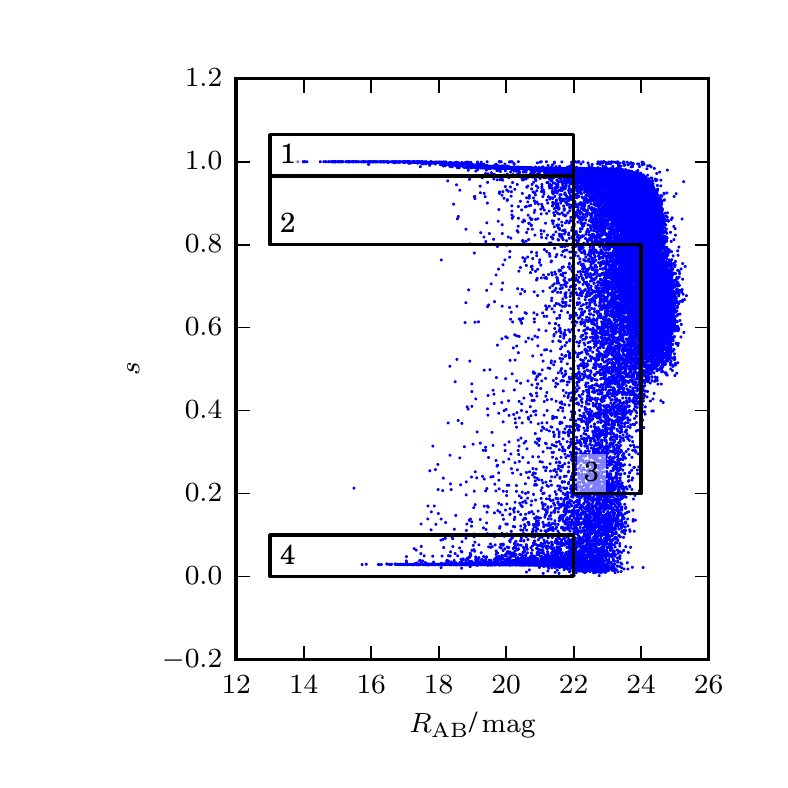}

}

\subfloat[\label{fig:Rogues-Gallery}Examples of the objects contained in the
boxes in \figref{Stellarity-vs-mag}.]{\includegraphics[width=8cm]{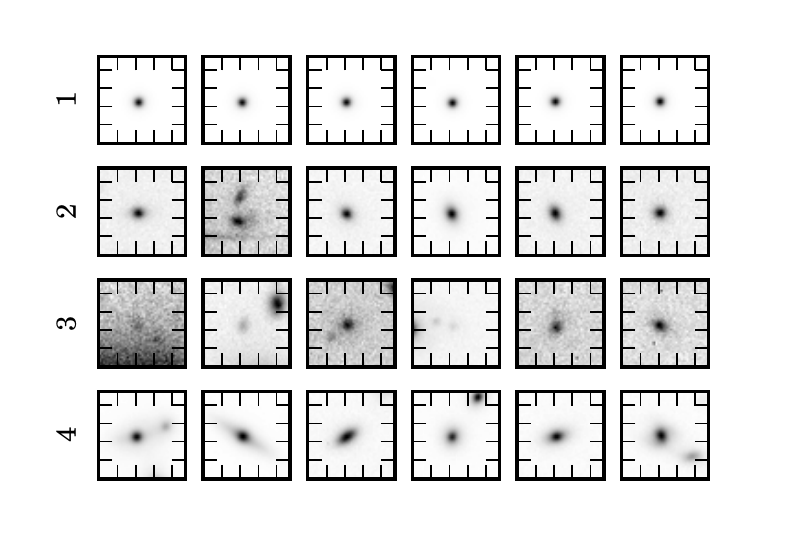}

}

\protect\caption{\label{fig:Point-source-discrimination}Point source discrimination
using \sextractor\unskip's object classifier. Any objects with $\stellarity<0.965$
are treated as galaxies. Objects in box 1 are stars, objects in box
2 were checked by eye, objects in box 3 ($\WFIABR>22$) could not
be classified and are assumed to be galaxies, and objects in box 4
are clearly extended. Stars $\WFIABR<13$ do not appear as they have
been masked (note that $\gscmagnitude\protect\neq\WFIABR$). These
examples come from the images of RXC~J0006.0\textendash 3443.}
\end{figure}

The \sextractor\ star-galaxy classifier is a neural network trained
on a sample of simulated point-source and non-point-source images
to return a stellarity $\stellarity$ in the range \numrange{0}{1},
where $1$ is a point-source and $0$ is not. We found that good star-galaxy
separation was achieved when the maximum stellarity from all three
un-degraded images was considered, and objects with $\stellarity>0.965$
in one of the bands were considered to be stars. Bright objects which
were not definitively classified $(\stellarity>0.8,\,\WFIABR<21)$
were checked by eye. An example stellarity-magnitude diagram with
images of some objects is shown in \figref{Point-source-discrimination}.
The figure shows good discrimination (a value close to either 1 or
0) of objects which are obviously point-like or extended at bright
magnitudes ($\WFIABR<20$). At dimmer magnitudes, classifications
become increasingly random. Because of the finite thickness of the
Galactic disc and a finite upper magnitude on the star population,
we expect the number of stars to drop rapidly above the magnitude
at which we can no longer properly classify stars. Objects at higher
magnitudes which are unclassified in the un-degraded images are always
assumed to be galaxies. By applying the degrading filter, the signal
to noise ratio of these objects is increased -- the spatially uncorrelated
background is suppressed, whereas spatially correlated objects on
the scale of the kernel used for degradation are enhanced -- so they
can be detected. 

\makeatletter{}

\subsection{Magnitude calibration}

Magnitudes from\noun{ }\sextractor~were converted into AB magnitudes
by making an atmospheric correction and a zero point correction. We
fixed the $\Bband$ band zero point $\Bzero$ using Data Release 7
of the AAVSO Photometric All-Sky Survey (APASS DR7) catalogue, which
has good coverage of almost all of our fields, and sufficient coverage
(>30\%) in those fields where coverage was incomplete. 

Using the observations and zero points from \citet{2012:Ziparo.F;Braglia.F;Pierini.D:In-the-whirlpools-coils:-tracing-substructure-from-combined-optical/X-ray-data-in-the-galaxy-cluster:article},
we generated a calibrated stellar locus. By minimizing the offset
of the stellar locus in each observation set from this calibrated
stellar locus, we found colour offsets $\colouroffset[\Bband-\Vband]=\Bzero-\Vzero$
and $\colouroffset[\Vband-\Rband]=\Vzero-\Rzero$. This method is
similar to that of \citet{2009:High.F;Stubbs.C;Rest.A:Stellar-Locus-Regression:-Accurate-Color-Calibration-and-the-Real-Time-Determination-of-Galaxy-Cluster:article}.
The resulting zero points are given in \tabref{Zero-points-for}.

No attempt was made to correct for galactic extinction whilst constructing
the stellar locus diagram as we found that the assumption that the
galactic dust could be modelled as a thin sheet was incorrect, and
that bluer stars (brighter and typically further away) tended to be
more reddened than redder stars (dimmer and typically closer). Instead,
we assumed that the observed stellar locus was independent of position
on the sky (at least away from the galactic plane, where all of our
targets are) and that we could cross-calibrate with similar observations.

\begin{table}
\protect\caption{\label{tab:Zero-points-for}Zero points for each of the observations.
We found $\Bzero$ using comparison to the APASS catalogue, and then
fitted the two colour offsets $\colouroffset[\Bband-\Vband]$ and
$\colouroffset[\Vband-\Rband]$ to a stellar locus from the literature
in order to constrain $\Vzero$ and $\Rzero$. The large variation
in the zero points is due the atmospheric absorption on the observation
nights; since the extinction coefficients $\extinctionterm$ were
not independently determined during observations it was impossible
to disentangle extinction from the zero point variation. }
\makeatletter{}
\begin{tabular}{llll}
\hline
Object & $\zeropoint{R}$ & $\zeropoint{V}$ & $\zeropoint{B}$ \\
 & \ensuremath{\si{\mag}} & \ensuremath{\si{\mag}} & \ensuremath{\si{\mag}} \\
\hline
RXC\,J0006.0--3443 & \ensuremath{24.37 \pm 0.02} & \ensuremath{24.17 \pm 0.02} & \ensuremath{24.70 \pm 0.02} \\
RXC\,J0049.4--2931 & \ensuremath{23.93 \pm 0.03} & \ensuremath{23.92 \pm 0.02} & \ensuremath{24.52 \pm 0.02} \\
RXC\,J0345.7--4112 & \ensuremath{24.52 \pm 0.05} & \ensuremath{24.31 \pm 0.05} & \ensuremath{24.87 \pm 0.05} \\
RXC\,J0547.6--3152 & \ensuremath{24.27 \pm 0.02} & \ensuremath{24.14 \pm 0.02} & \ensuremath{24.44 \pm 0.02} \\
RXC\,J0605.8--3518 & \ensuremath{24.38 \pm 0.02} & \ensuremath{24.11 \pm 0.02} & \ensuremath{24.61 \pm 0.02} \\
RXC\,J0616.8--4748 & \ensuremath{24.45 \pm 0.02} & \ensuremath{24.20 \pm 0.02} & \ensuremath{24.71 \pm 0.02} \\
RXC\,J0645.4--5413 & \ensuremath{24.41 \pm 0.02} & \ensuremath{24.13 \pm 0.02} & \ensuremath{24.61 \pm 0.02} \\
RXC\,J0821.8+0112 & \ensuremath{24.46 \pm 0.03} & \ensuremath{24.21 \pm 0.03} & \ensuremath{24.73 \pm 0.03} \\
RXC\,J2023.0--2056 & \ensuremath{24.31 \pm 0.03} & \ensuremath{24.10 \pm 0.03} & \ensuremath{24.65 \pm 0.03} \\
RXC\,J2048.1--1750 & \ensuremath{24.34 \pm 0.02} & \ensuremath{24.15 \pm 0.02} & \ensuremath{24.71 \pm 0.02} \\
RXC\,J2129.8--5048 & \ensuremath{24.39 \pm 0.02} & \ensuremath{24.19 \pm 0.02} & \ensuremath{24.76 \pm 0.02} \\
RXC\,J2218.6--3853 & \ensuremath{24.30 \pm 0.02} & \ensuremath{24.09 \pm 0.02} & \ensuremath{24.65 \pm 0.02} \\
RXC\,J2234.5--3744 & \ensuremath{24.35 \pm 0.02} & \ensuremath{24.16 \pm 0.02} & \ensuremath{24.69 \pm 0.02} \\
RXC\,J2319.6--7313 & \ensuremath{24.04 \pm 0.03} & \ensuremath{23.79 \pm 0.03} & \ensuremath{24.11 \pm 0.03} \\
\hline
\end{tabular}
 
\end{table}

Conversion of magnitudes and colour gradients to the Johnson system
(for comparisons with the literature) was carried out using colour
terms from the ESO WFI web page.%
\footnote{\url{https://www.eso.org/sci/facilities/lasilla/instruments/wfi/inst/zeropoints.html}
{[}Accessed: 2013-07-24{]}%
} The conversion between raw magnitudes $\WFIraw{\magnitude}$ and
Johnson magnitudes $\JC{\magnitude}$ is described by 

\begin{eqnarray}
\left(\begin{array}{c}
\WFIraw{\Rband}\\
\WFIraw{\Vband}\\
\WFIraw{\Bband}
\end{array}\right) & = & \left[\begin{array}{ccc}
1+\colourtermR & -\colourtermR & 0\\
-\colourtermV & 1+\colourtermV & 0\\
0 & -\colourtermB & 1+\colourtermB
\end{array}\right]\left(\begin{array}{c}
\JC{\Rband}\\
\JC{\Vband}\\
\JC{\Bband}
\end{array}\right)\nonumber \\
 &  & +\left(\begin{array}{c}
\airmass[\Rband]\,\extinctionterm[\Rband]\\
\airmass[\Vband]\,\extinctionterm[\Vband]\\
\airmass[\Bband]\,\extinctionterm[\Bband]
\end{array}\right)-\left(\begin{array}{c}
\Rzero\\
\Vzero\\
\Bzero
\end{array}\right)\,,\label{eq:Raw_JC_Transformation}
\end{eqnarray}
for $\zeropoint{\magnitude}$ the zero point in band $\magnitude$,
$\colourterm$ the colour term, $\airmass$ the airmass of the observation
and $\extinctionterm$ the extinction parameter. This equation can
be inverted in order to find standard magnitudes given raw magnitudes.
The conversion between raw magnitudes and AB magnitudes is described
by

\begin{eqnarray}
\left(\begin{array}{c}
\WFIAB{\Rband}\\
\WFIAB{\Vband}\\
\WFIAB{\Bband}
\end{array}\right) & = & \left(\begin{array}{c}
\WFIraw{\Rband}\\
\WFIraw{\Vband}\\
\WFIraw{\Bband}
\end{array}\right)-\left(\begin{array}{c}
\airmass[\Rband]\,\extinctionterm[\Rband]\\
\airmass[\Vband]\,\extinctionterm[\Vband]\\
\airmass[\Bband]\,\extinctionterm[\Bband]
\end{array}\right)\label{eq:AB_Raw_transformation}\\
 &  & +\left(\begin{array}{c}
\aboffset[\Rband]\\
\aboffset[\Vband]\\
\aboffset[\Bband]
\end{array}\right)+\left(\begin{array}{c}
\Rzero\\
\Vzero\\
\Bzero
\end{array}\right)\,,\nonumber 
\end{eqnarray}
for $\aboffset$ the AB offset for band $\magnitude$. The values
of the parameters are given in \tabref{Magnitude-conversion-parameters.},
and the AB corrections $\aboffset$ were taken from the ESO \magtoflux~tool%
\footnote{\url{http://archive.eso.org/mag2flux/}%
}.

\begin{table}
\protect\caption{\label{tab:Magnitude-conversion-parameters.}Magnitude conversion
parameters. The offsets $\aboffset$ were taken from the ESO \magtoflux~tool
at \protect\url{http://archive.eso.org/mag2flux/}.}
\makeatletter{}
\begin{tabular}{llll}
\hline
Band & \colourterm & \extinctionterm & \aboffset \\
 &  &  &  \\
\hline
\ensuremath{\band{R}} & \ensuremath{0.0 \pm 0} & \ensuremath{0.070 \pm 0.010} & $+0.23$ \\
\ensuremath{\band{V}} & \ensuremath{-0.13 \pm 0.01} & \ensuremath{0.11 \pm 0.01} & $+0.14$ \\
\ensuremath{\band{B}} & \ensuremath{0.25 \pm 0.02} & \ensuremath{0.22 \pm 0.01} & $-0.07$ \\
\hline
\end{tabular}
 
\end{table}

The results in the rest of this paper are based on AB system magnitudes.


\makeatletter{}

\subsection{Luminosity function analysis}

\subsubsection{Galaxy catalogue completeness from off-target observations}

\label{sec:off-target-number-counts}

The number counts of field galaxies arise from the summation of luminosity
functions at many redshifts, which themselves are influenced by galaxy
evolution. We use the number counts histogram to estimate catalogue
completeness at different magnitudes, and to compare the relative
over- or under-density of galaxies in the off-target regions of our
observations to other regions in the sky. 

\begin{figure}
\includegraphics[width=8cm]{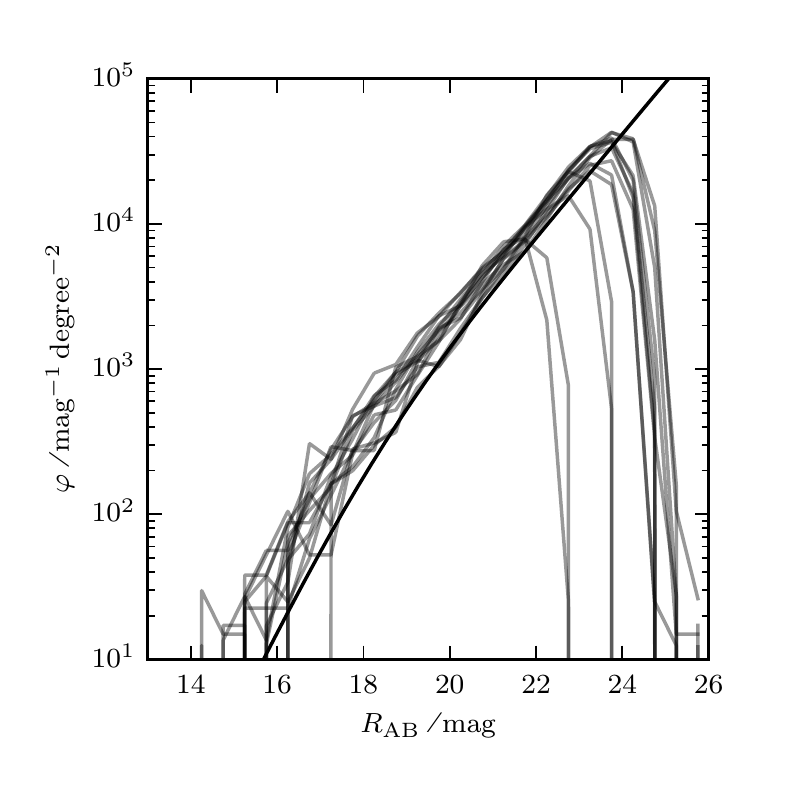}

\protect\caption{\label{fig:off-target-luminosity-function-plot}The off-target number
count surface densities $\surfacedensity$ for each cluster in the
sample are shown in grey, along with the expected number counts from
data compiled by Metcalfe in solid black (see \fnref{Metcalfe}).}
\end{figure}

Metcalfe compiled a set of number count measurements%
\footnote{\label{fn:Metcalfe}\url{http://astro.dur.ac.uk/~nm/pubhtml/counts/counts.html}
Metcalfe, (2010) {[}Accessed: 13-05-2013{]}%
} from \citet{2004MNRAS.354..991B,2003:Frith.W;Busswell.G;Fong.R:The-local-hole-in-the-galaxy-distribution:-evidence-from-2MASS:article,1991:Metcalfe.N;Shanks.T;Fong.R:Galaxy-number-counts.-II---CCD-observations-to-B--25-mag:article,1991:Jones.L;Fong.R;Shanks.T:Galaxy-number-counts.-I---Photographic-observations-to-B--23.5-mag:article,2001:Metcalfe.N;Shanks.T;Campos.A:Galaxy-number-counts---V.-Ultradeep-counts:-the-Herschel-and-Hubble-Deep-Fields:article,2000:McCracken.H;Metcalfe.N;Shanks.T:Galaxy-number-counts---IV.-Surveying-the-Herschel-Deep-Field-in-the-near-infrared:article,1995:Metcalfe.N;Shanks.T;Fong.R:Galaxy-number-counts---III.-Deep-CCD-observations-to-B27.5-mag:article,1996:Metcalfe.N;Shanks.T;Campos.A:Galaxy-formation-at-high-redshifts:article,2005:Metcalfe.N;Shanks.T;Weilbacher.P:An-H-band-survey-of-the-Herschel-Deep-Field:article}.
The dataset has an intrinsic scatter of $\SI{\sim0.3}{\dex}$, corresponding
to a factor of $\sim2.$ We use a 5th order polynomial spline fit
$\metcalfegalaxycountfunction\left(\Rband\right)$ as an empirical
shape for our field number counts histogram fitting. The number count
surface densities $\surfacedensity$ for all of the off-target regions
in the sample are plotted in \figref{off-target-luminosity-function-plot},
along with $\metcalfegalaxycountfunction$.

We assume that the fall-off at the magnitude limit can be described
by the logistic function $\fallofffunction\left(\magnitude\right)=\left(1+e^{\frac{\magnitude-\falloffmagnitude}{\falloffwidth}}\right)^{-1}$.
This function was chosen since it goes smoothly from $\sim1$ to $\sim0$
over a characterisable distance, but we make no claim that it precisely
describes the fall-off. The 50\% completeness limit is at $\falloffmagnitude$
in this model.

We fit the function $\combinedgalaxycountfunction=\galaxydensityfactor\,\metcalfegalaxycountfunction\,\fallofffunction$
where $\galaxydensityfactor$ is a normalisation factor, to the off-target
number counts histogram for each cluster in our sample (measured in
region $r>\SI{1.5}{\rfhsi}$), and the results are given in \tabref{Off-target-luminosity-function-fit_results}.
We also include the fall-off magnitude as a K-corrected absolute magnitude
at the cluster redshift, $\falloffmagnitude[\WFIABabsk{R}]$. The
K-corrections are made using the data of \citet{1997:Poggianti.B:K-and-evolutionary-corrections-from-UV-to-IR:article},
assuming that the galaxies are E-galaxies. 

The observed galaxy counts were all between $1\times$ and $2\times$
the empirically determined density, consistent with our galaxy clusters
occupying denser regions of the cosmic web, and within the expected
bounds of the scatter from Metcalfe's galaxy counts dataset. In addition,
the shapes of the off-target number counts in \figref{off-target-luminosity-function-plot}
do not closely trace the curve expected from the literature, suggesting
that the galaxy overdensity of the cluster extends beyond $\rfh$.

\begin{table*}
\protect\caption{\label{tab:Off-target-luminosity-function-fit_results}Off-target
number counts fit results. $\falloffmagnitude[\WFIAB{R}]$ is the
detection band falloff magnitude in the observer's magnitude system
and $\falloffmagnitude[\WFIABabsk{\Rband}]$ is the corresponding
K-corrected absolute magnitude. $\galaxydensityfactor$ is the normalisation
factor by which the literature number counts function was multiplied.}

\makeatletter{}
\begin{tabular}{lllll}
\hline
Object & \ensuremath{\falloffmagnitude[\WFIAB{R}]} & \ensuremath{\falloffmagnitude[\WFIABabsk{R}]} & \ensuremath{\falloffwidth} & \ensuremath{\galaxydensityfactor[\WFIAB{R}]} \\
 & \ensuremath{\si{\mag}} & \ensuremath{\si{\mag}} & \ensuremath{\si{\mag}} &  \\
\hline
RXC\,J0006.0--3443 & \ensuremath{24.07 \pm 0.03} & \ensuremath{-14.56 \pm 0.03} & \ensuremath{0.153 \pm 0.014} & \ensuremath{1.49 \pm 0.06} \\
RXC\,J0049.4--2931 & \ensuremath{23.10 \pm 0.03} & \ensuremath{-15.40 \pm 0.03} & \ensuremath{0.157 \pm 0.014} & \ensuremath{1.33 \pm 0.05} \\
RXC\,J0345.7--4112 & \ensuremath{24.40 \pm 0.04} & \ensuremath{-12.76 \pm 0.04} & \ensuremath{0.150 \pm 0.011} & \ensuremath{1.16 \pm 0.06} \\
RXC\,J0547.6--3152 & \ensuremath{23.34 \pm 0.03} & \ensuremath{-15.90 \pm 0.03} & \ensuremath{0.150 \pm 0.014} & \ensuremath{1.39 \pm 0.06} \\
RXC\,J0605.8--3518 & \ensuremath{23.99 \pm 0.03} & \ensuremath{-15.10 \pm 0.03} & \ensuremath{0.154 \pm 0.011} & \ensuremath{1.53 \pm 0.06} \\
RXC\,J0616.8--4748 & \ensuremath{23.74 \pm 0.02} & \ensuremath{-14.93 \pm 0.02} & \ensuremath{0.150 \pm 0.008} & \ensuremath{1.37 \pm 0.05} \\
RXC\,J0645.4--5413 & \ensuremath{23.74 \pm 0.03} & \ensuremath{-15.74 \pm 0.03} & \ensuremath{0.151 \pm 0.011} & \ensuremath{1.17 \pm 0.05} \\
RXC\,J0821.8+0112 & \ensuremath{24.07 \pm 0.03} & \ensuremath{-13.79 \pm 0.03} & \ensuremath{0.146 \pm 0.012} & \ensuremath{1.27 \pm 0.05} \\
RXC\,J2023.0--2056 & \ensuremath{24.34 \pm 0.03} & \ensuremath{-12.67 \pm 0.03} & \ensuremath{0.158 \pm 0.009} & \ensuremath{1.26 \pm 0.05} \\
RXC\,J2048.1--1750 & \ensuremath{24.25 \pm 0.02} & \ensuremath{-14.98 \pm 0.02} & \ensuremath{0.142 \pm 0.009} & \ensuremath{1.46 \pm 0.05} \\
RXC\,J2129.8--5048 & \ensuremath{24.02 \pm 0.04} & \ensuremath{-13.77 \pm 0.04} & \ensuremath{0.152 \pm 0.012} & \ensuremath{1.10 \pm 0.05} \\
RXC\,J2218.6--3853 & \ensuremath{22.25 \pm 0.03} & \ensuremath{-16.87 \pm 0.03} & \ensuremath{0.159 \pm 0.014} & \ensuremath{1.19 \pm 0.05} \\
RXC\,J2234.5--3744 & \ensuremath{24.11 \pm 0.03} & \ensuremath{-15.17 \pm 0.03} & \ensuremath{0.150 \pm 0.010} & \ensuremath{1.39 \pm 0.05} \\
RXC\,J2319.6--7313 & \ensuremath{21.93 \pm 0.05} & \ensuremath{-16.35 \pm 0.05} & \ensuremath{0.143 \pm 0.014} & \ensuremath{1.71 \pm 0.10} \\
\hline
\end{tabular}
 
\end{table*}

\subsubsection{Initial cluster luminosity function analysis}

\label{sec:cluster_luminosity_function_analysis}

\begin{figure}
\includegraphics[width=8cm]{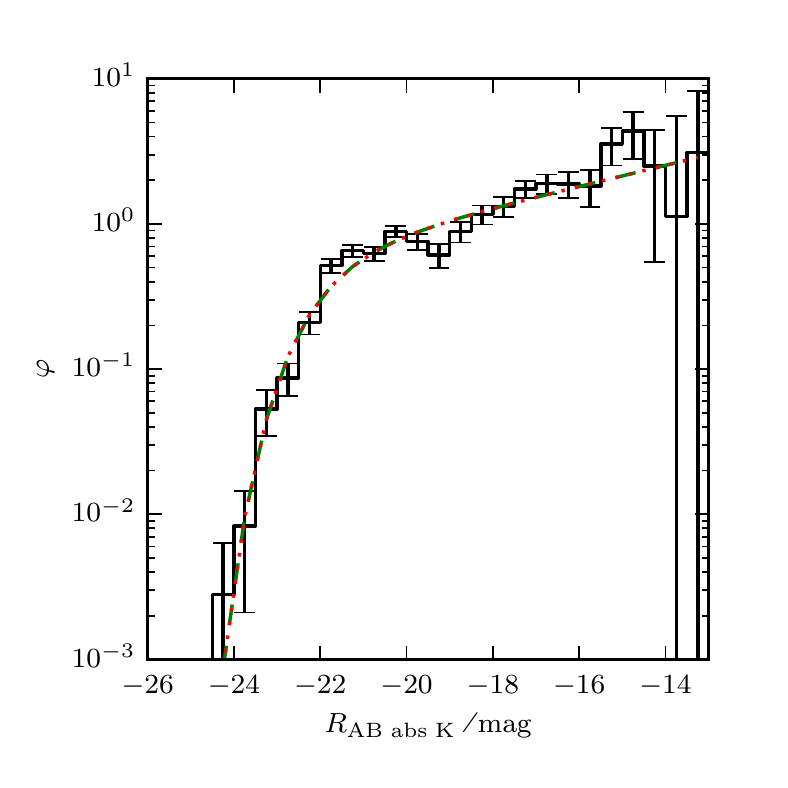}

\protect\caption{\label{fig:Mean-luminosity-function}Mean cluster luminosity function
in the detection band and best fitting Schechter function model. Before
stacking, the off-target number count density was subtracted, and
each profile divided by the remaining object count in $-21<\WFIABabsk{R}<-17$.}
\end{figure}

For each cluster, we measured the on-target luminosity function in
the region $r<\rfh$ in the $\Rband$ band, subtracting the off-target
number counts histogram measured from the region $r>\SI{1.5}{\rfhsi}$.
Each function was normalised by the mean density in $-21<\Rband<-17$.
We truncated each function well below the 50\% completeness limit,
at $\falloffmagnitude-4\falloffwidth$ as fitted in the off-target
region (see \secref{off-target-number-counts}). We fitted a Schechter
function $\schechterfunction(\luminosity)=\schechternormalisation\left(\frac{\luminosity}{\schechterluminosity}\right)^{\schechterslope}\exp\left(\frac{-\luminosity}{\schechterluminosity}\right)$
for $\schechterluminosity$ the Schechter luminosity (with the corresponding
magnitude $\mschechter{\magnitude}$), $\schechterslope$ the slope
parameter and $\schechternormalisation$ a normalisation factor to
the mean luminosity function. The function and its fit for the detection
band are shown in \figref{Mean-luminosity-function}, and the fitting
results for all three bands are given in \tabref{Schechter-function-fitting}.

\begin{table}
\protect\caption{\label{tab:Schechter-function-fitting}Schechter function fitting
results, valid for the MPG/ESO $\SI{2.2}{\metre}$ Telescope WFI filters.}
\makeatletter{}
\begin{tabular}{llll}
\hline
Band & \schechter{\magnitude} & \schechterslope & \redchi \\
 & \ensuremath{\si{\mag}} &  &  \\
\hline
\ensuremath{\WFIABabsk{R}} & \ensuremath{-22.26 \pm 0.17} & \ensuremath{-1.18 \pm 0.04} & 2.01 \\
\ensuremath{\WFIABabsk{V}} & \ensuremath{-21.79 \pm 0.24} & \ensuremath{-1.29 \pm 0.06} & 3.43 \\
\ensuremath{\WFIABabsk{B}} & \ensuremath{-20.87 \pm 0.28} & \ensuremath{-1.22 \pm 0.09} & 3.67 \\
\hline
\end{tabular}
 
\end{table}

Given the detection band Schechter magnitude $\mschechter{\Rband}$,
we define three groups of galaxies in each cluster: `bright' galaxies
are those satisfying $\left(\Rband<\brightcut\right)$, i.e. $\left(\Rluminosity/\schechter{\Rluminosity}>0.1\right)$;
`faint' galaxies satisfy $\left(\brightcut<\Rband<\faintcut\right)$,
i.e. $\left(0.1>\Rluminosity/\schechter{\Rluminosity}>0.01\right)$;
and `dwarf' galaxies satisfy $\left(\Rband>\faintcut\right)$, i.e.
$\left(\Rluminosity/\schechter{\Rluminosity}<0.01\right)$. Additionally,
we define the symbols $\brightcutsymbol=\brightcut$ and $\faintcutsymbol=\faintcut$.

The dwarf galaxy population (described in, e.g. \citealp{2006:Popesso.P;Biviano.A;Bohringer.H:RASS-SDSS-Galaxy-cluster-survey.-IV.-A-ubiquitous-dwarf-galaxy-population-in-clusters:article})
is not always obvious in our data due to different image depths in
different observations. The detection band data are sufficiently deep
to reach the faint limit, $\faintcut$, so in the rest of this paper
we use the `bright' and `faint' galaxies, ignoring the dwarfs. 

By imposing a magnitude cut at $\faintcut$, we lose a fraction of
the total luminosity in each cluster. By integrating the fitted luminosity
functions beyond the cut and extrapolating for very faint objects
we can estimate the ratio between the total luminosity in galaxies
we observe and the total integrated luminosity in the luminosity function.
Luminosities based on the `faint' cut need to be increased by \makeatletter{}
\ensuremath{\SI{2.5e+00}{\%}}
 ~in
$\Rband$, \makeatletter{}
\ensuremath{\SI{4.3e+00}{\%}}
 ~in
$\Vband$ and \makeatletter{}
\ensuremath{\SI{3.1e+00}{\%}}
 ~in
$\Bband$. 

A more exhaustive assessment of the luminosity functions, as well
as total luminosity measurements informed by the count density profiles
is given in \secref{exhausive_luminosity_measurements}.


\makeatletter{}

\subsection{Catalogue contamination by misidentified stars}

We estimate an upper bound on the number of stars at each magnitude
in our catalogues by fitting star counts with respect to magnitude
at magnitudes where star-galaxy separation is robust, and extrapolating
this to higher magnitudes. (These count estimates are upper bounds
since the star counts drop off at a faster rate at higher magnitudes.)
The number of stars which are expected given the power law, but not
seen, are assumed to have been misidentified as galaxies. The contamination
fraction $\cataloguecontamination$ measured in each catalogue at
selected magnitudes is given in \tabref{Contamination-statistics}.

\begin{table*}
\protect\caption{\label{tab:Contamination-statistics}Contamination fraction for each
catalogue measured at the bright limit, the faint limit, and the 50\%
detection limit. }

\makeatletter{}
\begin{tabular}{llll}
\hline
Object & \ensuremath{\cataloguecontamination_{\bright}} & \ensuremath{\cataloguecontamination_{\brightfaint}} & \ensuremath{\cataloguecontamination_{\falloffmagnitude}} \\
 &  &  &  \\
\hline
RXC\,J0006.0--3443 & \ensuremath{-0.00 \pm 0.05} & \ensuremath{0.004 \pm 0.014} & \ensuremath{0.003 \pm 0.005} \\
RXC\,J0049.4--2931 & \ensuremath{-0.00 \pm 0.08} & \ensuremath{-0.035 \pm 0.016} & \ensuremath{-0.002 \pm 0.008} \\
RXC\,J0345.7--4112 & \ensuremath{0.00 \pm 0.27} & \ensuremath{-0.01 \pm 0.05} & \ensuremath{-0.005 \pm 0.004} \\
RXC\,J0547.6--3152 & \ensuremath{-0.00 \pm 0.15} & \ensuremath{0.112 \pm 0.027} & \ensuremath{0.154 \pm 0.017} \\
RXC\,J0605.8--3518 & \ensuremath{-0.00 \pm 0.14} & \ensuremath{0.268 \pm 0.013} & \ensuremath{0.319 \pm 0.004} \\
RXC\,J0616.8--4748 & \ensuremath{-0.00 \pm 0.21} & \ensuremath{0.208 \pm 0.022} & \ensuremath{0.270 \pm 0.007} \\
RXC\,J0645.4--5413 & \ensuremath{0.02 \pm 0.09} & \ensuremath{0.331 \pm 0.010} & \ensuremath{0.365 \pm 0.007} \\
RXC\,J0821.8+0112 & \ensuremath{0.00 \pm 0.18} & \ensuremath{0.223 \pm 0.025} & \ensuremath{0.324 \pm 0.004} \\
RXC\,J2023.0--2056 & \ensuremath{0.0 \pm 0.5} & \ensuremath{0.15 \pm 0.09} & \ensuremath{0.474 \pm 0.004} \\
RXC\,J2048.1--1750 & \ensuremath{0.13 \pm 0.05} & \ensuremath{0.503 \pm 0.005} & \ensuremath{0.5358 \pm 0.0033} \\
RXC\,J2129.8--5048 & \ensuremath{0.00 \pm 0.32} & \ensuremath{0.05 \pm 0.06} & \ensuremath{0.183 \pm 0.005} \\
RXC\,J2218.6--3853 & \ensuremath{-0.00 \pm 0.15} & \ensuremath{0.368 \pm 0.011} & \ensuremath{0.277 \pm 0.021} \\
RXC\,J2234.5--3744 & \ensuremath{-0.02 \pm 0.04} & \ensuremath{-0.057 \pm 0.011} & \ensuremath{0.017 \pm 0.006} \\
RXC\,J2319.6--7313 & \ensuremath{-0.00 \pm 0.20} & \ensuremath{0.06 \pm 0.04} & \ensuremath{0.222 \pm 0.015} \\
\hline
Mean & \ensuremath{0.01 \pm 0.06} & \ensuremath{0.155 \pm 0.010} & \ensuremath{0.2241 \pm 0.0025} \\
\hline
\end{tabular}
 
\end{table*}


\makeatletter{}

\subsection{Red sequence selection}

\label{sec:red_sequence_selection}

The red sequence is a line in colour-magnitude space around which
elliptical galaxies in clusters tend to scatter (e.g. \citealp*{1992:Bower.R;Lucey.J;Ellis.R:Precision-Photometry-of-Early-Type-Galaxies-in-the-Coma-and-Virgo-Clusters---a-Test-of-the-Universality:article};
\citealp{2011:Valentinuzzi.T;Poggianti.B;Fasano.G:The-red-sequence-of-72-WINGS-local-galaxy-clusters:article}).
The origin of the red sequence and its relation to the mass-metallicity
relation was explored in a seminal paper by \citet{1987:Arimoto.N;Yoshii.Y:Chemical-and-photometric-properties-of-a-galactic-wind-model-for-elliptical-galaxies:article}.
The position of the line changes with redshift, and can be used to
detect new galaxy clusters and estimate cluster redshifts (e.g. \citealp{2000:Gladders.M;Yee.H:A-New-Method-For-Galaxy-Cluster-Detection.-I.-The-Algorithm:article}).
Spiral galaxies in clusters tend to be bluer than the red sequence
and migrate on to it as star formation fades. The scatter of field
galaxies in colour space is usually much larger than the scatter of
the red sequence galaxies and in this study we use this observation
to increase the signal to noise ratio of our radial profiles.

We fit a line to the red sequence described by $\cmrmodel\left(\magnitude\right)=\cmrgradient\,(\magnitude-\cmrpivot)+\cmrintercept$
where $\colour$ is a colour, $\cmrgradient$ is the gradient, $\magnitude$
is a magnitude, and $\cmrintercept$ is the colour at a `pivot point'
$\cmrpivot$. The distribution of cluster galaxies perpendicular to
the red sequence comprises a red sequence of galaxies with a relatively
narrow scatter ($\SI{\leq0.05}{\mag}$ in this case) centred on the
line, and a blue cloud of galaxies with a larger scatter ($\SI{\sim0.5}{\mag}$)
centred some distance below the red sequence. We model this as a distribution
of the form 
\begin{equation}
\scattermodel(\cmrresidual)=\scattermodelrednorm\, e^{-\frac{\left(\cmrresidual-\redseqoffset\right)^{2}}{2\redseqwidth^{2}}}+\scattermodelbluenorm\, e^{-\frac{\left(\cmrresidual-\bluecloudoffset\right)^{2}}{2\bluecloudwidth^{2}}}\,,\label{eq:red_sequence_Scatter_model}
\end{equation}
with $\cmrresidual\left(\colour,\magnitude\right)=\colour-\cmrmodel\left(\magnitude\right)$,
where $\redseqoffset$ is the offset of the distribution of red sequence
galaxies from the line \foreignlanguage{english}{$\cmrmodel$}, $\redseqwidth$
the width of the red sequence, and $\scattermodelrednorm$ the density
of the red sequence; $\bluecloudoffset$, $\bluecloudwidth$ and $\scattermodelbluenorm$
are equivalent quantities for the blue cloud. For well fitted red
sequences, $\redseqoffset/\redseqwidth$ should be small ($\ll1$).
We also define 
\begin{equation}
\rs=\frac{\cmrresidual-\redseqoffset}{\redseqwidth}\,,
\end{equation}
which is the scaled displacement of a point from the red sequence
in units of the width of the red sequence; galaxies with $-3<\rs<3$
are taken to be on the red sequence and those with $\rs<-3$ are blue
cloud objects.

We consider red sequences in the $\left(\Rband,\Bband-\Vband\right)=(\magnitude,\colour)$
colour-magnitude space (in the AB system), and fit to the galaxies
satisfying $\Rband<20$ and $r<\rfh$. 

\begin{figure*}
\includegraphics[width=17.2cm]{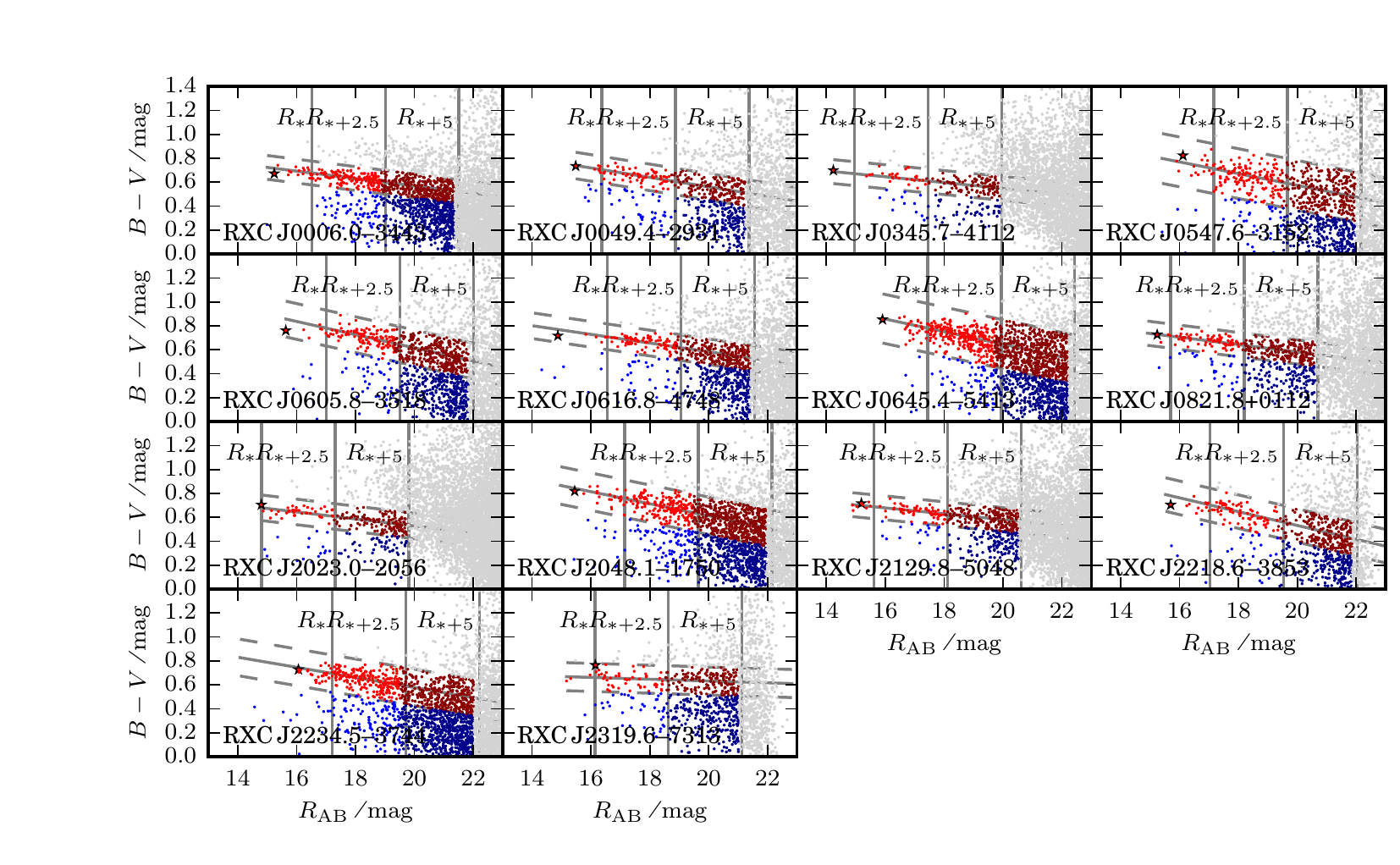}

\protect\caption{\label{fig:Fitted-red-sequences}Fitted red sequences for all the
clusters in the sample. The red sequence best-fitting line is shown
as a solid line, and the dashed lines show $\pm3\rs$. The vertical
lines show the Schechter magnitude, $\mschechter{\Rband}$, and the
bright and faint magnitude limits, $\brightcutsymbol$ and $\faintcutsymbol$
respectively. The star denotes the BCG. Objects below the lower dashed
line are `blue,' and objects between the dashed lines are `on the
red sequence.'}
\end{figure*}

The fitting procedure was:
\begin{enumerate}
\item \label{enu:first}\label{enu:point-selection}Fit of the straight
line $\cmrmodel$ to the colour-magnitude points, using a least squares
method. (On the first iteration, this step was skipped, and we assumed
a gradient of $-0.044$.) 
\item \label{enu:iterate_fitting} Compute the histogram of $\cmrresidual$
for all points within the magnitude limits. 
\item Fit of the scatter model $\scattermodel$ to the histogram.
\item \label{enu:last}Select points $\left|\rs\right|<3$ and use these
as the input for the straight line fit \enuref{first}.
\end{enumerate}
Steps \enuref{first} to \enuref{last} were repeated enough times
that the results stabilized ($\sim5$ iterations was usually sufficient).
In cases where the solutions oscillated, the solution with the narrowest
$\redseqwidth$ was selected. The fit results are shown in \figref{Fitted-red-sequences}. 

To compare our red sequence parameters to those in the literature
we refitted $\cmrgradient$ and $\cmrintercept$ in $\left(\Rband,\Bband-\Rband\right)$
and $\left(\Rband,\Vband-\Rband\right)$ using the same method, before
using (\ref{eq:Raw_JC_Transformation}) and (\ref{eq:AB_Raw_transformation})
to transform $\cmrgradient$ and $\cmrintercept$ into the Johnson-Cousins
$(\VJband,\BJband-\VJband)$ colour-magnitude space. K-corrections
were made assuming that red sequence objects are elliptical galaxies,
using values from \citet{1997:Poggianti.B:K-and-evolutionary-corrections-from-UV-to-IR:article}.
The parameters are shown in \tabref{Red-sequence-parameters.}. \citet[fig. 3]{2011:Valentinuzzi.T;Poggianti.B;Fasano.G:The-red-sequence-of-72-WINGS-local-galaxy-clusters:article}
give values for the K-corrected red sequence normalisation $\JCk{\cmrintercept}$
and gradient $\JC{\cmrgradient}$ for $72$ clusters in the $(\VJband,\BJband-\VJband)$
colour-magnitude space. The mean values are $\mean{\JCk{\cmrintercept}}=0.87\pm0.06$
(at $\Vabsband=-20$) and $\mean{\JC{\cmrgradient}}=-0.044\pm0.009$.
These values are in reasonable agreement with our results in \tabref{Red-sequence-parameters.}. 

\begin{table*}
\protect\caption{\label{tab:Red-sequence-parameters.}Red sequence parameters in the
Johnson-Cousins colour-magnitude space $(\VJband,\BJband-\VJband)$,
where $\JC{\cmrgradient}$ is the gradient, and $\JC{\cmrintercept}=\VJband,\BJband-\VJband$
at $\Vabsband=-20$. $\JCk{\cmrintercept}$ are compensated for evolution
and redshift dependence on colour using values from \citet{1997:Poggianti.B:K-and-evolutionary-corrections-from-UV-to-IR:article}
assuming that the red-sequence comprises of E-galaxies.}
\makeatletter{}
\begin{tabular}{llll}
\hline
Object & \JC{\rsgradient} & \JC{\rsintercept} & \JCk{\rsintercept} \\
 &  & \ensuremath{\si{\mag}} & \ensuremath{\si{\mag}} \\
\hline
RXC\,J0006.0--3443 & \ensuremath{-0.063 \pm 0.024} & \ensuremath{1.12 \pm 0.16} & \ensuremath{0.74 \pm 0.16} \\
RXC\,J0049.4--2931 & \ensuremath{-0.035 \pm 0.031} & \ensuremath{1.23 \pm 0.16} & \ensuremath{0.87 \pm 0.16} \\
RXC\,J0345.7--4112 & \ensuremath{-0.053 \pm 0.068} & \ensuremath{0.995 \pm 0.155} & \ensuremath{0.84 \pm 0.16} \\
RXC\,J0547.6--3152 & \ensuremath{-0.059 \pm 0.015} & \ensuremath{1.18 \pm 0.16} & \ensuremath{0.71 \pm 0.16} \\
RXC\,J0605.8--3518 & \ensuremath{-0.077 \pm 0.017} & \ensuremath{1.11 \pm 0.16} & \ensuremath{0.66 \pm 0.16} \\
RXC\,J0616.8--4748 & \ensuremath{-0.045 \pm 0.022} & \ensuremath{0.963 \pm 0.150} & \ensuremath{0.57 \pm 0.15} \\
RXC\,J0645.4--5413 & \ensuremath{-0.071 \pm 0.012} & \ensuremath{1.15 \pm 0.16} & \ensuremath{0.64 \pm 0.16} \\
RXC\,J0821.8+0112 & \ensuremath{-0.028 \pm 0.010} & \ensuremath{1.11 \pm 0.15} & \ensuremath{0.85 \pm 0.15} \\
RXC\,J2023.0--2056 & \ensuremath{-0.038 \pm 0.042} & \ensuremath{1.10 \pm 0.15} & \ensuremath{0.92 \pm 0.15} \\
RXC\,J2048.1--1750 & \ensuremath{-0.064 \pm 0.021} & \ensuremath{1.03 \pm 0.16} & \ensuremath{0.56 \pm 0.16} \\
RXC\,J2129.8--5048 & \ensuremath{-0.032 \pm 0.009} & \ensuremath{1.12 \pm 0.15} & \ensuremath{0.86 \pm 0.15} \\
RXC\,J2218.6--3853 & \ensuremath{-0.055 \pm 0.059} & \ensuremath{1.22 \pm 0.19} & \ensuremath{0.77 \pm 0.19} \\
RXC\,J2234.5--3744 & \ensuremath{-0.066 \pm 0.007} & \ensuremath{0.997 \pm 0.143} & \ensuremath{0.51 \pm 0.14} \\
RXC\,J2319.6--7313 & \ensuremath{+0.007 \pm 0.034} & \ensuremath{1.15 \pm 0.16} & \ensuremath{0.83 \pm 0.16} \\
\hline
Mean & \ensuremath{-0.049 \pm 0.009} & \ensuremath{1.11 \pm 0.04} & \ensuremath{0.74 \pm 0.04} \\
St. dev. & \ensuremath{+0.022 \pm 0.008} & \ensuremath{0.0828 \pm 0.0441} & \ensuremath{0.13 \pm 0.04} \\
\hline
\end{tabular}
 
\end{table*}

The red sequence for RXC~J2319.6--7313 is unclear and achieving a
reasonable fit for this object depended on the selection of the solution
with the narrowest $\redseqwidth$. This is the target with the most
southerly declination in the REXCESS sample, is only visible from
ESO La Silla at relatively high airmass ($\sim1.5$) and the observation
image depth was somewhat lower than for most of the other targets
(as shown in \tabref{Off-target-luminosity-function-fit_results}).
We suspect that the uncertainties on the measured colours may be underestimated,
making the red sequence more difficult to see. Before finding the
fitting method described above, we investigated several alternative
fitting algorithms in an attempt to get robust fits for RXC~J2319.6--7313.
Of particular interest was fitting in several bands simultaneously,
i.e. iteration of line fitting in $\left(\Rband,\,\Bband-\Rband,\,\Vband-\Rband\right)$
space with an iterative cut based on perpendicular distance from the
line. We found that there is a substantial population of galaxies
which are identified as `on the red sequence' using the single band-pair
method, but which are excluded from it in the multi-band method. Apart
from scatter in the colour measurements, possible reasons for the
offsets in the other band pairs are that the galaxies may lie at a
substantially different redshift, have substantial dust attenuation
or AGN activity, or be influenced by the spread in star formation
histories. The improved discrimination of cluster and background galaxies
in this case would lead to lower background levels overall and higher
signal-to-noise ratios on the radial profiles. However, this method
required excellent starting values, and was particularly susceptible
to blue objects at fainter magnitudes. For RXC~J2319.6--7313, with
its broad red sequence which is poorly separated from the blue cloud,
the method failed. 

There is no significant change in the the red sequence results here
-- or the subsequent results based on the red sequence selection --
for the other clusters when using the other fitting algorithms we
investigated.


\makeatletter{}

\subsection{Radial density model}

We assume a generalised NFW model for the radial density profiles,
namely 
\begin{equation}
\volumedensity(r)=\frac{\volumedensity_{0}}{\left(\frac{r}{\NFWscalelength}\right)^{\gamma}\left(1+\left(\frac{r}{\NFWscalelength}\right)^{\alpha}\right)^{\frac{\beta-\gamma}{\alpha}}},\label{eq:generalisedNFW_form}
\end{equation}
where $\volumedensity_{0}$ is a density normalisation, $\halooverdensity$
is the factor by which the halo is overdense with respect to the critical
density of the universe at the object redshift, $\NFWscalelength=1/\concentrationcritical{\halooverdensity}$
is a characteristic scale length measured in units of $r_{\halooverdensity}$,
$\gamma$ is the inner slope, $\beta$ is the outer slope and $\alpha=1$.
All parameters are calculated in terms of $\halooverdensity=500$.

There is a strong degeneracy between $\beta$ and $\NFWscalelength$
when fitting the generalised NFW profile, so we estimated a value
of $\concentrationcritical{500}$ for each cluster using equation
(12) from \citet{2004:Dolag.K;Bartelmann.M;Perrotta.F:Numerical-study-of-halo-concentrations-in-dark-energy-cosmologies:article}
to be used in subsequent calculations. The values are shown in \tabref{concentrations}.
When analysing stacked profiles, we assume the mean value of $\concentrationcritical{500}$
from all the contributing clusters.

\begin{table*}
\protect\caption{\label{tab:concentrations} Concentration parameters $c_{\halooverdensity}$
with respect to the critical density. The values of $\concentrationmatter{200}$
were calculated using equation (12) in \citet{2004:Dolag.K;Bartelmann.M;Perrotta.F:Numerical-study-of-halo-concentrations-in-dark-energy-cosmologies:article}.
$\concentrationcritical{200}$ and $\concentrationcritical{500}$
were derived from $\concentrationmatter{200}$ assuming our fiducial
cosmology.}
\makeatletter{}
\begin{tabular}{llll}
\hline
Object & \concentrationmatter{200} & \concentrationcritical{200} & \concentrationcritical{500} \\
 &  &  &  \\
\hline
RXC\,J0006.0--3443 & $7.46 \pm 0.07$ & $4.43 \pm 0.04$ & $2.907 \pm 0.030$ \\
RXC\,J0049.4--2931 & $8.16 \pm 0.06$ & $4.87 \pm 0.04$ & $3.217 \pm 0.028$ \\
RXC\,J0345.7--4112 & $9.04 \pm 0.07$ & $5.43 \pm 0.04$ & $3.609 \pm 0.029$ \\
RXC\,J0547.6--3152 & $7.04 \pm 0.07$ & $4.17 \pm 0.04$ & $2.724 \pm 0.031$ \\
RXC\,J0605.8--3518 & $7.29 \pm 0.07$ & $4.32 \pm 0.04$ & $2.833 \pm 0.030$ \\
RXC\,J0616.8--4748 & $7.72 \pm 0.06$ & $4.60 \pm 0.04$ & $3.024 \pm 0.029$ \\
RXC\,J0645.4--5413 & $6.67 \pm 0.07$ & $3.93 \pm 0.05$ & $2.560 \pm 0.032$ \\
RXC\,J0821.8+0112 & $8.57 \pm 0.06$ & $5.14 \pm 0.04$ & $3.400 \pm 0.028$ \\
RXC\,J2023.0--2056 & $8.89 \pm 0.07$ & $5.33 \pm 0.04$ & $3.539 \pm 0.029$ \\
RXC\,J2048.1--1750 & $7.16 \pm 0.07$ & $4.24 \pm 0.04$ & $2.775 \pm 0.030$ \\
RXC\,J2129.8--5048 & $8.15 \pm 0.07$ & $4.87 \pm 0.04$ & $3.213 \pm 0.029$ \\
RXC\,J2218.6--3853 & $7.10 \pm 0.07$ & $4.21 \pm 0.04$ & $2.751 \pm 0.031$ \\
RXC\,J2234.5--3744 & $6.76 \pm 0.07$ & $3.99 \pm 0.05$ & $2.600 \pm 0.033$ \\
RXC\,J2319.6--7313 & $8.31 \pm 0.06$ & $4.97 \pm 0.04$ & $3.284 \pm 0.028$ \\
\hline
Mean & $7.737 \pm 0.018$ & $4.608 \pm 0.011$ & $3.031 \pm 0.008$ \\
St. dev. & $0.785 \pm 0.019$ & $0.496 \pm 0.012$ & $0.346 \pm 0.008$ \\
\hline
\end{tabular}
 
\end{table*}


\makeatletter{}

\subsection{Brightest cluster galaxy properties and positions\label{sec:Brightest-cluster-galaxy-properties-positions-BCG}}

Large cD galaxies in the cluster central region are expected to have
a significant impact on the distribution of fainter galaxies, which
are less tightly bound and more susceptible to disruption than larger
bright galaxies. Their distances from the cluster centres and sizes
are given in \tabref{BCG-parameters}. 

\begin{table*}
\protect\caption{\label{tab:BCG-parameters}BCG parameters. $\ensuremath{\WFIABabsk{R}}$
is the absolute magnitude of the BCG and $\ensuremath{\luminosity_{R}}$
is the corresponding luminosity. $\ensuremath{\rccen}$ is the distance
of the BCG from the X-ray peak. $\areafractionincentralbin$ is the
ratio of the area of the BCG to $\pi(0.1\,\rfh)^{2}$. $\akron$ is
the semi-major axis of the elliptical aperture used to measure the
BCG luminosity.}
\makeatletter{}
\begin{tabular}{llllll}
\hline
Object & \ensuremath{\WFIABabsk{R}} & \ensuremath{ \luminosity_{R} } & \ensuremath{\rccen} & \areafractionincentralbin & \akron \\
 & \ensuremath{\si{\mag}} & \ensuremath{\SI{1e+11}{\Lsun}} & \ensuremath{\si{\rfhsi}} &  & \ensuremath{\si{\kilo\parsec}} \\
\hline
RXC\,J0006.0--3443 & \ensuremath{-23.393 \pm 0.025} & $1.65 \pm 0.04$ & $0.0033$ & \ensuremath{0.11267 \pm 0.00008} & \ensuremath{55.590 \pm 0.028} \\
RXC\,J0049.4--2931 & \ensuremath{-23.019 \pm 0.026} & $1.168 \pm 0.028$ & $0.0056$ & \ensuremath{0.06321 \pm 0.00008} & \ensuremath{21.180 \pm 0.020} \\
RXC\,J0345.7--4112 & \ensuremath{-22.92 \pm 0.05} & $1.07 \pm 0.05$ & $0.0017$ & \ensuremath{0.099439 \pm 0.000031} & \ensuremath{23.026 \pm 0.005} \\
RXC\,J0547.6--3152 & \ensuremath{-23.124 \pm 0.023} & $1.287 \pm 0.027$ & $0.0241$ & \ensuremath{0.05545 \pm 0.00009} & \ensuremath{28.507 \pm 0.034} \\
RXC\,J0605.8--3518 & \ensuremath{-23.455 \pm 0.022} & $1.745 \pm 0.035$ & $0.0049$ & \ensuremath{0.08959 \pm 0.00008} & \ensuremath{35.869 \pm 0.023} \\
RXC\,J0616.8--4748 & \ensuremath{-23.791 \pm 0.023} & $2.38 \pm 0.05$ & $0.0068$ & \ensuremath{0.14889 \pm 0.00010} & \ensuremath{57.391 \pm 0.028} \\
RXC\,J0645.4--5413 & \ensuremath{-23.576 \pm 0.024} & $1.95 \pm 0.04$ & $0.0095$ & \ensuremath{0.10326 \pm 0.00011} & \ensuremath{53.77 \pm 0.04} \\
RXC\,J0821.8+0112 & \ensuremath{-22.625 \pm 0.025} & $0.813 \pm 0.019$ & $0.0057$ & \ensuremath{0.042928 \pm 0.000019} & \ensuremath{18.431 \pm 0.005} \\
RXC\,J2023.0--2056 & \ensuremath{-22.204 \pm 0.028} & $0.551 \pm 0.014$ & $0.0093$ & \ensuremath{0.064183 \pm 0.000027} & \ensuremath{19.517 \pm 0.006} \\
RXC\,J2048.1--1750 & \ensuremath{-23.777 \pm 0.023} & $2.35 \pm 0.05$ & $0.1729$ & \ensuremath{0.07459 \pm 0.00005} & \ensuremath{30.889 \pm 0.016} \\
RXC\,J2129.8--5048 & \ensuremath{-22.601 \pm 0.024} & $0.795 \pm 0.018$ & $0.1130$ & \ensuremath{0.041114 \pm 0.000023} & \ensuremath{21.526 \pm 0.009} \\
RXC\,J2218.6--3853 & \ensuremath{-23.424 \pm 0.023} & $1.70 \pm 0.04$ & $0.0283$ & \ensuremath{0.06235 \pm 0.00013} & \ensuremath{32.02 \pm 0.05} \\
RXC\,J2234.5--3744 & \ensuremath{-23.218 \pm 0.023} & $1.404 \pm 0.030$ & $0.1475$ & \ensuremath{0.041763 \pm 0.000029} & \ensuremath{29.832 \pm 0.014} \\
RXC\,J2319.6--7313 & \ensuremath{-22.127 \pm 0.031} & $0.514 \pm 0.015$ & $0.0061$ & \ensuremath{0.06545 \pm 0.00027} & \ensuremath{24.68 \pm 0.07} \\
\hline
\end{tabular}
 
\end{table*}

The BCGs in \REXCESS~were studied by \citet{2010:Haarsma.D;Leisman.L;Donahue.M:Brightest-Cluster-Galaxies-and-Core-Gas-Density-in-REXCESS-Clusters:article}.
In this paper we use variable aperture elliptical magnitudes, whereas
\citeauthor{2010:Haarsma.D;Leisman.L;Donahue.M:Brightest-Cluster-Galaxies-and-Core-Gas-Density-in-REXCESS-Clusters:article}
used magnitudes measured in a $\SI{12}{\hparam^{-1}\kilo\parsec}$
fixed aperture metric radius, and as a consequence the magnitudes
we measure here are higher.


\makeatletter{}

\subsection{Radial density profiles}

By measuring the shape of the radial density profiles of galaxy clusters,
we can assess the extent to which galaxies (effectively collisionless
particles) trace gas (a collisional fluid) and dark matter (which
we assume to be a collisionless fluid) in galaxy clusters of different
morphological types and masses. Since the evolutionary history of
a galaxy may be quite different if it is on the red-sequence as opposed
to in the blue cloud, or bright as opposed to faint, we must consider
each sub-population of galaxies (red and bright, red and faint, blue
and bright, and blue and faint) separately. 

We produced radial surface density profiles $\surfacedensity$ for
individual clusters by summing object counts in radial bins around
the cluster centres (see \secref{X-ray_sample_description} for details
on the centring method), with annular bin edges at \makeatletter{}
\SIlist{0;0.1;0.15;0.2;0.25;0.35;0.5;0.7;1;1.25;1.5;2.5;5}{\rfhsi}
 \unskip.
The number counts were divided by the total exposed area in each annulus
to get the surface density.

\makeatletter{}

\subsubsection{Count density profile fitting}

\label{sec:profile_fitting_method}

We fitted the generalised NFW model profile defined in \eqref{generalisedNFW_form}
to each of the individual galaxy count density profiles. This measurement
is used to constrain the background count density and to provide an
estimate of the total galaxy count within a particular radius.

When fitting the model profile to projected quantities like the observed
galaxy count density profiles, we integrate $\volumedensity(r)$ numerically
along the line of sight, and allow an additional uniform surface density
component $\NFWareaconstant$ as background, which is not projected.
We fixed $\alpha=\gamma=1$ because the statistics of the individual
profiles were too poor to constrain these parameters. 

An additional limit, $\volumedensity(r>\rcutoff)=0$ is imposed so
that the integrals converge and to break the degeneracy between $\beta$
and $\NFWareaconstant$. We found that values of $\NFWareaconstant$
were consistent within their uncertainties for $\rcutoff=$\ \SIlist{2.5;5;10}{\rfhsi}
for the red populations which have good statistics, but that the poor
statistics of the blue populations led to some cases of negative $\NFWareaconstant$
if $\rcutoff>\SI{2.5}{\rfhsi}$. Whilst there are indications that
the cluster extends beyond $\SI{2.5}{\rfhsi}$, our data outside that
radius are too sparse to make reasonable measurements, so we set $\rcutoff=\SI{2.5}{\rfhsi}$.
This limit is also used where we estimate the projected total mass
profile within the cluster, and the projected gas density profile.

To generate estimates of the total number of galaxies $\totalgalaxycount$
in each cluster, we integrated the fitted density profiles out to
the relevant $\rfl{\halooverdensity}$. This was done for each galaxy
population (combinations of red/blue and bright/faint) as well as
for the combined total population (red and blue, down to the faint
limit).

Count density profiles for all of the individual clusters are shown
in Figures \ref{fig:Population-radial-density-1} to \ref{fig:Population-radial-density-14}
(online material) and their best fitting radial density profile parameters
are given in \tabref{Individual-NFW-fitting-results-short} (full
length table is given in \tabref{Individual-NFW-fitting-results.},
online material). 

\begin{table*}
\protect\caption{\label{tab:Individual-NFW-fitting-results-short}NFW fitting results
for each of the galaxy population filters -- bright and faint; red
sequence and blue. The full length table is given in \tabref{Individual-NFW-fitting-results.}
(online material). }

\makeatletter{}
\begin{tabular}{lllll}
\hline
Object & Galaxy filter & \ensuremath{\beta} & \ensuremath{\surfacedensity_0} & \NFWareaconstant \\
 &  &  & \ensuremath{\si{\rfhsi\tothe{-3}}} & \ensuremath{\si{\rfhsi\tothe{-2}}} \\
\hline
RXC\,J0006.0--3443 & Bright, red & \ensuremath{2.7 \pm 0.5} & \ensuremath{49.7 \pm 9.0} & \ensuremath{7.2 \pm 2.1} \\
RXC\,J0049.4--2931 & Bright, red & \ensuremath{3.4 \pm 0.5} & \ensuremath{59.0 \pm 10.9} & \ensuremath{1.9 \pm 0.5} \\
RXC\,J0345.7--4112 & Bright, red & \ensuremath{3.2 \pm 1.0} & \ensuremath{12.2 \pm 4.5} & \ensuremath{2.3 \pm 0.6} \\
RXC\,J0547.6--3152 & Bright, red & \ensuremath{2.67 \pm 0.35} & \ensuremath{50.1 \pm 6.8} & \ensuremath{5.4 \pm 1.4} \\
RXC\,J0605.8--3518 & Bright, red & \ensuremath{3.2 \pm 0.4} & \ensuremath{55.5 \pm 8.8} & \ensuremath{7.2 \pm 0.8} \\
RXC\,J0616.8--4748 & Bright, red & \ensuremath{2.58 \pm 0.26} & \ensuremath{30.1 \pm 3.4} & \ensuremath{2.7 \pm 0.5} \\
RXC\,J0645.4--5413 & Bright, red & \ensuremath{2.74 \pm 0.29} & \ensuremath{96.2 \pm 13.0} & \ensuremath{9.8 \pm 1.5} \\
RXC\,J0821.8+0112 & Bright, red & \ensuremath{2.5 \pm 0.4} & \ensuremath{25.8 \pm 4.0} & \ensuremath{1.8 \pm 0.9} \\
\hline
\end{tabular}
 
\end{table*}


\makeatletter{}

\subsubsection{\label{sec:Background-count-density}Background count density analysis}

Before stacking the count density profiles of the clusters, the density
of galaxies which are in the same line of sight as the cluster but
not physically bound to it must be estimated and subtracted. Having
already established that the objects may be embedded in more dense
regions of the large scale structure (in \secref{off-target-number-counts}),
we made a number of different estimates of the local background for
each observation set, with the aim of assessing whether the background
level was high because the cluster itself is extended, or due to a
uniform surface density of objects across the observation field.

The simplest estimate $\simplebackground$ was found by taking the
count density in the region bounded by \makeatletter{}
$r>\SI{1.5 }{\rfhsi}$
 \unskip.
This method is susceptible to contamination by the wings of the galaxy
density distribution, including virialised galaxies and structures
like infalling subclusters and filaments.

We control for infalling structures which are not azimuthally symmetric
by making $n$ independent estimates $\sectorbackground$ (with standard
deviation $\standarddeviation{\sectorbackground}$) of the background
level in annular sectors in the same bounded region. If the objects
in the sectors are spatially uncorrelated and obey Poisson statistics,
we would expect that the combined uncertainty of the $n$ independent
samples of the background density, $\sectorbackgroundalpha=\standarddeviation{\sectorbackground}/\sqrt{n}$,
would be the same as the Poissonian value for the complete region,
i.e. $\sectorbackgroundalpha\sim\standarderroron{\simplebackground}$.
On the other hand, spatially correlated objects are not drawn from
a single Poisson distribution and appear significantly more frequently
in some the sectors, leading to a higher $\sectorbackgroundalpha$,
i.e. we would expect $\sectorbackgroundalpha>\standarderroron{\simplebackground}$
if local structures are present. 

To control for broad wings of the galaxy density distribution, we
fitted the NFW model described in \eqref{generalisedNFW_form} - including
a constant surface density component $\NFWareaconstant$ - to the
individual cluster profiles. $\NFWareaconstant$ is sensitive to uniform
density across the field, whereas $\simplebackground$ includes the
cluster wings and infalling structures as well. $\simplebackground/\NFWareaconstant>1$
indicates that $\simplebackground$ is contaminated by the wings of
the cluster and overestimates the background level.

The results of these measurements for the red galaxies down to the
faint magnitude limit are given in \tabref{Background-count-density-table},
and \tabref{Background-count-density-table-short} includes comprehensive
results for all of the galaxy populations used in the radial profiles
(the full length table is given in \tabref{Background-count-density-table-long},
online material).

\begin{table*}
\protect\caption{\label{tab:Background-count-density-table}Background count density
measurements for the red bright and faint galaxy population: simple
count density $\simplebackground$ including Poisson uncertainties;
standard error measured from sectors $\sectorbackgroundalpha$; constant
density component from an NFW fit $\NFWareaconstant$. Comprehensive
results for red-bright, red-faint, blue-bright and blue-faint galaxy
populations are given in \tabref{Background-count-density-table-short}.}

\makeatletter{}
\begin{tabular}{llllll}
\hline
Object & \simplebackground & \sectorbackgroundalpha & \NFWareaconstant & \ensuremath{\simplebackground/\NFWareaconstant} & \ensuremath{\sectorbackgroundalpha/\simplebackgroundalpha} \\
 & \ensuremath{\si{\arcminuteword\tothe{-2}}} & \ensuremath{\si{\arcminuteword\tothe{-2}}} & \ensuremath{\si{\arcminuteword\tothe{-2}}} &  & \ensuremath{\si{}} \\
\hline
RXC\,J0006.0--3443 & $0.86 \pm 0.05$ & $0.04$ & $0.74 \pm 0.08$ & $1.17$ & $0.9$ \\
RXC\,J0049.4--2931 & $0.490 \pm 0.029$ & $0.04$ & $0.471 \pm 0.019$ & $1.04$ & $1.4$ \\
RXC\,J0345.7--4112 & $0.124 \pm 0.021$ & $0.02$ & $0.090 \pm 0.035$ & $1.38$ & $1.1$ \\
RXC\,J0547.6--3152 & $0.89 \pm 0.05$ & $0.08$ & $0.78 \pm 0.08$ & $1.15$ & $1.6$ \\
RXC\,J0605.8--3518 & $1.11 \pm 0.04$ & $0.08$ & $1.07 \pm 0.06$ & $1.04$ & $1.9$ \\
RXC\,J0616.8--4748 & $0.73 \pm 0.04$ & $0.04$ & $0.67 \pm 0.06$ & $1.09$ & $1.1$ \\
RXC\,J0645.4--5413 & $1.49 \pm 0.05$ & $0.13$ & $1.30 \pm 0.10$ & $1.15$ & $2.7$ \\
RXC\,J0821.8+0112 & $0.356 \pm 0.029$ & $0.05$ & $0.28 \pm 0.05$ & $1.27$ & $1.7$ \\
RXC\,J2023.0--2056 & $0.13 \pm 0.04$ & $0.04$ & $0.05 \pm 0.10$ & $2.41$ & $1.0$ \\
RXC\,J2048.1--1750 & $1.61 \pm 0.06$ & $0.10$ & $1.38 \pm 0.16$ & $1.17$ & $1.8$ \\
RXC\,J2129.8--5048 & $0.34 \pm 0.04$ & $0.07$ & $0.284 \pm 0.031$ & $1.20$ & $1.8$ \\
RXC\,J2218.6--3853 & $0.80 \pm 0.04$ & $0.07$ & $0.81 \pm 0.05$ & $0.99$ & $1.8$ \\
RXC\,J2234.5--3744 & $1.15 \pm 0.05$ & $0.05$ & $1.02 \pm 0.06$ & $1.13$ & $1.1$ \\
RXC\,J2319.6--7313 & $0.519 \pm 0.030$ & $0.02$ & $0.47 \pm 0.05$ & $1.11$ & $0.7$ \\
\hline
\end{tabular}
 
\end{table*}

\begin{table*}
\protect\caption{\label{tab:Background-count-density-table-short}Background count
density measurements. In the case of the RXC~J2023.0-2056 bright
blue filter, no objects are detected in the region used for measuring
$\simplebackground$. The full length table is provided in \tabref{Background-count-density-table-long}
(online material).}

\makeatletter{}
\begin{tabular}{llllll}
\hline
Object & Galaxy filter & \simplebackground & \sectorbackgroundalpha & \NFWareaconstant & \ensuremath{\simplebackground/\NFWareaconstant} \\
 &  & \ensuremath{\si{\arcminuteword\tothe{-2}}} & \ensuremath{\si{\arcminuteword\tothe{-2}}} & \ensuremath{\si{\arcminuteword\tothe{-2}}} &  \\
\hline
RXC\,J0006.0--3443 & Bright, red & $0.166 \pm 0.020$ & $0.02$ & $0.12 \pm 0.04$ & $1.34$ \\
RXC\,J0006.0--3443 & Faint, red & $0.70 \pm 0.04$ & $0.04$ & $0.62 \pm 0.06$ & $1.12$ \\
RXC\,J0006.0--3443 & Bright, blue & $0.128 \pm 0.017$ & $0.02$ & $0.102 \pm 0.029$ & $1.25$ \\
RXC\,J0006.0--3443 & Faint, blue & $1.14 \pm 0.05$ & $0.07$ & $0.66 \pm 0.18$ & $1.73$ \\
RXC\,J0049.4--2931 & Bright, red & $0.064 \pm 0.011$ & $0.01$ & $0.051 \pm 0.013$ & $1.25$ \\
RXC\,J0049.4--2931 & Faint, red & $0.427 \pm 0.027$ & $0.03$ & $0.411 \pm 0.019$ & $1.04$ \\
RXC\,J0049.4--2931 & Bright, blue & $0.074 \pm 0.011$ & $0.01$ & $0.054 \pm 0.022$ & $1.38$ \\
RXC\,J0049.4--2931 & Faint, blue & $0.81 \pm 0.04$ & $0.04$ & $0.795 \pm 0.024$ & $1.02$ \\
\hline
\end{tabular}
 
\end{table*}

Comparing $\sectorbackgroundalpha$ and $\standarderroron{\simplebackground}$,
we see that 7 of the clusters have $\sectorbackgroundalpha/\simplebackgroundalpha>1.5$,
consistent with the existence of localised substructures within some
of the sectors. We conclude that in half of the sample, filamentary
structure or infalling objects contribute to the high background in
the cluster outskirts.

Values of $>1$ for the ratio $\simplebackground/\NFWareaconstant$
are suggestive that, in most cases, $\simplebackground$ is an overestimate
contaminated by the wings of the cluster profile. (In the case of
RXC J2023.0-2056, $\NFWareaconstant$ is consistent with zero.) Measurements
of the remnant luminosity function in this region, once a model of
the field galaxy density is removed, are presented in \secref{Residual-cluster-luminosity-offtarget}.

We can also compare $\simplebackground/\NFWareaconstant$ with $\galaxydensityfactor$
from \tabref{Off-target-luminosity-function-fit_results}. In RXC J0345.7--4112,
RXC J0645.4--5413, RXC J0821.8+0112 and RXC J2129.8--5048, $\simplebackground/\NFWareaconstant\geq\galaxydensityfactor$,
so at least a portion of the overdensity seen in the number counts
function is likely due to contamination from the cluster itself, rather
than the large scale structure along the line of sight.

The least contaminated estimate of the background galaxy count density
appears to be $\NFWareaconstant$, so we use that to generate background
subtracted count density profiles. Whilst relying on a background
measurement which is dependent on a prior cluster shape rather than
one which is shape independent is not ideal, there is sufficient scope
in the NFW model to make a reasonable characterisation of most of
the clusters we see.


\makeatletter{}

\subsubsection{Comparison profiles and normalisation}

With the given count statistics and binning, a deprojection of the
count density profiles produces very noisy profiles and therefore
comparisons are best made with other projected profiles. 

\label{sec:electron_density_profiles}

The deprojected electron density profiles of \citet{2008:Croston.J;Pratt.G;Bohringer.H:Galaxy-cluster-gas-density-distributions-of-the-representative-XMM-Newton-cluster-structure-survey-REXCESS:article}
do not extend to $\rcutoff$ (the edge of our galaxy density model)
because the cluster X-ray emission becomes too faint to observe, so
these were extrapolated using a generalised NFW profile (described
in \eqref{generalisedNFW_form}) fit to Croston's data, assuming $c_{500}$
from \tabref{concentrations} and allowing $\beta$, $\gamma$ and
$\volumedensity_{0}$ to vary. 

\label{sec:total_matter_density_profiles}

We also wished to compare the galaxy count density profiles with some
model of the total mass of the cluster. Since this is dominated by
dark matter, we assumed the NFW density profile form described in
\eqref{generalisedNFW_form} with $\beta=3$ and $\alpha=\gamma=1$,
and the relevant $\concentrationcritical{500}$ from \tabref{concentrations}. 

The electron and total mass profiles were projected using the same
numerical integrator as used for the galaxy count density profile
fitting, in the same radial bins and with the same $\rcutoff$. Typically,
the projected electron density profiles are dominated outside of $\SI{\sim0.8}{\rfhsi}$
by extrapolated densities.

Since the electron, total mass and galaxy count density profiles each
have different normalisation, we plot $\surfacedensity/\sum_{r<\rfh}\surfacedensity\,\annulusarea$,
where $\annulusarea$ is the total area in the annulus, wherever they
are compared.


\makeatletter{}

\subsubsection{\label{sec:Stacked-profiles}Stacked profiles}

\begin{figure*}
\includegraphics[width=17.2cm]{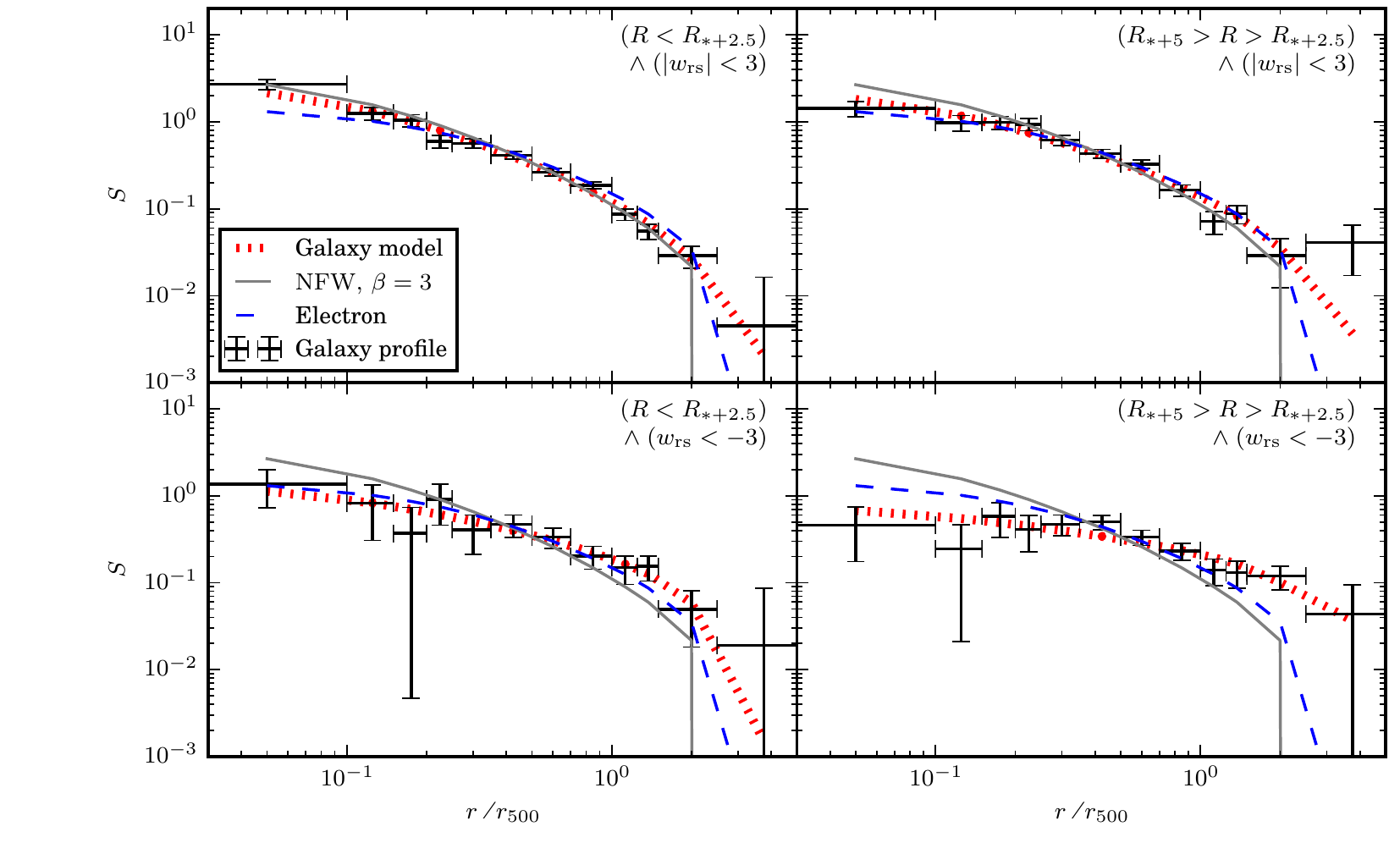}

\protect\caption{\label{fig:Stacked-radial-count-profiles}\label{fig:first_stacked_count_density_profile}\label{fig:Stacked-radial-profiles}Mean
projected radial profile of all of the clusters. Galaxy number density
profiles are shown as black points with error bars, with the best
fitting NFW model shown as a red dotted line. The projected electron
density is shown as a blue dashed line. The NFW profile with parameters
$\left(\alpha=1,\beta=3,\gamma=1\right)$ and $c_{500}$ from \tabref{concentrations}
is shown in grey. The upper panels represent galaxies on the red sequence,
and the lower panels include only galaxies bluer than the red sequence.
The panels on the left are `bright' and the panels on the right are
`faint' galaxies. The radial densities $\surfacedensity$ shown are
unitless, having been normalised by the mean $\surfacedensity$ of
all of the annuli within $\rfh$, such that they overlay each other
as closely as possible. }
\end{figure*}

We produce stacked galaxy density profiles for all clusters and each
subsample by taking the mean of $\galaxydensitynorm{\surfacedensity}=\left(\surfacedensity-\NFWareaconstant\right)/\totalgalaxycountredblue$
where $\totalgalaxycountredblue$ is the total galaxy count for the
cluster measured from the radial profile of all red and blue galaxies
down the faint limit (see \secref{profile_fitting_method}). The units
of $\galaxydensitynorm{\surfacedensity}$ and $\surfacedensity$ are
$\si{\rfhsi\tothe{-2}}$. We produce stacked electron density profiles
by taking the mean of $\electrondensitynorm{\surfacedensity}=\electronsurfacedensity/\sum_{r<\rfh}\electronsurfacedensity\,\annulusarea$,
where $\electrondensitynorm{\surfacedensity}$ and $\electronsurfacedensity$
have units $\si{\rfhsi\tothe{-2}}$. Similarly, we produce stacked
total mass profiles by taking the mean of $\massdensitynorm{\surfacedensity}=\masssurfacedensity/\sum_{r<\rfh}\masssurfacedensity\,\annulusarea$,
where $\massdensitynorm{\surfacedensity}$ has units $\si{\rfhsi\tothe{-2}}$
and $\masssurfacedensity$ has units $\si{\solarmass\rfhsi\tothe{-2}}$.
The mean profiles for the whole sample are shown in \figref{Stacked-radial-count-profiles}. 

The bright red profile follows the total mass profile reasonably well.
The faint red profile is a little broader and there is a $3\sigma$
($3\times$ the galaxy count density uncertainty) deficit of faint
red galaxies in the central bin. The blue profiles are both substantially
broader than the total mass profile. There appears to be suppression
of faint blue objects compared to the overall trend which is limited
to the region $\SI{<0.15}{\rfhsi}$; with respect to the NFW total
mass curve (shown in grey) the deficit is around $5\sigma$. The presence
of a cD brightest cluster galaxy in the cluster core could make the
detection of faint objects more difficult, but since the BCG occupies
a small fraction of the area of the central bin -- if indeed it is
positioned there -- this effect is negligible (see \tabref{BCG-parameters}
for BCG sizes relative to the central bin area, and the distance of
the BCG from the X-ray peak). Additionally, we would expect to see
a similar effect for the red sequence galaxies, for which there is
no evidence. 

\begin{figure*}
\subfloat[\label{fig:Massive-clusters:}\Massive\ clusters: \protect\makeatletter{}
RXC\,J0006.0--3443, RXC\,J0547.6--3152, RXC\,J0605.8--3518, RXC\,J0645.4--5413, RXC\,J2048.1--1750, RXC\,J2218.6--3853, RXC\,J2234.5--3744
 ]{\includegraphics[width=8.2cm]{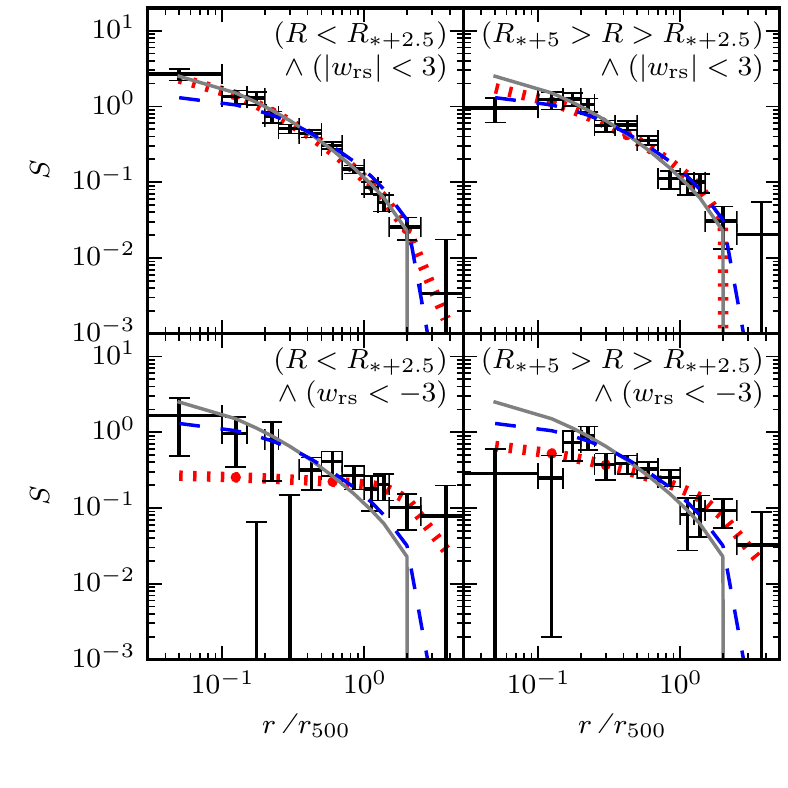}}\colfigspacer\subfloat[\label{fig:Low-mass-clusters:}\Lowmass\ clusters: \protect\makeatletter{}
RXC\,J0049.4--2931, RXC\,J0345.7--4112, RXC\,J0616.8--4748, RXC\,J0821.8+0112, RXC\,J2023.0--2056, RXC\,J2129.8--5048, RXC\,J2319.6--7313
 ]{\includegraphics[width=8.2cm]{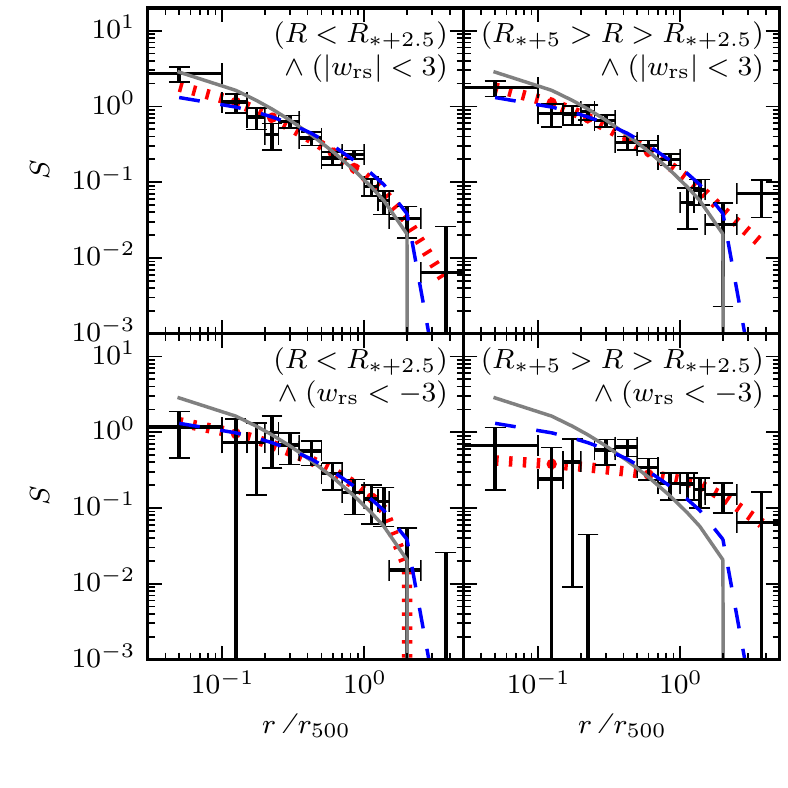}}\protect\caption{\label{fig:Profiles-massive_not}Stacked, projected radial number
density profiles for clusters above and below the median mass in the
population. The lines are the same as described in \figref{Stacked-radial-count-profiles}.}
\end{figure*}

\begin{figure*}
\subfloat[\label{fig:Disturbed-clusters:-}\Disturbed\ clusters: \protect\makeatletter{}
RXC\,J0006.0--3443, RXC\,J0616.8--4748, RXC\,J2023.0--2056, RXC\,J2048.1--1750, RXC\,J2129.8--5048, RXC\,J2218.6--3853, RXC\,J2319.6--7313
  ]{\includegraphics[width=8.2cm]{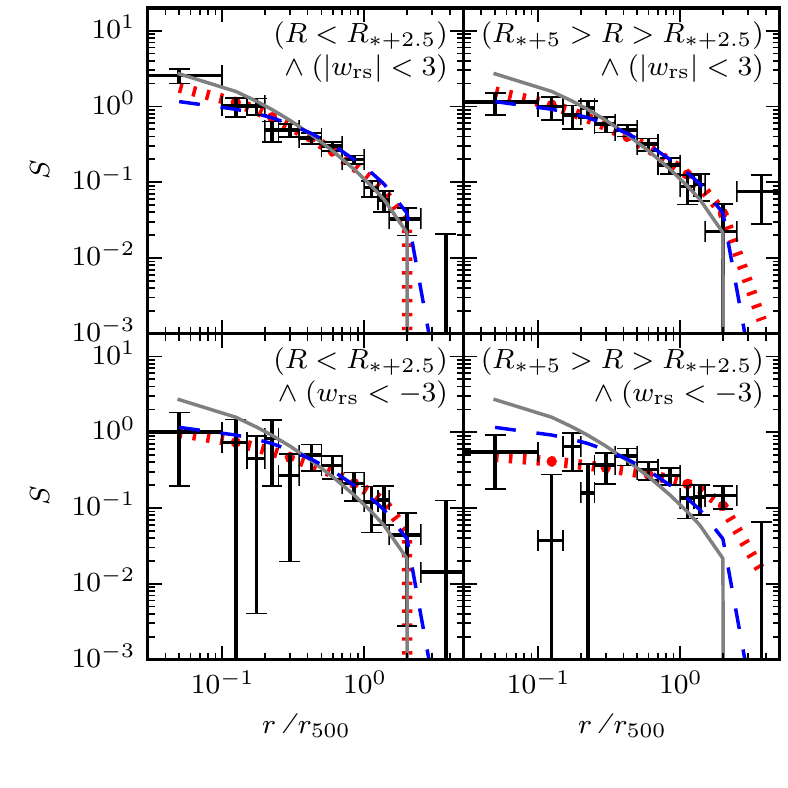}}\colfigspacer\subfloat[\label{fig:Regular-clusters:}\Regular\ clusters: \protect\makeatletter{}
RXC\,J0049.4--2931, RXC\,J0345.7--4112, RXC\,J0547.6--3152, RXC\,J0605.8--3518, RXC\,J0645.4--5413, RXC\,J0821.8+0112, RXC\,J2234.5--3744
 ]{\includegraphics[width=8.2cm]{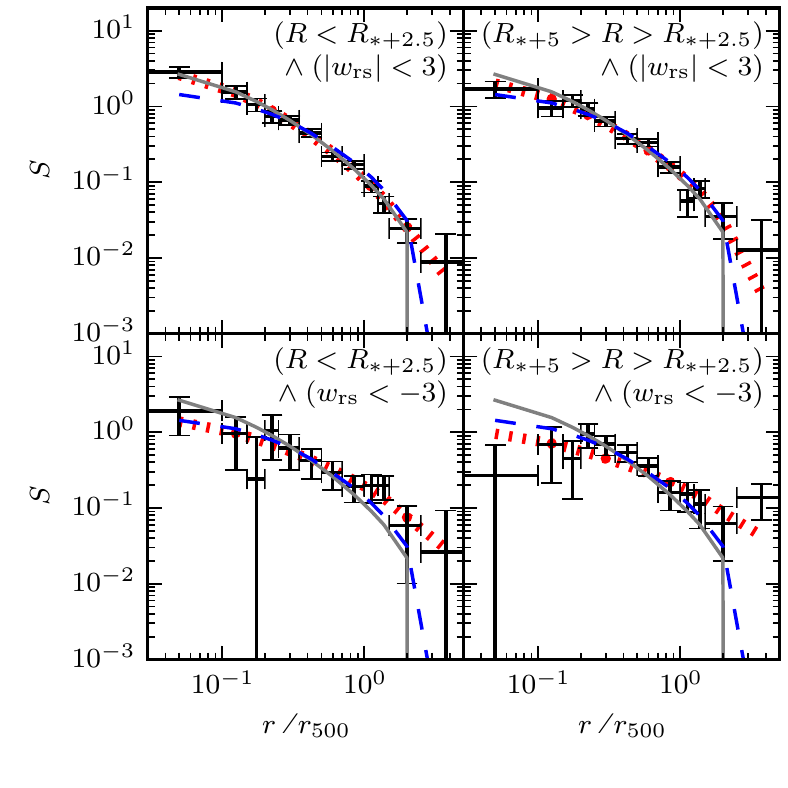}}\protect\caption{\label{fig:Profiles-for-disturbed_not}Stacked, projected radial number
density profiles for disturbed and regular clusters. The lines are
the same as described in \figref{Stacked-radial-count-profiles}. }
\end{figure*}

\begin{figure*}
\subfloat[\label{fig:Cool-core-clusters:}\Coolcore\ clusters: \protect\makeatletter{}
RXC\,J0345.7--4112, RXC\,J0605.8--3518, RXC\,J2319.6--7313
 ]{\includegraphics[width=8.2cm]{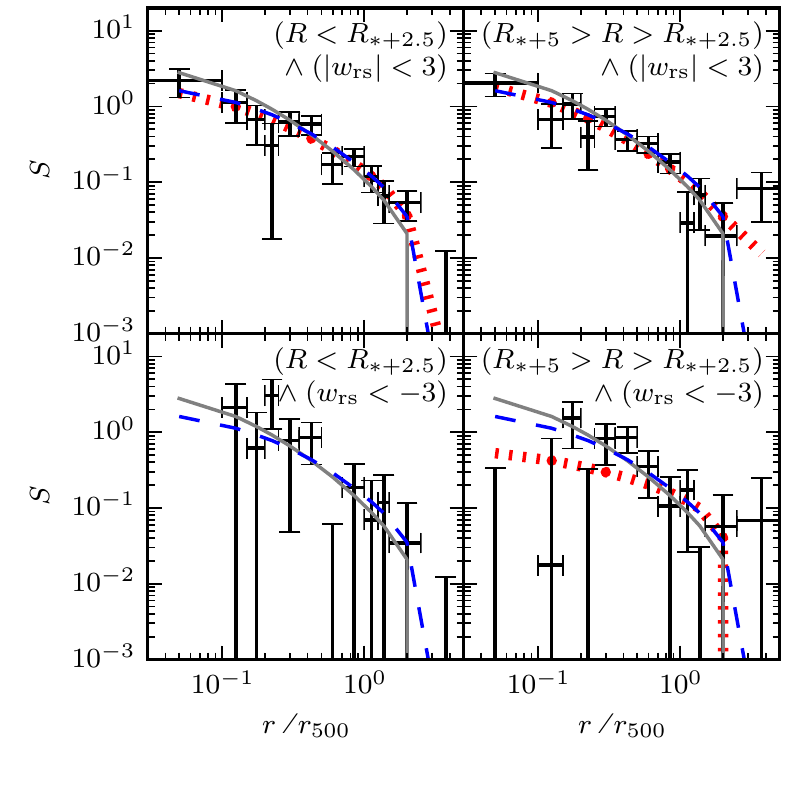}}\colfigspacer\subfloat[\label{fig:Non-cool-core-clusters:}\Noncoolcore\ clusters: \protect\makeatletter{}
RXC\,J0006.0--3443, RXC\,J0049.4--2931, RXC\,J0547.6--3152, RXC\,J0616.8--4748, RXC\,J0645.4--5413, RXC\,J0821.8+0112, RXC\,J2023.0--2056, RXC\,J2048.1--1750, RXC\,J2129.8--5048, RXC\,J2218.6--3853, RXC\,J2234.5--3744
 ]{\includegraphics[width=8.2cm]{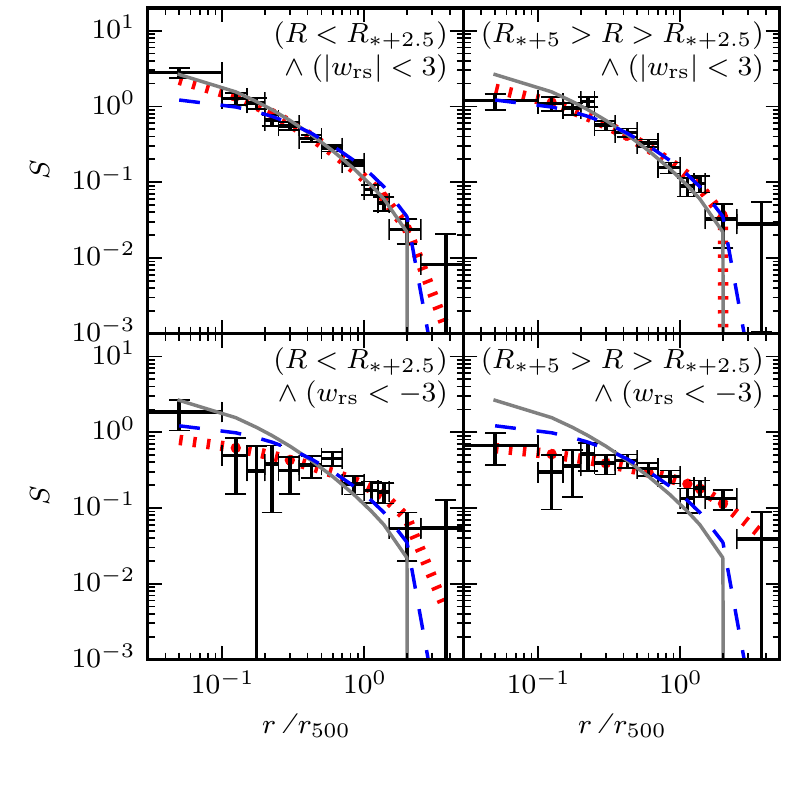}}\protect\caption{\label{fig:Profiles-cool_not}\label{fig:last_stacked_count_density_profile}Stacked,
projected radial number density profiles for \coolcore\  and \noncoolcore\ clusters.
The lines are the same as described in \figref{Stacked-radial-count-profiles}. }
\end{figure*}

In \secref{Subsamples} we described subsamples of the \rexcess\ dataset
based on morphological classifications and and cluster total mass.
Mean profiles for the \massive\ and \lowmass\ subsamples are shown
in \figref{Profiles-massive_not}, for the \disturbed\ and \regular\ subsamples
in \figref{Profiles-for-disturbed_not}, and for the \coolcore\ and
\noncoolcore\ subsamples in \figref{Profiles-cool_not}. 

The subsample profiles appear to show that the suppression of faint
blue galaxies in the cluster cores is driven by the massive subsample
where this effect is marked, and is absent in the low mass subsample
(both in \figref{Profiles-massive_not}). The massive subsample also
shows suppression of the faint red galaxies in the innermost radial
bin ($\SI{<0.1}{\rfhsi}$), not seen in the mean profile for the whole
sample. \figref{Profiles-cool_not} shows that there is even stronger
suppression of faint blue galaxies in the cores of the \coolcore\ clusters
(only one of which is classified as massive), and this effect appears
in all three \coolcore\ clusters' individual profiles (see \secref{IndividualRadialProfiles},
online material). Additionally, suppression of bright blue galaxies
is noted in the \coolcore\ cluster cores, an effect not seen at all
in the stacked profile for the full sample. In the \coolcore\ clusters
the red populations appear to be unaffected. 

There is some suppression of faint blue galaxies in the centres of
the regular clusters, but other than that the profiles for the regular
and disturbed clusters appear qualitatively very similar to one another.
In particular, we find no evidence from this analysis that there is
any substantial difference between the two subsamples which might
give a measure of the dynamical state to complement the X-ray based
centre shifts parameter used for the disturbed/regular classification
(see \secref{Subsamples}). The difference which is seen could be
statistical noise.

There is strong evidence that the profiles extend at least up to the
$\SI{2.5}{\rfhsi}$ limit, as the outer bins of the stacked profiles
have a significant positive residual even after background subtraction. 

We fitted projected generalised NFW models to the stacked galaxy count
density profiles using the same method as used for the individual
clusters to yield $\betagal$, allowing for a constant density component
$\NFWareaconstant$ in case the background subtraction before stacking
was incomplete. We include for comparison the mean electron density
profile outer slope $\mean{\betael}$, which is calculated by taking
the mean $\betael$ of all the clusters in the subsample.  The parameters
are tabulated in \tabref{Stacked-cluster-best} and the fits are shown
in Figures \ref{fig:first_stacked_count_density_profile} to \ref{fig:last_stacked_count_density_profile}.

The best fitting profiles for both bright and faint red sequence galaxies
have outer slopes which are flatter than, but nevertheless in rough
agreement with $\beta=3$, the slope of the assumed total mass profile.
The bright blue profile is substantially broader than the NFW, and
the best fitting has outer slope inconsistent with $\beta=3$. The
faint blue profile is similar, with a best fitting outer slope consistent
with the outer slope for the bright blue galaxies but with large uncertainties.

\begin{table}
\protect\caption{\label{tab:Stacked-cluster-best}Stacked cluster best fitting parameters.
$^{1}$No error is quoted where the fitting routine failed to estimate
the covariance matrix. }

\makeatletter{}
\begin{tabular}{llll}
\hline
Object & Galaxy filter & \betagal & \mean{\betael} \\
 &  &  &  \\
\hline
\Allclusters & Bright, red & \ensuremath{2.78 \pm 0.16} & \ensuremath{2.936 \pm 0.026} \\
\Massive & Bright, red & \ensuremath{2.97 \pm 0.16} & \ensuremath{3.131 \pm 0.032} \\
\Lowmass & Bright, red & \ensuremath{2.57 \pm 0.30} & \ensuremath{2.74 \pm 0.04} \\
\Disturbed & Bright, red & \ensuremath{2.53 \pm 0.24} & \ensuremath{2.853 \pm 0.034} \\
\Regular & Bright, red & \ensuremath{3.05 \pm 0.13} & \ensuremath{3.02 \pm 0.04} \\
\Coolcore & Bright, red & \ensuremath{2.3 \pm 0.4} & \ensuremath{2.70 \pm 0.06} \\
\Noncoolcore & Bright, red & \ensuremath{2.86 \pm 0.17} & \ensuremath{3.001 \pm 0.028} \\
\Allclusters & Faint, red & \ensuremath{2.51 \pm 0.23} & \ensuremath{2.936 \pm 0.026} \\
\Massive & Faint, red & \ensuremath{2.4 \pm 0.4} & \ensuremath{3.131 \pm 0.032} \\
\Lowmass & Faint, red & \ensuremath{2.55 \pm 0.32} & \ensuremath{2.74 \pm 0.04} \\
\Disturbed & Faint, red & \ensuremath{2.29 \pm 0.28} & \ensuremath{2.853 \pm 0.034} \\
\Regular & Faint, red & \ensuremath{2.68 \pm 0.24} & \ensuremath{3.02 \pm 0.04} \\
\Coolcore & Faint, red & \ensuremath{2.6 \pm 0.5} & \ensuremath{2.70 \pm 0.06} \\
\Noncoolcore & Faint, red & \ensuremath{2.40 \pm 0.25} & \ensuremath{3.001 \pm 0.028} \\
\Allclusters & Bright, blue & \ensuremath{1.79 \pm 0.26} & \ensuremath{2.936 \pm 0.026} \\
\Massive & Bright, blue & \ensuremath{-0.1 \pm 1.6} & \ensuremath{3.131 \pm 0.032} \\
\Lowmass & Bright, blue & \ensuremath{1.84 \pm 0.20} & \ensuremath{2.74 \pm 0.04} \\
\Disturbed & Bright, blue & \ensuremath{1.63 \pm 0.34} & \ensuremath{2.853 \pm 0.034} \\
\Regular & Bright, blue & \ensuremath{2.07 \pm 0.34} & \ensuremath{3.02 \pm 0.04} \\
\Coolcore & Bright, blue & \ensuremath{1.29^1} & \ensuremath{2.70 \pm 0.06} \\
\Noncoolcore & Bright, blue & \ensuremath{1.4 \pm 0.5} & \ensuremath{3.001 \pm 0.028} \\
\Allclusters & Faint, blue & \ensuremath{1.3 \pm 0.4} & \ensuremath{2.936 \pm 0.026} \\
\Massive & Faint, blue & \ensuremath{1.4 \pm 0.6} & \ensuremath{3.131 \pm 0.032} \\
\Lowmass & Faint, blue & \ensuremath{0.6 \pm 1.2} & \ensuremath{2.74 \pm 0.04} \\
\Disturbed & Faint, blue & \ensuremath{0.7 \pm 0.6} & \ensuremath{2.853 \pm 0.034} \\
\Regular & Faint, blue & \ensuremath{1.8 \pm 0.6} & \ensuremath{3.02 \pm 0.04} \\
\Coolcore & Faint, blue & \ensuremath{1.2 \pm 1.8} & \ensuremath{2.70 \pm 0.06} \\
\Noncoolcore & Faint, blue & \ensuremath{1.14 \pm 0.35} & \ensuremath{3.001 \pm 0.028} \\
\hline
\end{tabular}
 
\end{table}

\begin{figure}
\includegraphics{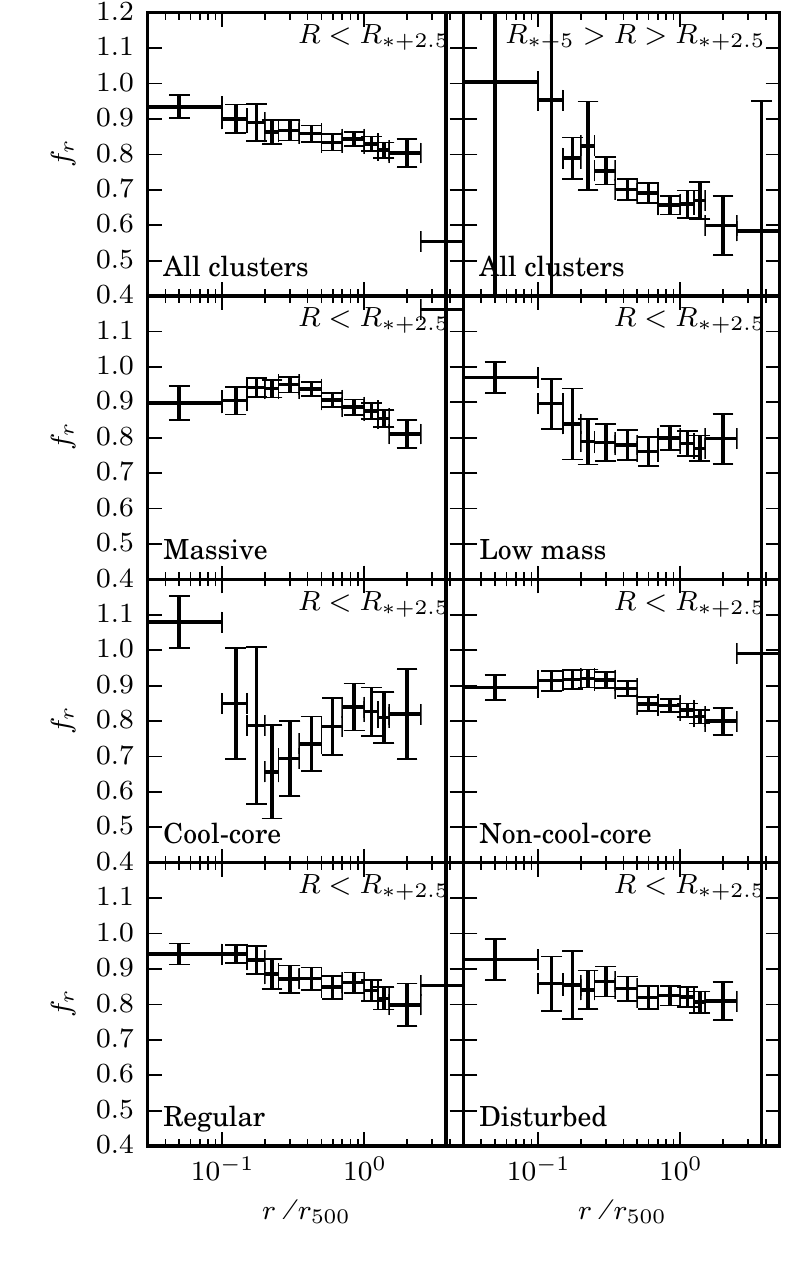}

\protect\caption{\label{fig:Cumulative-red-fraction}Cumulative fraction of red galaxies.
The top two panels show (left) bright and faint galaxies for the whole
sample and (right) only bright galaxies for the whole sample. The
remaining panels show bright and faint galaxies for the specified
subsamples.}
\end{figure}

The cumulative fraction of red galaxies for the full sample, and
for the high and low mass clusters is shown in \figref{Cumulative-red-fraction}.
These measurements reflect the morphology-density relation for ellipticals
and spirals \citep[e.g.][]{1984:Dressler.A:The-Evolution-of-Galaxies-in-Clusters:article},
and our measured blue fraction at the limit of our observations --
well within the cluster region of influence -- is substantially lower
than the field spiral population. Comparing the low mass and massive
clusters, we see that outside $\SI{0.2}{\rfhsi}$, the red fraction
reaches a plateau in the low mass clusters, but in the massive clusters
it is still higher than 90\% and doesn't reach the same plateau level
even at the limit of our observations. Even in low mass clusters,
the red fraction approaches 100\% in the central regions.

\makeatletter{}

\subsection{Luminosity measurements}

\label{sec:exhausive_luminosity_measurements}

\makeatletter{}

\subsubsection{\label{sec:Variation-of-lumfunc-with-radius}Variation of the cluster
luminosity function with radius}

\begin{figure*}
\subfloat[\label{fig:Stacked-luminosity-functions-full-rexcess_red_sequence}Stacked
luminosity functions for all the clusters in the dataset, including
red sequence galaxies satisfying $\rso{3}$.]{\includegraphics{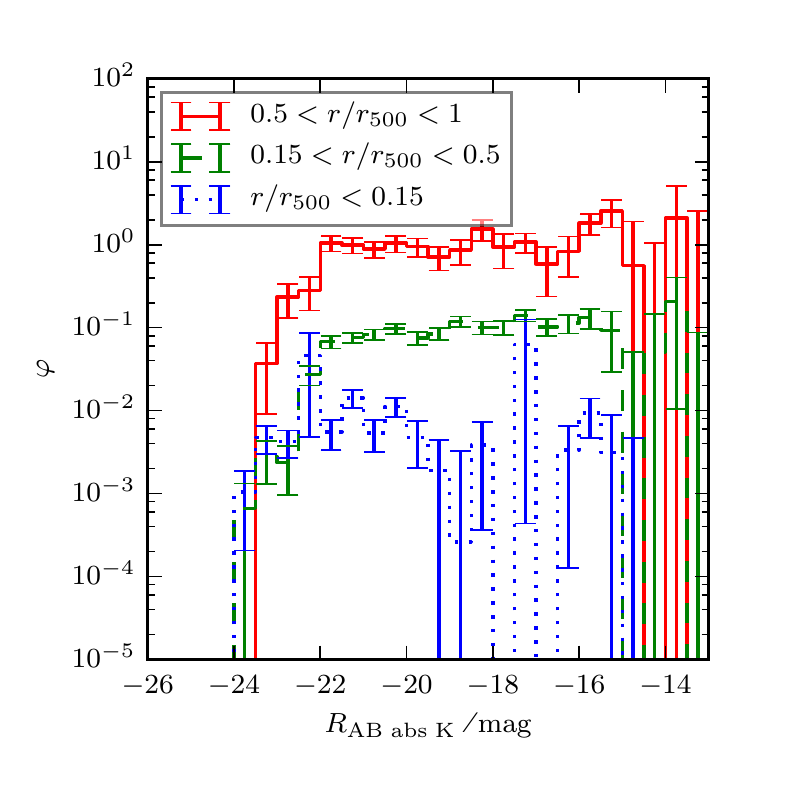}

}\colfigspacer\subfloat[\label{fig:Stacked-luminosity-functions-blue-galaxies}Stacked luminosity
functions for for all the clusters in the dataset, including blue
galaxies satisfying $\rsul{-3}$.]{\includegraphics{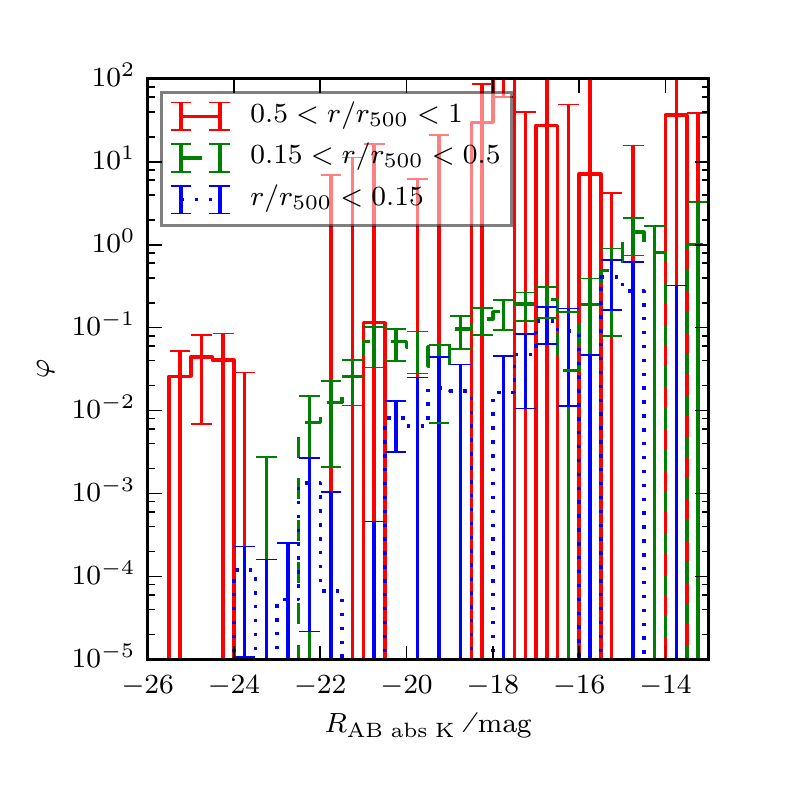}

}

\subfloat[Stacked luminosity functions for \massive~clusters, including red
sequence galaxies satisfying $\rso{3}$.]{\includegraphics{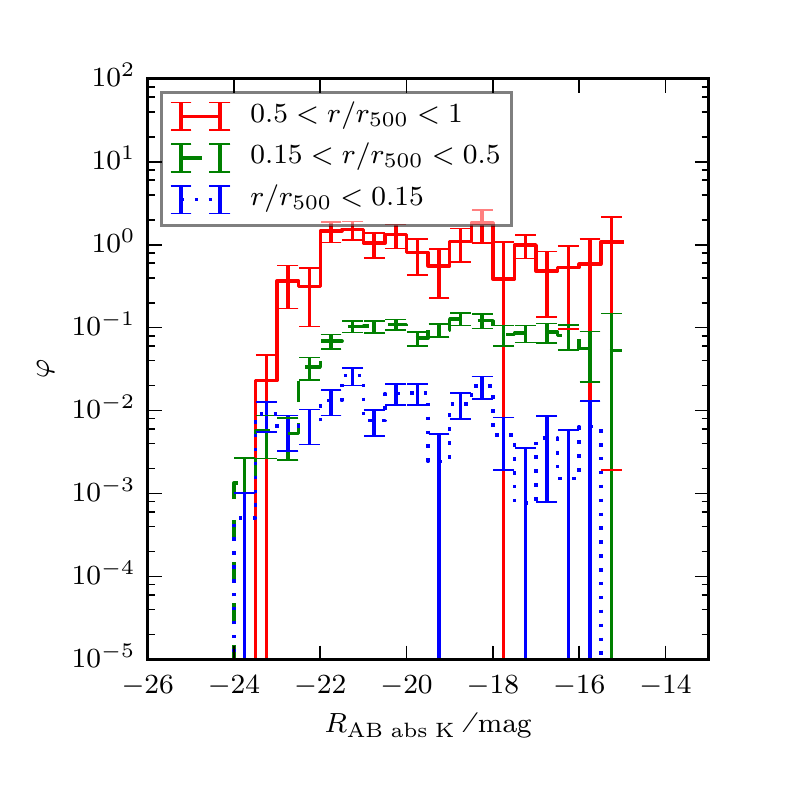}

}\colfigspacer\subfloat[Stacked luminosity functions for \lowmass~clusters, including red
sequence galaxies satisfying $\rso{3}$.]{\includegraphics{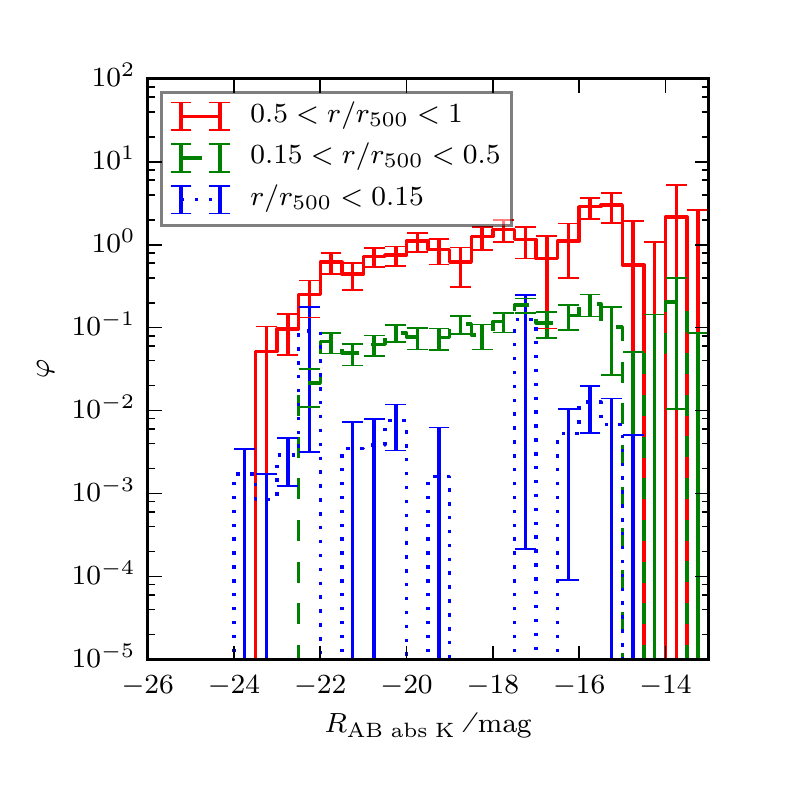}

}

\protect\caption{\label{fig:Stacked-annuli-luminosity-functions}Stacked luminosity
functions for the annuli bounded by \foreignlanguage{english}{\SIlist[list-units=brackets]{0;0.15;0.5;1.0}{\rfhsi}}.
The functions are normalised and then artificially separated by a
factor of 10. }
\end{figure*}

Since we see a reduction in the faint galaxy counts in cluster centres,
we produced background subtracted luminosity functions for the projected
annuli with edges at \SIlist[list-units=brackets]{0;0.15;0.5;1.0}{\rfhsi}
for the full galaxy population and for the red sequence galaxies.
A selection of these luminosity functions are shown in \figref{Stacked-annuli-luminosity-functions}.
The method of generation, normalisation and stacking is described
in \secref{cluster_luminosity_function_analysis}; the only difference
is that we now impose an additional catalogue selection based on the
red sequence fit.

\begin{figure*}
\subfloat[Stacked luminosity functions for \disturbed~clusters, including red
sequence galaxies satisfying $\rso{3}$.]{\includegraphics{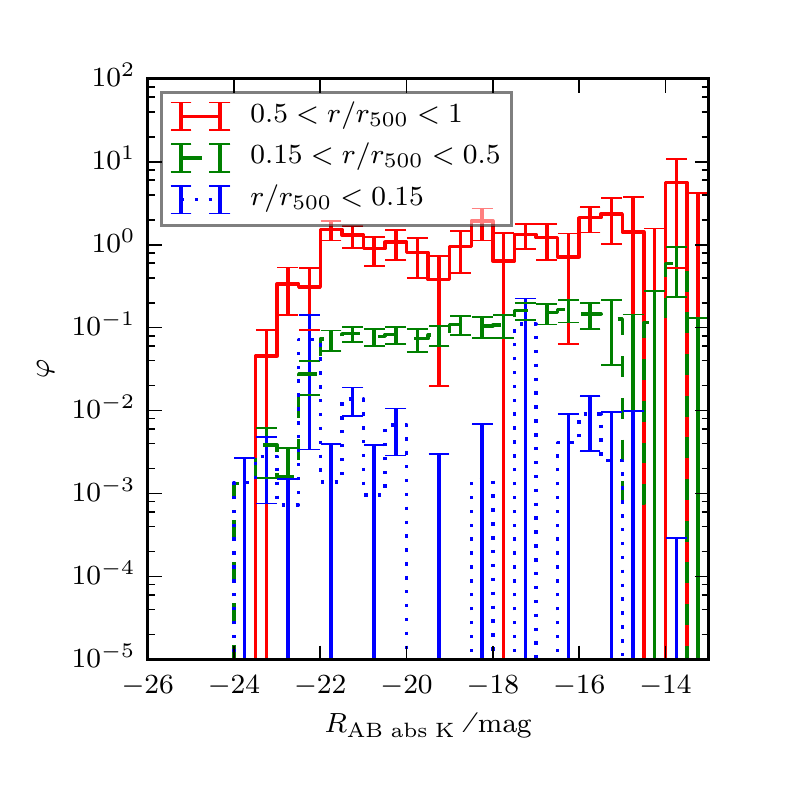}

}\colfigspacer\subfloat[Stacked luminosity functions for \regular~clusters, including red
sequence galaxies satisfying $\rso{3}$.]{\includegraphics{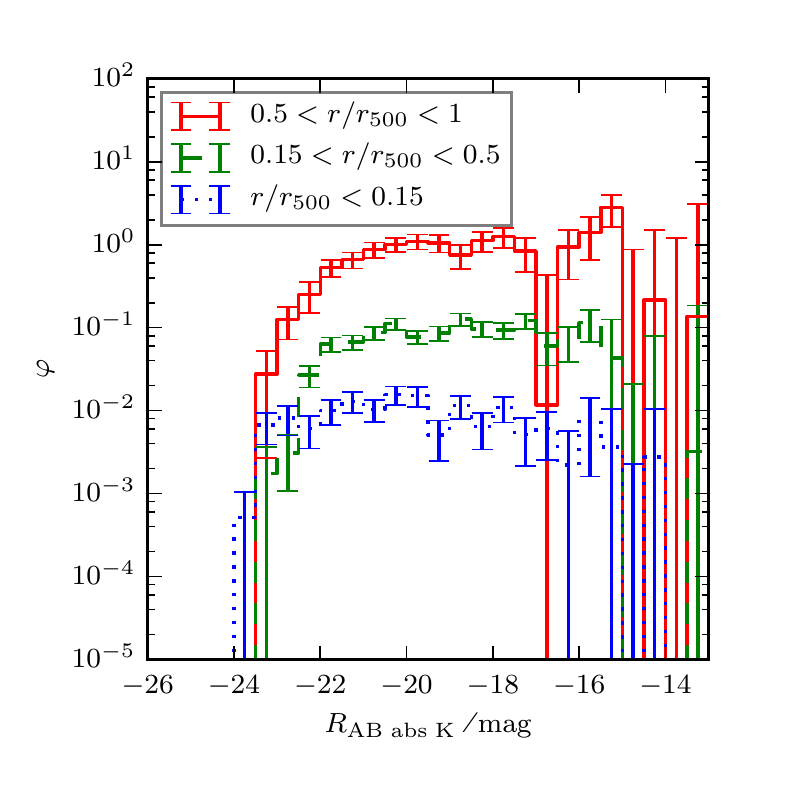}

}\protect\caption{\label{fig:Stacked-luminosity-functions-disturbed_regular}Stacked
luminosity functions for the annuli bounded by \foreignlanguage{english}{\SIlist[list-units=brackets]{0;0.15;0.5;1.0}{\rfhsi}}.
The functions are normalised and then artificially separated by a
factor of 10. }

\end{figure*}

\figref{Stacked-annuli-luminosity-functions} shows that the red sequence
luminosity functions outside of $\SI{0.15}{\rfhsi}$ are all extremely
similar. The inner luminosity functions have a break at around $\WFIABabsk{\Rband}=-18$,
above which the function drops below the trend. This suppression is
largely due to the massive clusters. The uncertainties on the inner
luminosity function for the low mass clusters are too large to conclude
that there is suppression; within the uncertainties it appears that
the faint galaxy count continues the trend seen at brighter magnitudes.
We note that the massive clusters are more distant on average, and
that the magnitude limit is lower for these observations, but not
sufficiently low that it explains the break at $\WFIABabsk{\Rband}=-18$.
Additionally, the stacking procedure -- normalising to a complete
part of the luminosity function and ensuring that truncated magnitude
bins do not contribute to the mean -- should minimize any influence
of the completeness limit on the final luminosity function shapes.
There is some evidence that the massive sample luminosity function
is a little flatter than the low mass sample one, but given the size
of the uncertainties it is difficult to be certain. 

The blue object luminosity functions vary strongly with respect to
cluster-centric distance. \figref{Stacked-luminosity-functions-blue-galaxies}
shows an excess of bright galaxies in the outer cluster regions ($0.5<r/\rfh<1$)
which isn't seen at smaller radii, and we found that there was an
excess of bright blue galaxies in the off-target region as well. In
the two inner regions sampled ($r/\rfh<0.15$ and $0.15<r/\rfh<0.5$)
there is strong variation in shape of the luminosity function, away
from a simple Schechter function. 

There is no evidence of a difference between the luminosity function
of disturbed and regular clusters, shown in \figref{Stacked-luminosity-functions-disturbed_regular}.
Any differences which are apparent are consistent with being statistical
effects.


\makeatletter{}

\subsubsection{\label{sec:Residual-cluster-luminosity-offtarget}Residual cluster
luminosity function in the off-target region}

Given the evidence of structures in the off-target region in \secref{Background-count-density}
and \secref{Stacked-profiles}, we re-analysed the luminosity function
in the $\rth<r$ region to try to find a residual cluster luminosity
function, once our assumed field galaxy function was subtracted. 

The possible excess of cluster galaxies outside $\rfh$ means that
the normalisation factor $\galaxydensityfactor$ found for the assumed
field galaxy function $\metcalfegalaxycountfunction$ (see \secref{off-target-number-counts})
may be slightly overestimated. However, since the cluster luminosity
function is largely invariant with distance from the cluster centre,
we can attempt to fit both cluster and field simultaneously for the
full image. This should give an improved estimate of \foreignlanguage{english}{$\galaxydensityfactor$},
which can be used with $\metcalfegalaxycountfunction$ as the background. 

\begin{figure*}
\includegraphics{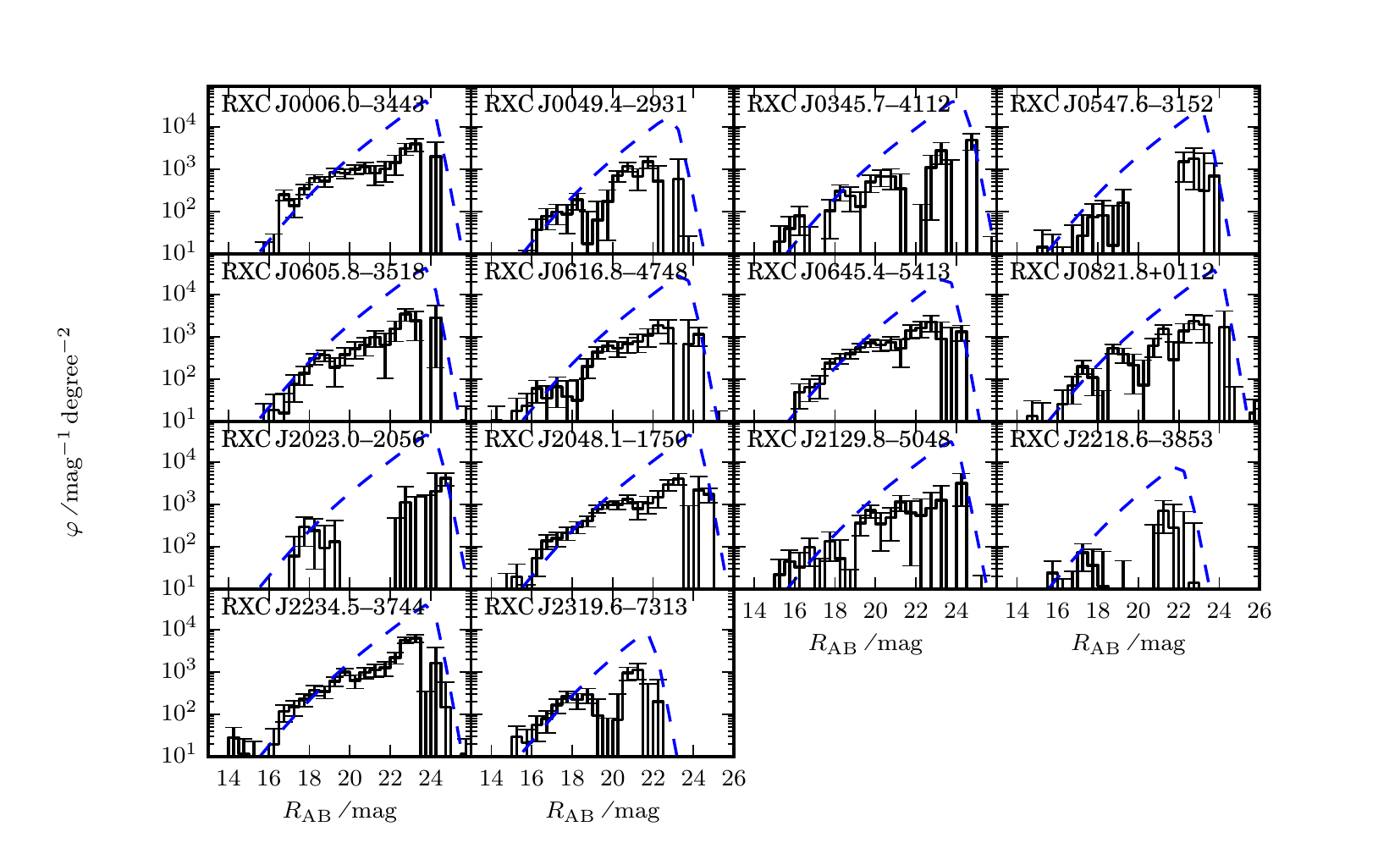}

\protect\caption{\label{fig:Off-target-residual-cluster-lumfunc}Off-target region
($\rth<r$) residual cluster luminosity function. The assumed background
model (including the falloff component) is shown as a dashed line
for comparison.}

\end{figure*}

Assuming values for $\schechterslope$ and $\mschechter{\magnitude}$
from \tabref{Schechter-function-fitting}%
\footnote{$\schechterslope$ and $\mschechter{\magnitude}$ were fitted to the
entire sample. Since the cluster luminosity function is invariant
with distance from the cluster centre, the background subtraction
performed before the stacking operation in \secref{cluster_luminosity_function_analysis}
should not bias the shape of the final luminosity function, despite
the small residual of cluster galaxies in the region used as the background.

} we fitted a combined model $\fallofffunction\left(\schechterfunction(\luminosity)+\galaxydensityfactor\,\metcalfegalaxycountfunction\right)$
to the luminosity histogram of each full field, making no magnitude
or red-sequence based selections since these may alter the field number
counts. The $\metcalfegalaxycountfunction$ component was subtracted
from the count histogram in the off-target region, and the results
are shown in \figref{Off-target-residual-cluster-lumfunc}.

The shape of the residual in \figref{Off-target-residual-cluster-lumfunc}
appears, in most cases, inconsistent with the shape of the background
number counts function which is also shown. The residual often resembles
the cluster luminosity function. This is consistent with the tentative
conclusion drawn in \secref{Background-count-density}, that the cluster
does extend some distance outside of $\rth$. Compared with the on-target
luminosity functions shown in \secref{Variation-of-lumfunc-with-radius},
these residuals have denser faint components, with an upturn at fainter
magnitudes similar to the dwarf upturn seen in \citet{2006:Popesso.P;Biviano.A;Bohringer.H:RASS-SDSS-Galaxy-cluster-survey.-IV.-A-ubiquitous-dwarf-galaxy-population-in-clusters:article},
but are also consistent with being due to some remnant field contamination.


\makeatletter{}

\subsubsection{Total cluster luminosity}

The total cluster luminosity and the related mass-to-light ratio are
useful parameters when assessing the efficiency or disruption of star
formation in different types of galaxy clusters. Since we are dealing
with projected data, a correction needed to be made for the cluster
galaxies outside of $\rfl{500}$ or $\rfl{200}$ but, when seen in
projection, in one of the annular bins.

We considered two populations of galaxies when calculating the total
cluster luminosities: $\rso{3}$ which includes only galaxies on the
red sequence, and $\rsul{3}$ which includes both the red sequence
and the blue cloud, but excludes objects redder than the red sequence.
To ensure that the correction to the total luminosity required due
to the magnitude limit was approximately equal for all the clusters
in the sample, we considered galaxies satisfying $\Rband<\faintcut$.

Using the best fitting NFW models -- which include a background estimate
-- for each of the galaxy populations, we assign a weight $\regionweight$
which represents the probability that a galaxy seen in a particular
radial bin is within \foreignlanguage{english}{$\rfl{\halooverdensity}$}.
If the total best fitting model count density for a particular radial
bin is $\surfacedensity_{\mathrm{total}}=\NFWareaconstant+\surfacedensity_{\mathrm{NFW,}\rcutoff=2.5\rfh}$
(as described in \secref{profile_fitting_method}), the weight assigned
to each galaxy in that radial bin is $\regionweight_{\halooverdensity}=\surfacedensity_{\mathrm{NFW,}\rcutoff=\rfl{\halooverdensity}}/\surfacedensity_{\mathrm{total}}$,
where $\surfacedensity_{\mathrm{NFW}}$ is found by integrating the
volume density model $\volumedensity$ in annuli along the line of
sight, and setting $\volumedensity=0$ where $r>\rcutoff$. The total
luminosity within $\rfl{\halooverdensity}$ is then the sum $\sum_{i}\,\regionweight_{\halooverdensity\, i}\, L_{i}$,
and the total count is $\sum_{i}\,\regionweight_{i}$ for all galaxies
$i$ in the particular population. The major source of uncertainty
in this calculation is the uncertainty on $\regionweight$, arising
from the uncertainties on the fitting parameters in the model. The
BCG is assigned $\regionweight=1$, but other galaxies are not specially
treated. Typical values of $\regionweight$ are around $0.9$ in the
innermost radial bins.

Since the radial count density profiles for the bright red, faint
red, bright blue and faint blue populations are different, we calculate
luminosities for all four subpopulations separately, and then sum
the relevant sub-populations to get total red or red plus blue luminosities.
In cases where the best fitting model is consistent with there being
no overdensity for a particular population and has very large uncertainties
on the relevant fit parameters this subpopulation is not included
in the final total luminosity. Only the RXC~J0345.7--4112 faint blue
population is affected by this step.

The total luminosities for each cluster are given in \tabref{Total-luminosities.}.

\begin{table*}
\protect\caption{\label{tab:Total-luminosities.}Total $\Rband$ band luminosities.}

\makeatletter{}
\begin{tabular}{llll}
\hline
Object & Galaxy filter & \ensuremath{\luminosity_{500}} & \ensuremath{\luminosity_{200}} \\
 &  & \ensuremath{\SI{1e+12}{\Lsun}} & \ensuremath{\SI{1e+12}{\Lsun}} \\
\hline
RXC\,J0006.0--3443 & Bright and faint, red & \ensuremath{1.49 \pm 0.16} & \ensuremath{2.21 \pm 0.25} \\
RXC\,J0006.0--3443 & Bright and faint, red and blue & \ensuremath{1.78 \pm 0.17} & \ensuremath{2.83 \pm 0.28} \\
RXC\,J0049.4--2931 & Bright and faint, red & \ensuremath{1.02 \pm 0.12} & \ensuremath{1.30 \pm 0.15} \\
RXC\,J0049.4--2931 & Bright and faint, red and blue & \ensuremath{1.23 \pm 0.16} & \ensuremath{1.74 \pm 0.24} \\
RXC\,J0345.7--4112 & Bright and faint, red & \ensuremath{0.266 \pm 0.028} & \ensuremath{0.33 \pm 0.04} \\
RXC\,J0345.7--4112 & Bright and faint, red and blue & \ensuremath{0.31 \pm 0.04} & \ensuremath{0.38 \pm 0.05} \\
RXC\,J0547.6--3152 & Bright and faint, red & \ensuremath{2.08 \pm 0.17} & \ensuremath{2.87 \pm 0.25} \\
RXC\,J0547.6--3152 & Bright and faint, red and blue & \ensuremath{2.20 \pm 0.22} & \ensuremath{3.08 \pm 0.31} \\
RXC\,J0605.8--3518 & Bright and faint, red & \ensuremath{1.13 \pm 0.08} & \ensuremath{1.46 \pm 0.11} \\
RXC\,J0605.8--3518 & Bright and faint, red and blue & \ensuremath{1.16 \pm 0.09} & \ensuremath{1.51 \pm 0.12} \\
RXC\,J0616.8--4748 & Bright and faint, red & \ensuremath{0.96 \pm 0.06} & \ensuremath{1.43 \pm 0.10} \\
RXC\,J0616.8--4748 & Bright and faint, red and blue & \ensuremath{1.17 \pm 0.11} & \ensuremath{1.79 \pm 0.17} \\
RXC\,J0645.4--5413 & Bright and faint, red & \ensuremath{3.59 \pm 0.26} & \ensuremath{5.1 \pm 0.4} \\
RXC\,J0645.4--5413 & Bright and faint, red and blue & \ensuremath{3.61 \pm 0.26} & \ensuremath{5.1 \pm 0.4} \\
RXC\,J0821.8+0112 & Bright and faint, red & \ensuremath{0.75 \pm 0.07} & \ensuremath{1.18 \pm 0.12} \\
RXC\,J0821.8+0112 & Bright and faint, red and blue & \ensuremath{0.97 \pm 0.09} & \ensuremath{1.63 \pm 0.17} \\
RXC\,J2023.0--2056 & Bright and faint, red & \ensuremath{0.49 \pm 0.11} & \ensuremath{0.70 \pm 0.18} \\
RXC\,J2023.0--2056 & Bright and faint, red and blue & \ensuremath{0.63 \pm 0.15} & \ensuremath{0.88 \pm 0.23} \\
RXC\,J2048.1--1750 & Bright and faint, red & \ensuremath{2.14 \pm 0.23} & \ensuremath{3.4 \pm 0.4} \\
RXC\,J2048.1--1750 & Bright and faint, red and blue & \ensuremath{2.63 \pm 0.24} & \ensuremath{4.5 \pm 0.4} \\
RXC\,J2129.8--5048 & Bright and faint, red & \ensuremath{0.91 \pm 0.13} & \ensuremath{1.14 \pm 0.19} \\
RXC\,J2129.8--5048 & Bright and faint, red and blue & \ensuremath{0.94 \pm 0.13} & \ensuremath{1.22 \pm 0.19} \\
RXC\,J2218.6--3853 & Bright and faint, red & \ensuremath{1.04 \pm 0.21} & \ensuremath{1.30 \pm 0.25} \\
RXC\,J2218.6--3853 & Bright and faint, red and blue & \ensuremath{1.19 \pm 0.21} & \ensuremath{1.55 \pm 0.27} \\
RXC\,J2234.5--3744 & Bright and faint, red & \ensuremath{3.34 \pm 0.13} & \ensuremath{4.38 \pm 0.17} \\
RXC\,J2234.5--3744 & Bright and faint, red and blue & \ensuremath{3.81 \pm 0.19} & \ensuremath{5.4 \pm 0.4} \\
RXC\,J2319.6--7313 & Bright and faint, red & \ensuremath{0.45 \pm 0.05} & \ensuremath{0.77 \pm 0.09} \\
RXC\,J2319.6--7313 & Bright and faint, red and blue & \ensuremath{0.56 \pm 0.07} & \ensuremath{0.91 \pm 0.11} \\
\hline
\end{tabular}
 
\end{table*}

\subsubsection{Mass-to-light ratio relation\label{sec:Mass_to_light_ratio}}

\begin{figure*}
\includegraphics{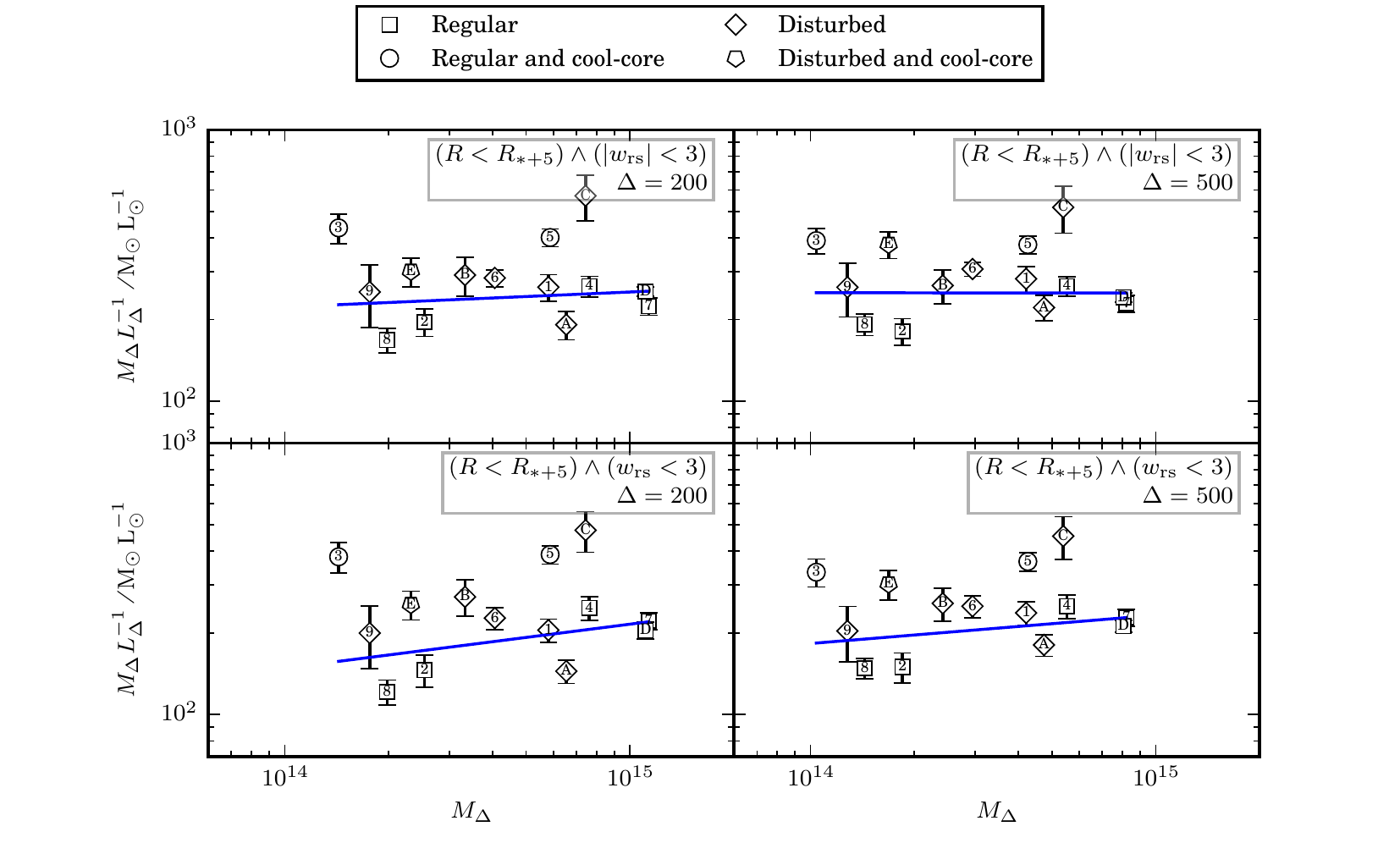}

\protect\caption{\label{fig:Mass-to-light-relations-plots}Mass-to-light relations.
The marker labels are the IDs given in \tabref{Overview-of-the-clusters}.
The uncertainties only take into account the luminosities; mass uncertainties
have not been considered. }
\end{figure*}

$\massl{500}$ and $\massl{200}$ were calculated using $\rfh$ from
\tabref{Overview-of-the-clusters}, concentration parameters from
\tabref{concentrations} (to transform between $\rfh$ and $\rth$)
and the fiducial cosmology. We calculated mass-to-light ratios $\massl{\halooverdensity}/\luml{\halooverdensity}$
for each of the clusters, which are plotted against $\massl{\halooverdensity}$
in \figref{Mass-to-light-relations-plots}. Using $\massl{\halooverdensity}/\luml{\halooverdensity}=\mlnorm\left(\massl{\halooverdensity}/\mlpivotmass\right)^{\mlindex}$
as a model, with $\mlpivotmass=\SI{5e14}{\solarmass}$, we found best
fitting parameters to the mass-to-light vs. mass relation, which are
given in \tabref{Mass-to-light-relation-best-fit-params}.

\begin{table*}
\protect\caption{\label{tab:Mass-to-light-relation-best-fit-params}Best fitting parameters
to the mass-to-light ratio relation $\massl{\halooverdensity}/\luml{\halooverdensity}=\mlnorm\left(\massl{\halooverdensity}/\mlpivotmass\right)^{\mlindex}$
where $\mlpivotmass=\SI{5e14}{\solarmass}$. The values of the $\massl{\halooverdensity}\luml{\halooverdensity}^{-1}$
measured at $\SI{1e14}{\Msun}$ and $\SI{1e15}{\Msun}$ are corrected
for the magnitude cut at $\Rband<\faintcut$ using the correction
factor described in \secref{cluster_luminosity_function_analysis}.
$\mlnorm$ is not corrected for the magnitude cut.}

\makeatletter{}
\begin{tabular}{lllllll}
\hline
Galaxy filter & \halooverdensity & \mlnorm & \mlindex & \ensuremath{\massl{\halooverdensity}\luml{\halooverdensity}^{-1}(10^{14}\,\MSol)} & \ensuremath{\massl{\halooverdensity}\luml{\halooverdensity}^{-1}(10^{15}\,\MSol)} & \ensuremath{\massl{\halooverdensity}\luml{\halooverdensity}^{-1}\si{\hparam^{-1}}(10^{15}\,\MSol)} \\
 &  & \ensuremath{\si{\Msun\Lsun\tothe{-1}}} &  & \ensuremath{\si{\Msun\Lsun\tothe{-1}}} & \ensuremath{\si{\Msun\Lsun\tothe{-1}}} & \ensuremath{\si{\Msun\Lsun\tothe{-1}}} \\
\hline
Bright and faint, red & \ensuremath{200} & \ensuremath{242.9 \pm 16.7} & \ensuremath{0.06 \pm 0.10} & \ensuremath{\left(2.2 \pm 0.4\right) \times 10^{2}} & \ensuremath{\left(2.5 \pm 0.2\right) \times 10^{2}} & \ensuremath{\left(3.5 \pm 0.3\right) \times 10^{2}} \\
Bright and faint, red & \ensuremath{500} & \ensuremath{250.4 \pm 15.4} & \ensuremath{-0.00 \pm 0.09} & \ensuremath{\left(2.4 \pm 0.4\right) \times 10^{2}} & \ensuremath{\left(2.4 \pm 0.2\right) \times 10^{2}} & \ensuremath{\left(3.5 \pm 0.3\right) \times 10^{2}} \\
Bright and faint, red and blue & \ensuremath{200} & \ensuremath{192.5 \pm 18.0} & \ensuremath{0.16 \pm 0.14} & \ensuremath{\left(1.4 \pm 0.4\right) \times 10^{2}} & \ensuremath{\left(2.1 \pm 0.3\right) \times 10^{2}} & \ensuremath{\left(3.0 \pm 0.4\right) \times 10^{2}} \\
Bright and faint, red and blue & \ensuremath{500} & \ensuremath{216.2 \pm 15.6} & \ensuremath{0.10 \pm 0.10} & \ensuremath{\left(1.8 \pm 0.3\right) \times 10^{2}} & \ensuremath{\left(2.3 \pm 0.2\right) \times 10^{2}} & \ensuremath{\left(3.2 \pm 0.3\right) \times 10^{2}} \\
\hline
\end{tabular}
 
\end{table*}

Three objects -- RXC~J0345.7--4112 (\coolcore), RXC~J00605.8--3518
(\coolcore\  and \massive) and RXC~J2218.6--3853 (\massive\  and
\disturbed) -- lie slightly above the fitted mass-to-light relationship,
but do not significantly affect the fit.

We find a slope $\mlindex$ of \makeatletter{}
\ensuremath{0.06 \pm 0.10}
 ~for
the red sequence within $\rth$, and \makeatletter{}
\ensuremath{0.16 \pm 0.14}
 ~for
the red plus blue galaxy population within $\rth$. The increase in
slope when the blue luminosity is included compared with the case
with just the red sequence luminosity is consistent with a decrease
in blue fraction at high masses, already noted in \secref{Stacked-profiles}
and shown in \figref{Cumulative-red-fraction}. The increase in the
mean luminosity of blue galaxies as cluster mass increases and faint
galaxies are disrupted/otherwise suppressed is insufficient to compensate
for the decreased blue fraction.

\citet{2007:Popesso.P;Biviano.A;Bohringer.H:RASS-SDSS-galaxy-cluster-survey.-VII.-On-the-cluster-mass-to-light-ratio-and-the-halo-occupation:article}
measured $M_{200}/L_{200}$ for red sequence objects and quote a slope
$\mlindex=0.18\pm0.04$, once projection effects are taken into account.
Whilst this result is in fair agreement with our measurements given
the uncertainties, both of our best estimates for the red sequence
are somewhat flatter.

\citet{2009:Sheldon.E;Johnston.D;Scranton.R:Cross-correlation-Weak-Lensing-of-SDSS-Galaxy-Clusters.-I.-Measurements:article}
quote a logarithmic slope on the mass-to-light ratio of $\mlindex=0.33\pm0.02$
for objects in the MaxBCG catalogue of galaxy clusters in the Sloan
Digital Sky Survey measured in the $i$ band, a value in fair agreement
with our measurement for the red and blue populations, but in this
case too, our result is flatter. 

The \foreignlanguage{english}{$\massl{200}$} range available to both
\citeauthor{2007:Popesso.P;Biviano.A;Bohringer.H:RASS-SDSS-galaxy-cluster-survey.-VII.-On-the-cluster-mass-to-light-ratio-and-the-halo-occupation:article}
and \citeauthor{2009:Sheldon.E;Johnston.D;Scranton.R:Cross-correlation-Weak-Lensing-of-SDSS-Galaxy-Clusters.-I.-Measurements:article}
is $\sim$\SIrange{5e12}{1e15}{\hparam\tothe{-1}\solarmass}, substantially
larger than the single order of magnitude mass range in \rexcess\unskip,
lending their analyses greater power to resolve mass dependent effects. 

\citet{1997:Carlberg.R;Yee.H;Ellingson.E:The-Average-Mass-and-Light-Profiles-of-Galaxy-Clusters:article}
quote an asymptotic value $\SI{289(50)}{\hparam\solMass\per\solLum}$
for the Gunn $r$ band, in excellent agreement with our value of \makeatletter{}
\ensuremath{\SI{3.0(4)e+02}{\hparam\solarmass\per\solarluminosity}}
 for
the red and blue galaxies in the $R$ band, measured at $\SI{1e15}{\solarmass}$.




\makeatletter{}

\section{Discussion}

\label{sec:Discussion}

Both the invariance of the cluster luminosity function with respect
to radius (outside of cluster centres, i.e. $r>\SI{0.15}{\rfhsi}$)
and suppression of faint galaxies in the central regions of galaxy
clusters have been noted before \citep{2006:Popesso.P;Biviano.A;Bohringer.H:RASS-SDSS-Galaxy-cluster-survey.-IV.-A-ubiquitous-dwarf-galaxy-population-in-clusters:article}.
\citeauthor{2006:Popesso.P;Biviano.A;Bohringer.H:RASS-SDSS-Galaxy-cluster-survey.-IV.-A-ubiquitous-dwarf-galaxy-population-in-clusters:article}
found significant suppression in the late type luminosity function
(corresponding to our blue population) for small cluster-centric distances;
we find that there is substantial change in the red luminosity function
close to the cluster core as well.

The colour-magnitude relation parameters drawn from the WINGS clusters
\citep{2011:Valentinuzzi.T;Poggianti.B;Fasano.G:The-red-sequence-of-72-WINGS-local-galaxy-clusters:article}
are quite similar to the values we see here, although the gradient
scatter from our sample is twice as large as that from WINGS. That
sample is also X-ray selected, but from clusters with lower redshifts.
Whilst it may be the case that there is a tightening of the distribution
of red sequence parameters at redshifts approaching $z=0$, it is
difficult to distinguish this effect from the increased measurement
uncertainties introduced by increasing numbers of field galaxies in
the same region of colour-magnitude space.

Of particular interest to us was any indication that the galaxy density
profile of disturbed clusters is also disturbed. Both the slope of
the galaxy count density distribution ($\betagal$) and the luminosity
functions could have shown differences, but there is no significant
evidence of a difference in either of these two properties in the
disturbed and regular subsamples.  The similarity in the luminosity
functions echoes the findings of \citet{2013:De-Propris.R;Phillipps.S;Bremer.M:Deep-luminosity-functions-and-colour-magnitude-relations-for-cluster-galaxies-at-0.2-lt-z-lt-0.6:article}
where luminosity functions of collisional and normal clusters in a
sample selected by X-ray, optical and weak and strong lensing were
studied.

We suggest that two main processes can be invoked to explain the distribution
of red and blue, bright and faint galaxies in clusters. Ram pressure
stripping occurs as a galaxy moves with velocity $v$ through the
intracluster medium (ICM) with density $\volumedensity$, and the
gas in the galaxy is subjected to pressure $P\propto\volumedensity v^{2}$
\citep{1972:Gunn.J;Gott.J:On-the-Infall-of-Matter-Into-Clusters-of-Galaxies-and-Some-Effects-on-Their-Evolution:article}.
The pressure ablates cool gas from the halo, slowing star formation
and turning blue galaxies redder. This effect should be more pronounced
in regions of galaxy clusters with high gas densities, in particular
in \coolcores. The galaxy infall velocity is related to the cluster
mass $M$ by $v^{2}\propto M$, so ram pressure stripping should also
be stronger for more massive clusters. Because this process affects
bright (as well as faint) galaxies, which dominate the total luminosity
of the cluster, it should lead to a decrease in fraction of the cluster
luminosity provided by the blue galaxy population as cluster mass
increases. As it affects star formation as a whole, it should also
lead to decreased overall star formation efficiency in more massive
clusters and to a positive slope on the mass-to-light ratio relation
measured using just red sequence galaxies.

The second main process, harassment, occurs as weakly bound galaxies
interact tidally with more massive objects. Parts of the weakly bound
galaxy are stripped away, or the galaxy is completely disrupted. The
remnants are a source of intracluster light (ICL). This process is
strongest in regions where  galaxy count densities and velocities
are highest and affects more weakly bound (less massive/lower luminosity)
galaxies more. Because the galaxy count densities in the central regions
of the clusters are not strongly dependent on mass, this effect is
expected to be less mass dependent than ram pressure stripping.

There are several key pieces of evidence we can use to disentangle
the two processes. The suppression of faint galaxies independent of
the galaxy colour in the densest regions of the galaxy clusters, with
the strongest effect in the most massive clusters, suggests harassment
-- a gravitational process independent of gas density and star formation
in the affected galaxy -- is important. The steeper mass-to-light
ratio relation for the blue plus red galaxies vs. the red sequence
alone, as well as decreasing blue galaxy fraction with higher mass,
is evidence that ram pressure stripping -- a process which primarily
affects blue galaxies -- is increasingly effective in reducing star
formation rates as infall velocities of galaxies rise. From the flatter
blue galaxy count density profiles in all of the clusters, it is clear
that the blue galaxy population does not survive long enough to relax
into the cluster potential before being stripped of its cold gas and
becoming part of the red population. There is some evidence that the
suppression of blue galaxies is most pronounced in the regions with
the highest gas densities at the centres of \coolcore\ clusters,
but given the small sample size we cannot be certain that this is
not a statistical anomaly.

The \REXCESS\ sample   was selected by X-ray luminosity, ensuring
that only clusters which are well evolved and have deep gravitational
potential wells with hot, dense ICM are selected. This is in contrast
to clusters in optically selected samples which are not always as
well evolved, and consequently may not have a sufficiently dense ICM
for efficient ram pressure stripping. \citet{2004:Bohringer.H;Matsushita.K;Churazov.E:Implications-of-the-central-metal-abundance-peak-in-cooling-core-clusters-of-galaxies:article}
note that \coolcores\ in clusters are long-lived, which may allow
more time for processes which disrupt galaxies and stop star formation
from taking place.

Both the red sequence and red plus blue mass-to-light ratio relation
slopes we measure are flatter than in the literature, compared to
both X-ray selected samples \citep{2007:Popesso.P;Biviano.A;Bohringer.H:RASS-SDSS-galaxy-cluster-survey.-VII.-On-the-cluster-mass-to-light-ratio-and-the-halo-occupation:article}
or optically selected samples \citep{2009:Sheldon.E;Johnston.D;Masjedi.M:Cross-correlation-Weak-Lensing-of-SDSS-Galaxy-Clusters.-III.-Mass-to-Light-Ratios:article}.
Given the scatter in the relation and the relatively large uncertainties
on the best fitting parameters, as well as the fact that the \rexcess\ sample
contains only clusters spanning one order of magnitude at the highest
masses, it is impossible using these data to distinguish between the
case where the differences between the slopes measured here and in
the literature are purely statistical in nature, or due to different
physical processes in the two samples -- e.g. stronger ICM effects.

The ICL has not been taken into account in this work, but if the relative
density of the ICL in the centres  of massive clusters were higher
than in low mass clusters, then this would be further evidence for
increased harassment. If we assume that 10\% of the light of galaxy
clusters is ICL \citep[e.g.][]{2005:Zibetti.S;White.S;Schneider.D:Intergalactic-stars-in-z-0.25-galaxy-clusters:-systematic-properties-from-stacking-of-Sloan:article},
our mass-to-light ratio normalisations may be overestimated by a factor
of $\sim1.1$, leading to a correction of comparable size to the normalisation
uncertainties. However, based on the measurements of the BCG sizes
and luminosities as compared with \citet{2010:Haarsma.D;Leisman.L;Donahue.M:Brightest-Cluster-Galaxies-and-Core-Gas-Density-in-REXCESS-Clusters:article}
described in \secref{Brightest-cluster-galaxy-properties-positions-BCG},
it seems likely that a substantial fraction of the intracluster light
is included in the BCG luminosity we measure, so the correction may
well be smaller. 

\citet{2007:Gonzalez.A;Zaritsky.D;Zabludoff.A:A-Census-of-Baryons-in-Galaxy-Clusters-and-Groups:article}
and \citet{2013:Gonzalez.A;Sivanandam.S;Zabludoff.A:Galaxy-Cluster-Baryon-Fractions-Revisited:article}
discuss the reduced efficiency of ICL generation in more massive objects
which is coupled with a higher X-ray gas fraction. \citet{2005:Zibetti.S;White.S;Schneider.D:Intergalactic-stars-in-z-0.25-galaxy-clusters:-systematic-properties-from-stacking-of-Sloan:article}
find that the ICL surface brightness is correlated with BCG luminosity,
but that the total fraction of light contributed by the ICL is almost
independent of cluster richness and BCG luminosity.  Given the open
discussion on the ICL light fraction as a function of cluster mass,
it is too premature to include the effect of the ICL in the mass-to-light
ratio in our results.


\makeatletter{}

\section{Summary and conclusions}

\label{sec:Conclusions}

We have used a sample of 14 galaxy clusters from the REXCESS survey
to investigate radial density profiles of galaxies and intra-cluster
medium.

\begin{itemize}
\item The red galaxy density traces the dark matter density closely outside
of the cluster centres (in the region $r>\SI{0.15}{\rfhsi}$). The
best fitting NFW model outer-slopes $\betagal$ are roughly consistent
with $\betagal=3$, with a best estimate $\betagal=$ \makeatletter{}
\ensuremath{2.78 \pm 0.16}
 \unskip,
fitted to the stacked bright red sequence galaxy density profile of
all the clusters. 
\item The blue sequence count density profile slopes are substantially shallower
than the $\beta=3$ total mass model, with a best estimate $\betagal=$
\makeatletter{}
\ensuremath{1.79 \pm 0.26}
 \unskip,
fitted to the stacked bright blue galaxy density profile of all the
clusters. 
\item The mean outer slope for the gas density profiles of the full sample
is $\betael=$ \makeatletter{}
\ensuremath{2.936 \pm 0.026}
 \unskip.
Within the cluster centres the gas and dark matter profiles tend to
diverge.
\item We find that faint blue galaxies are suppressed in the centres of
massive and regular clusters. Faint red galaxies are also suppressed
in the centres of massive clusters. Both bright and faint blue galaxies
are heavily suppressed in the centres of cool-core clusters, but the
faint red galaxies are unaffected. This is consistent with the idea
that the suppression of star formation is driven by ram pressure stripping
of gas from galaxies, but that wholesale disruption of galaxies is
caused by galaxy interactions in regions with high galaxy densities.
\item Our measurement of the logarithmic slope $\mlindex$ of the galaxy
cluster mass-to-light relation within $\rth$ of \makeatletter{}
\ensuremath{0.16 \pm 0.14}
 
for all galaxies, measured in the $\Rband$ band, is in fair agreement
with $\mlindex=0.33\pm0.02$ from \citet{2009:Sheldon.E;Johnston.D;Scranton.R:Cross-correlation-Weak-Lensing-of-SDSS-Galaxy-Clusters.-I.-Measurements:article},
measured in the $i$ band. Our measurement of the mass to light ratio
normalisation of \makeatletter{}
\ensuremath{\SI{3.0(4)e+02}{\hparam\solarmass\per\solarluminosity}}
 ~(evaluated
at $\SI{1e15}{\solarmass}$) in the $\Rband$ band is in excellent
agreement with \citet{1997:Carlberg.R;Yee.H;Ellingson.E:The-Average-Mass-and-Light-Profiles-of-Galaxy-Clusters:article}
measured in the Gunn $r$ band. 
\item There is no evidence of any difference in the galaxy count density
profiles when comparing clusters classified as having disturbed X-ray
morphology with those which are regular. 
\end{itemize}


\makeatletter{}

\section*{Acknowledgements}

We acknowledge support from the DFG Transregio Program TR33 `Dark
Universe' and the Munich Excellence Cluster `Structure and Evolution
of the Universe.' GC acknowledges support from DLR through project
50~OR~1305. DP acknowledges the financial support from Labex OCEVU.
The X-ray data used here are based on observations with \emph{XMM-Newton},
an ESA science mission with instruments and contributions directly
funded by ESA member states and NASA. This research has made use of
the SIMBAD database, operated at CDS, Strasbourg, France. This research
has made use of the NASA/IPAC Extragalactic Database (NED) which is
operated by the Jet Propulsion Laboratory, California Institute of
Technology, under contract with the National Aeronautics and Space
Administration. This research has made use of NASA's Astrophysics
Data System. This research made use of \astropy, a community-developed
core \python\  package for astronomy \citep{2013:Astropy-Collaboration;Robitaille.T;Tollerud.E:Astropy:-A-community-Python-package-for-astronomy:article}.
Additional analysis was carried out using \scipy\ and plots were
made using \matplotlib\ \citep{2007:Hunter.J:Matplotlib:-A-2D-graphics-environment:article}.
This research was made possible through the use of the AAVSO Photometric
All-Sky Survey (APASS), funded by the Robert Martin Ayers Sciences
Fund.


\bibliographystyle{mn2e}
\bibliography{rexcess_profile_paper_pub_expand_shortbib}

\clearpage{}

\makeatletter{}
\appendix
\pagenumbering{roman}

\begin{titlepage}
\centering 
\vspace*{\fill} 
\LARGE Supplementary material (online)
\vspace*{\fill} 
\end{titlepage}

\section{Additional figures}

\begin{figure}
\includegraphics[width=8cm]{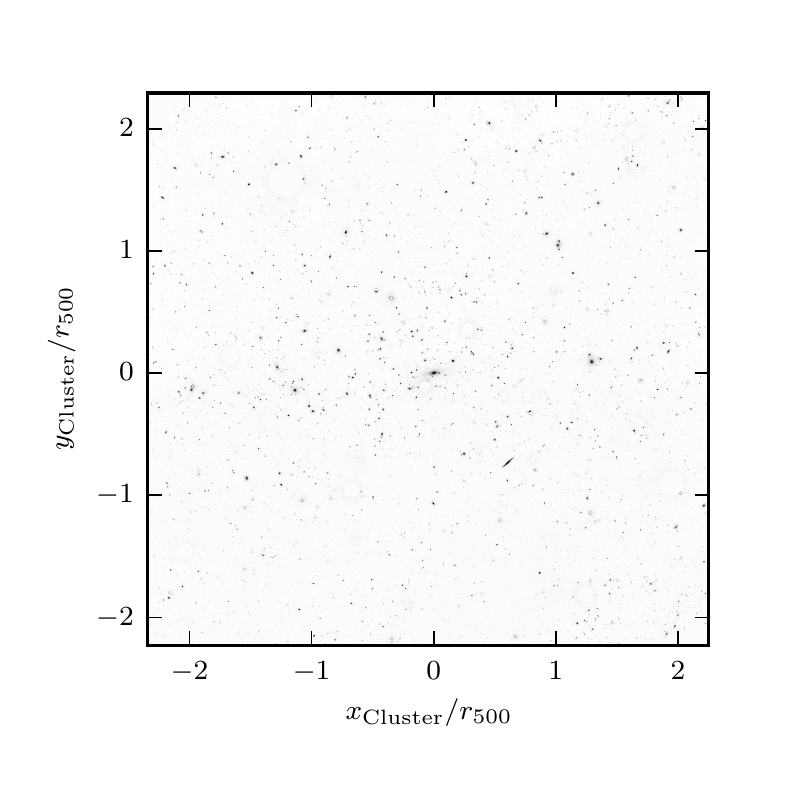}\protect\caption{\label{fig:RXC-J0006.0-3443-in-R}RXC J0006.0--3443 in the $R$ band.
Stars have been excised from this image.}
\end{figure}

\begin{figure}
\includegraphics[width=8cm]{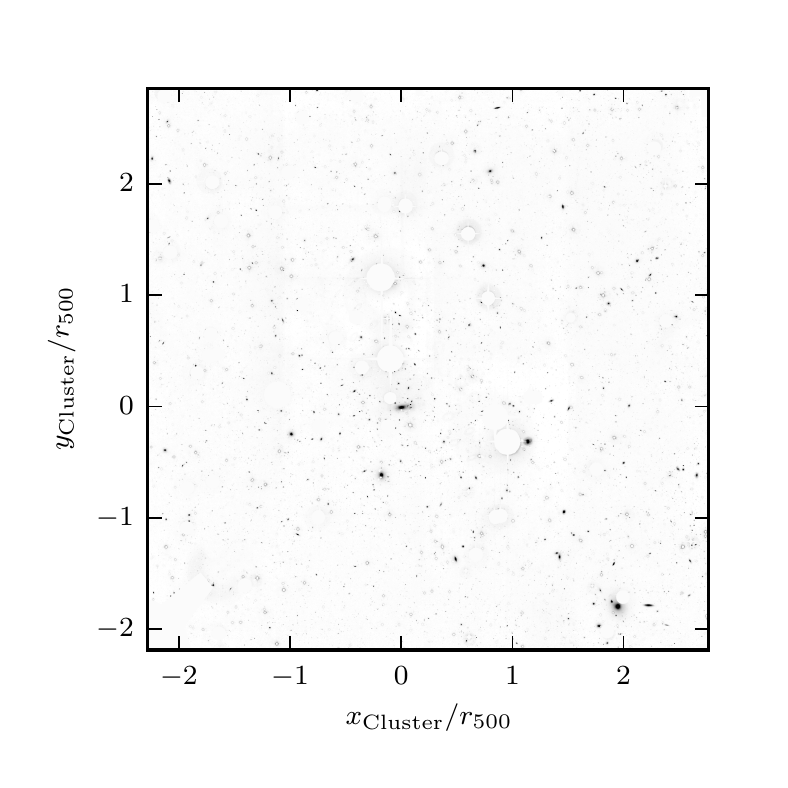}\protect\caption{\label{fig:RXC-J0616.8-4748-in-R}RXC J0616.8--4748 in the $R$ band.
Stars have been excised from this image.}
\end{figure}

\begin{figure}
\includegraphics[width=8cm]{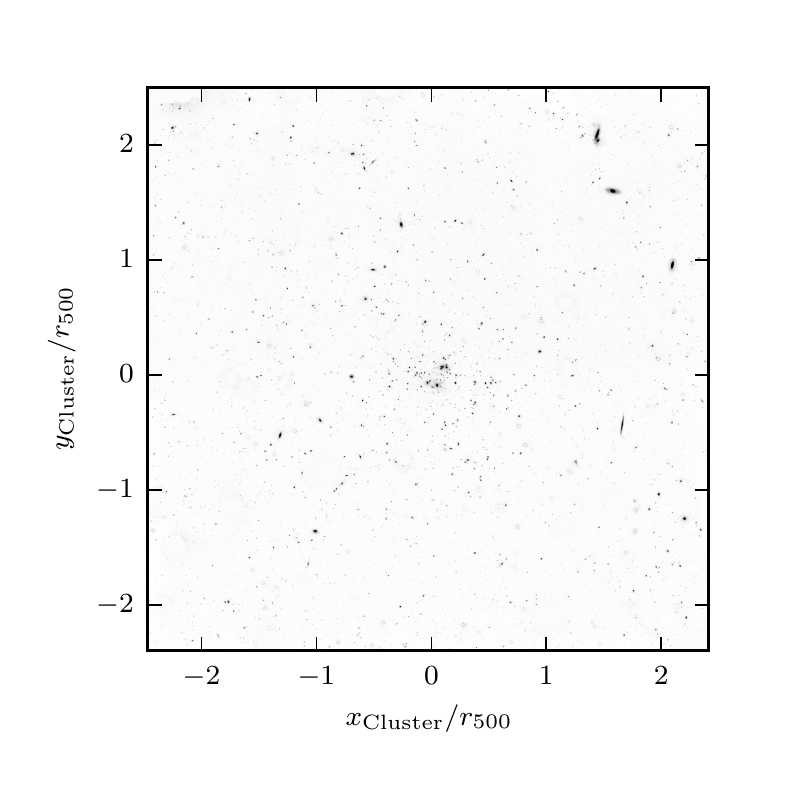}\protect\caption{\label{fig:RXC-J2234.5-3744-in-R}RXC J2234.5--3744 in the $R$ band.
Stars have been excised from this image.}
\end{figure}

\begin{figure}
\includegraphics[width=8cm]{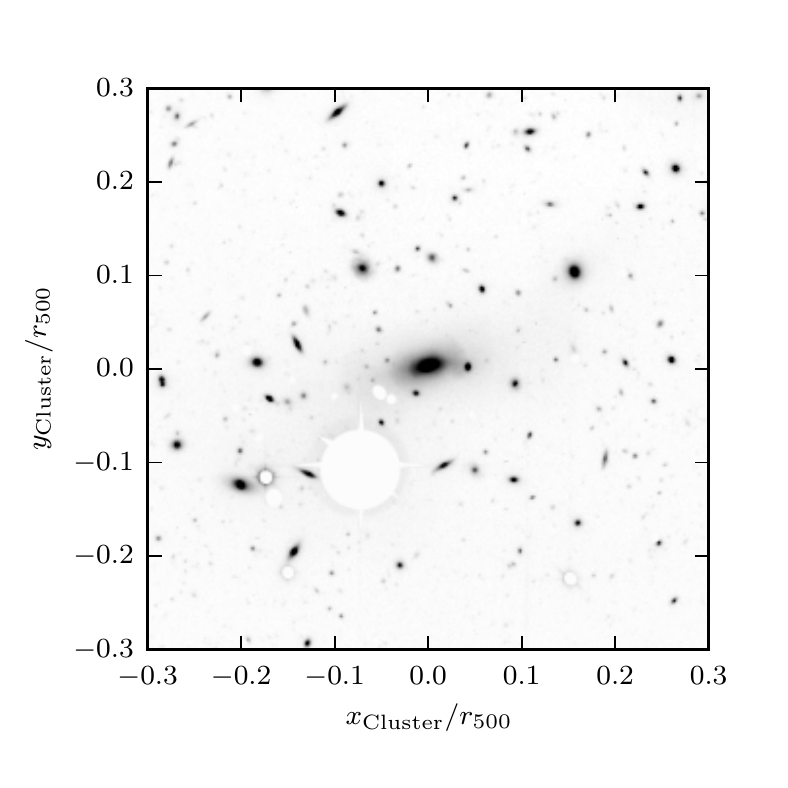}\protect\caption{\label{fig:RXC-J0006.0-3443-in-R-Detail}The centre of RXC J0006.0--3443
in the $R$ band. Stars have been excised from this image.}
\end{figure}

\figref{RXC-J0006.0-3443-in-R}, \figref{RXC-J0616.8-4748-in-R},
\figref{RXC-J2234.5-3744-in-R} and \figref{RXC-J0006.0-3443-in-R-Detail}
show $R$ band images of a selection of our targets, where the stars
have been excised from the images.

\makeatletter{}

\clearpage{}

\section{Individual cluster radial profiles}

\label{sec:IndividualRadialProfiles}

\begin{figure}
\includegraphics[width=8.2cm]{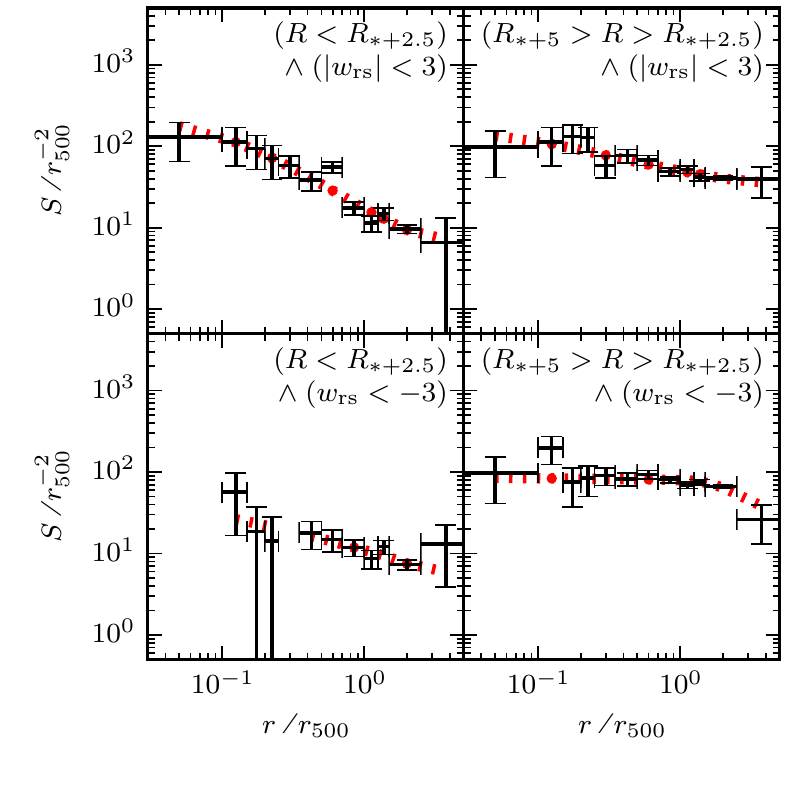}

\protect\caption{\label{fig:Population-radial-density-1}Radial density profiles for
RXC J0006.0--3443. The points with uncertainties represent the galaxy
count density profile, normalised by $\rfh^{2}$, and the red dashed
line is the best fitting model with parameters given in \tabref{Individual-NFW-fitting-results.}.}
\end{figure}

Figures \ref{fig:Population-radial-density-1} to \ref{fig:Population-radial-density-14}
show radial count density profiles and best fitting models, before
background subtraction, of individual clusters in the \rexcess\ sample.
The best fitting radial profile parameters for each galaxy population
in individual clusters in the \rexcess\ sample are given in \tabref{Individual-NFW-fitting-results.}. 

\makeatletter{}

\begin{figure}
\includegraphics[width=8.2cm]{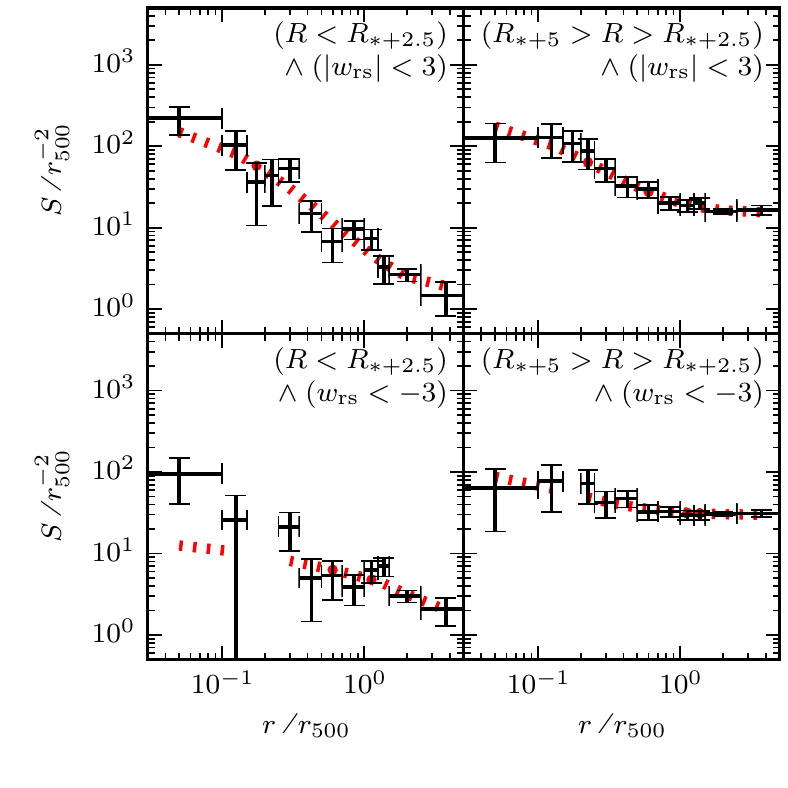}

\protect\caption{Radial density profiles for RXC J0049.4--2931. The lines are the same
as described in \figref{Population-radial-density-1}. }
\end{figure}

\begin{figure}
\includegraphics[width=8.2cm]{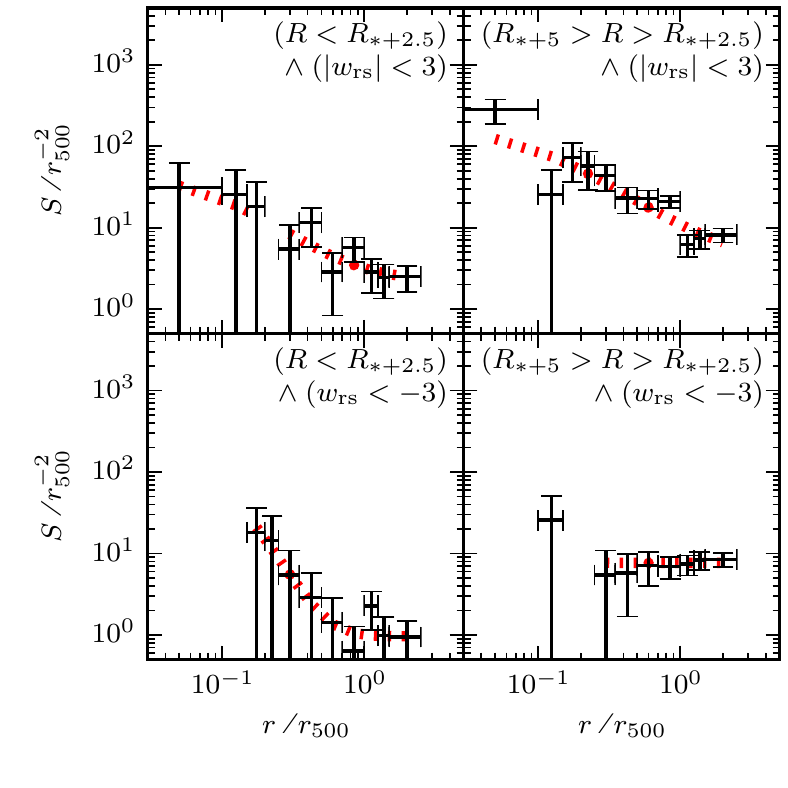}\protect\caption{Radial density profiles for RXC J0345.7--4112. The lines are the same
as described in \figref{Population-radial-density-1}. }
\end{figure}

\begin{figure}
\includegraphics[width=8.2cm]{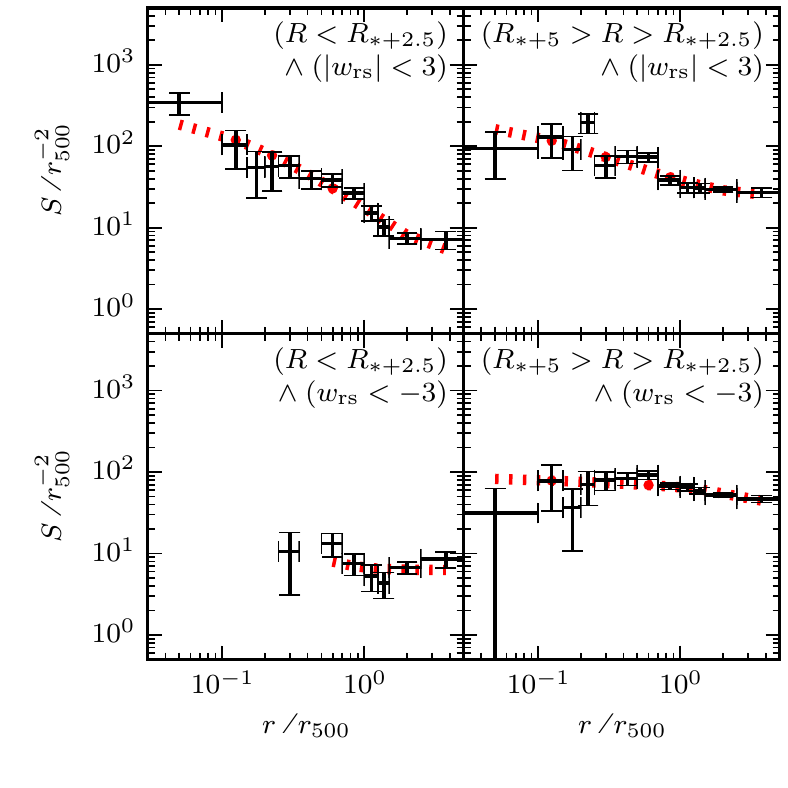}\protect\caption{Radial density profiles for RXC J0547.6--3152. The lines are the same
as described in \figref{Population-radial-density-1}. }
\end{figure}

\begin{figure}
\includegraphics[width=8.2cm]{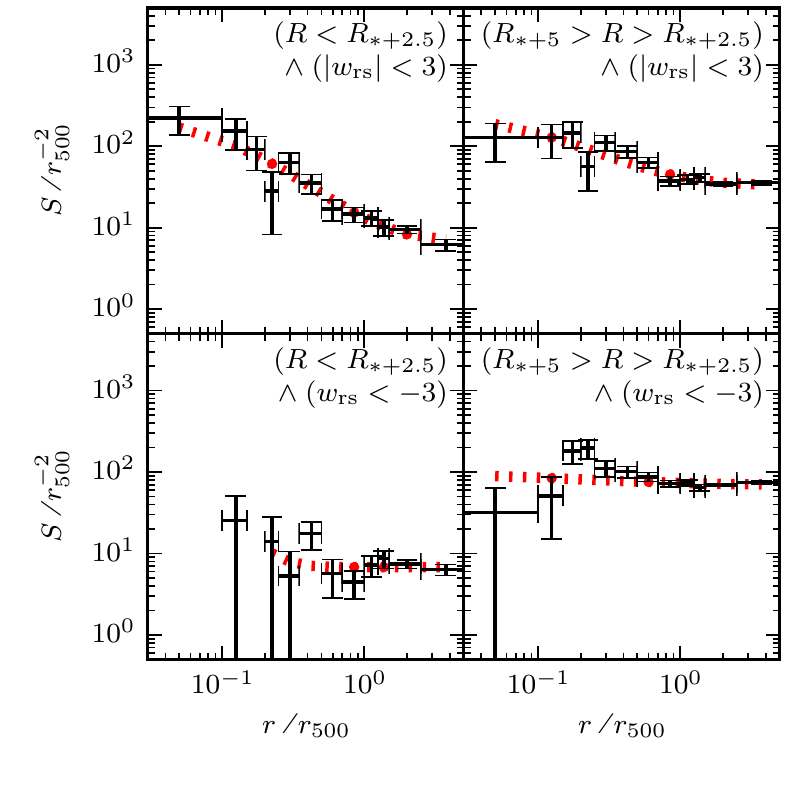}

\protect\caption{Radial density profiles for RXC J0605.8--3518. The lines are the same
as described in \figref{Population-radial-density-1}. }
\end{figure}

\begin{figure}
\includegraphics[width=8.2cm]{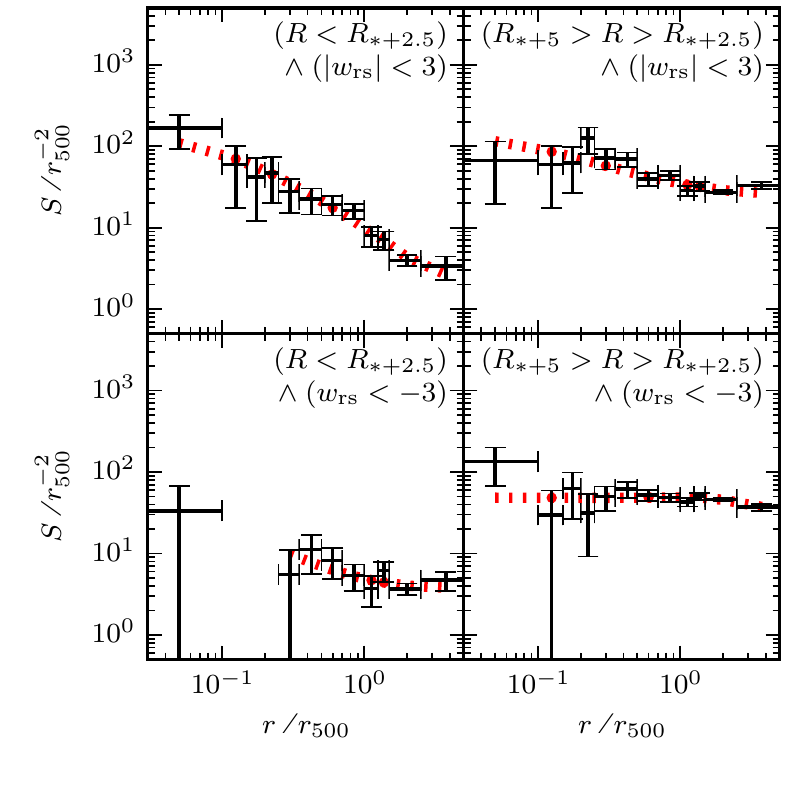}\protect\caption{Radial density profiles for RXC J0616.8--4748. The lines are the same
as described in \figref{Population-radial-density-1}. }
\end{figure}

\begin{figure}
\includegraphics[width=8.2cm]{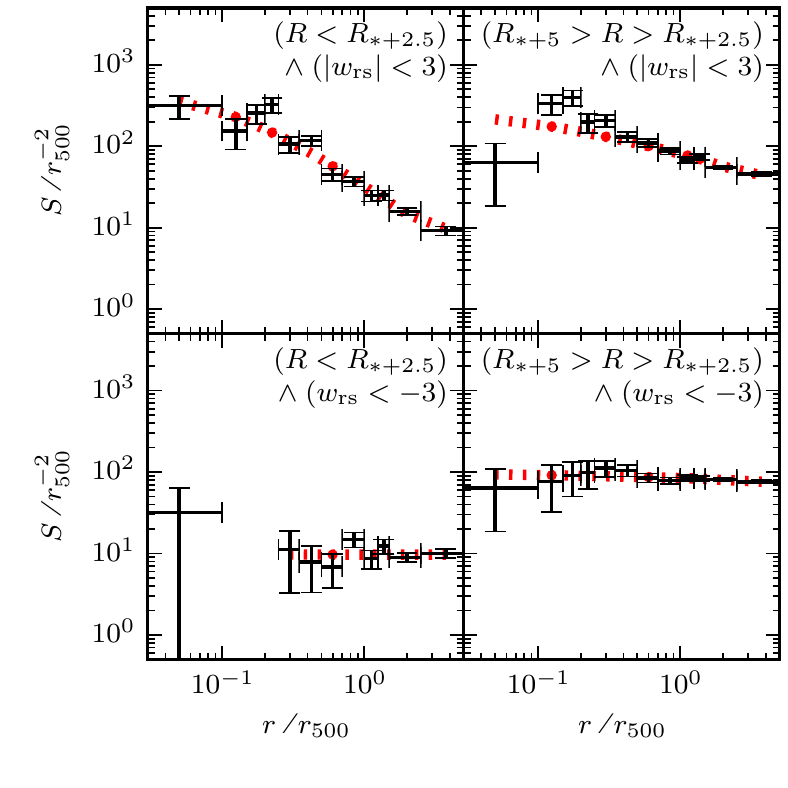}\protect\caption{Radial density profiles for RXC J0645.4--5413. The lines are the same
as described in \figref{Population-radial-density-1}. }
\end{figure}

\begin{figure}
\includegraphics[width=8.2cm]{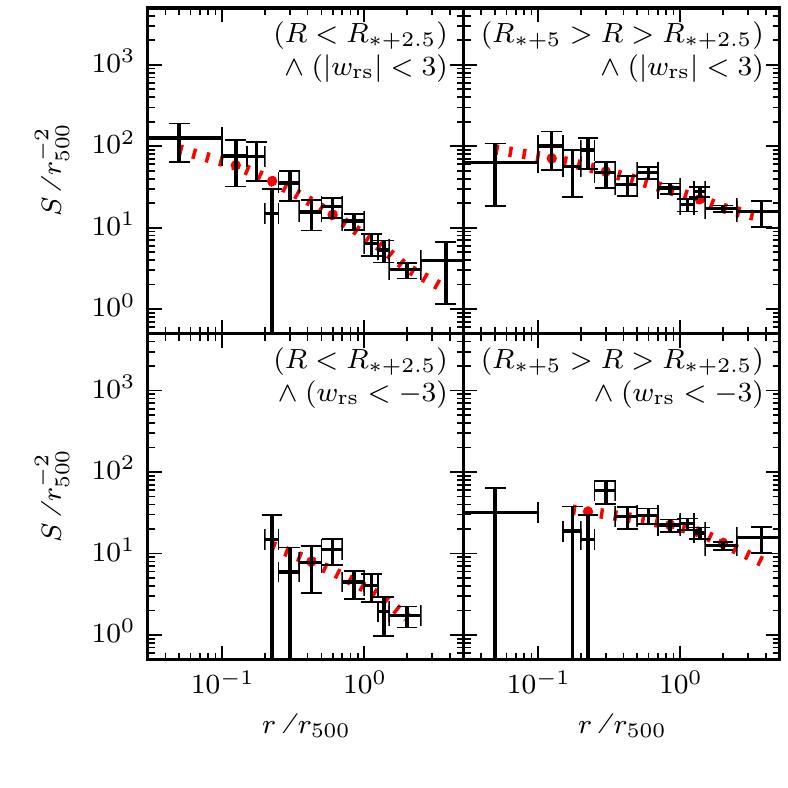}\protect\caption{\label{fig:0821-profiles}Radial density profiles for RXC J0821.8+0112.
The lines are the same as described in \figref{Population-radial-density-1}. }
\end{figure}

\begin{figure}
\includegraphics[width=8.2cm]{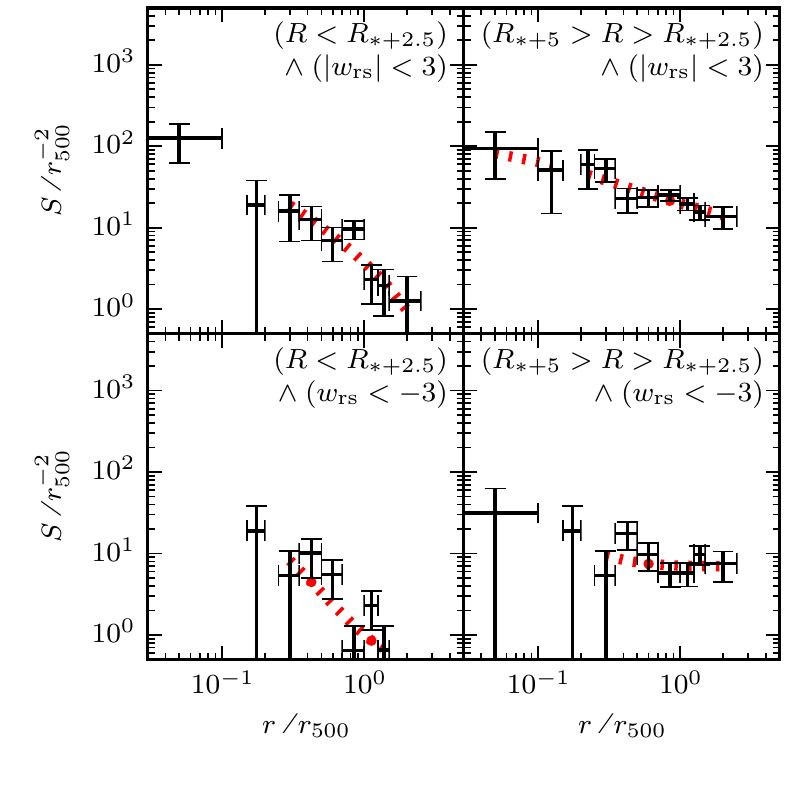}

\protect\caption{Radial density profiles for RXC J2023.0--2056. The lines are the same
as described in \figref{Population-radial-density-1}. }
\end{figure}

\begin{figure}
\includegraphics[width=8.2cm]{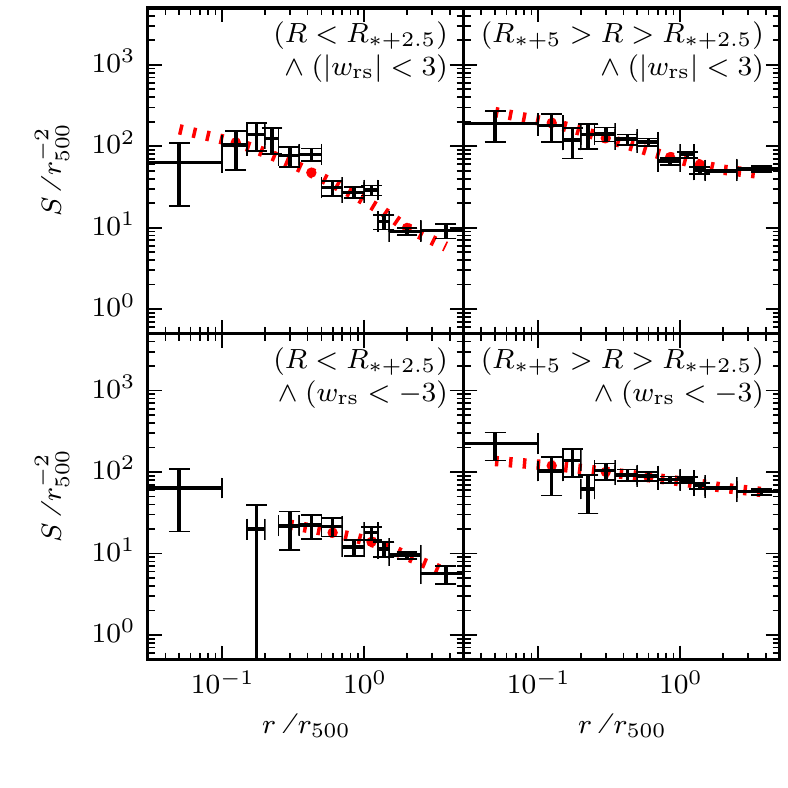}\protect\caption{Radial density profiles for RXC J2048.1--1750. The lines are the same
as described in \figref{Population-radial-density-1}. }
\end{figure}

\begin{figure}
\includegraphics[width=8.2cm]{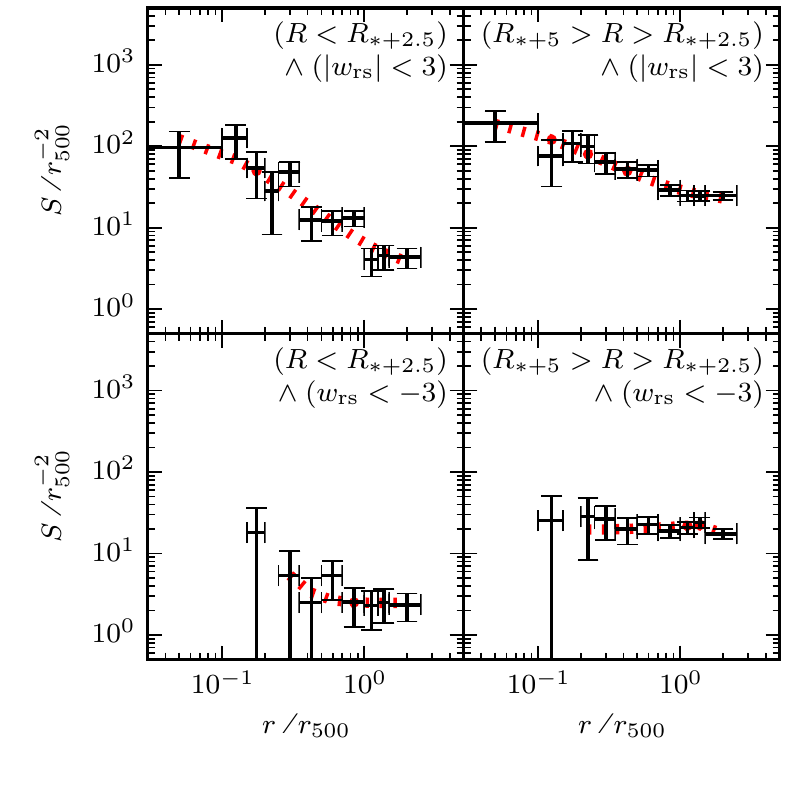}\protect\caption{Radial density profiles for RXC J2129.8--5048. The lines are the same
as described in \figref{Population-radial-density-1}. }
\end{figure}

\begin{figure}
\includegraphics[width=8.2cm]{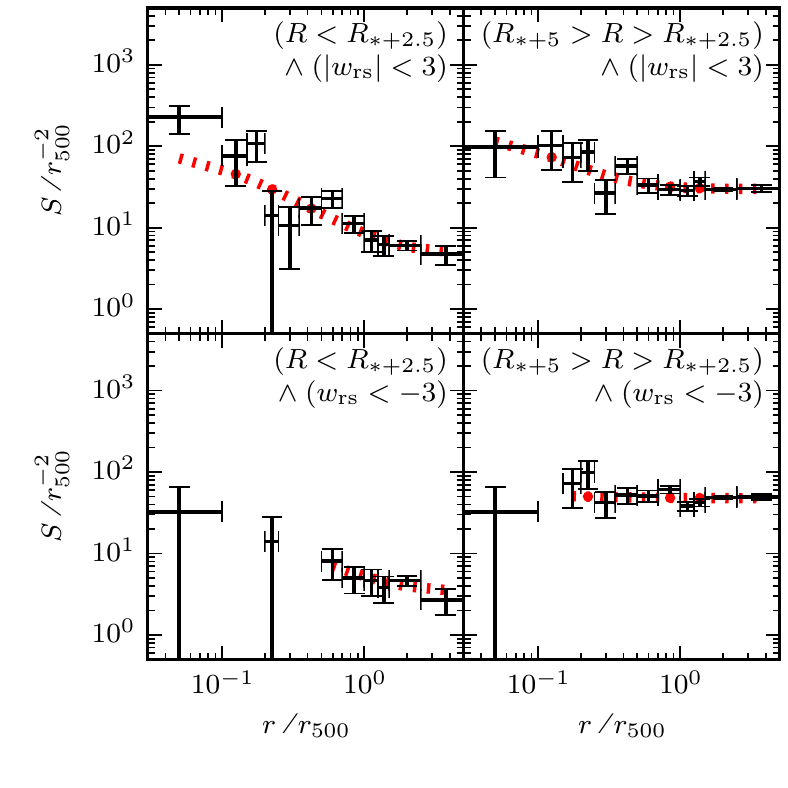}

\protect\caption{Radial density profiles for RXC J2218.6--3853. The lines are the same
as described in \figref{Population-radial-density-1}. }
\end{figure}

\begin{figure}
\includegraphics[width=8.2cm]{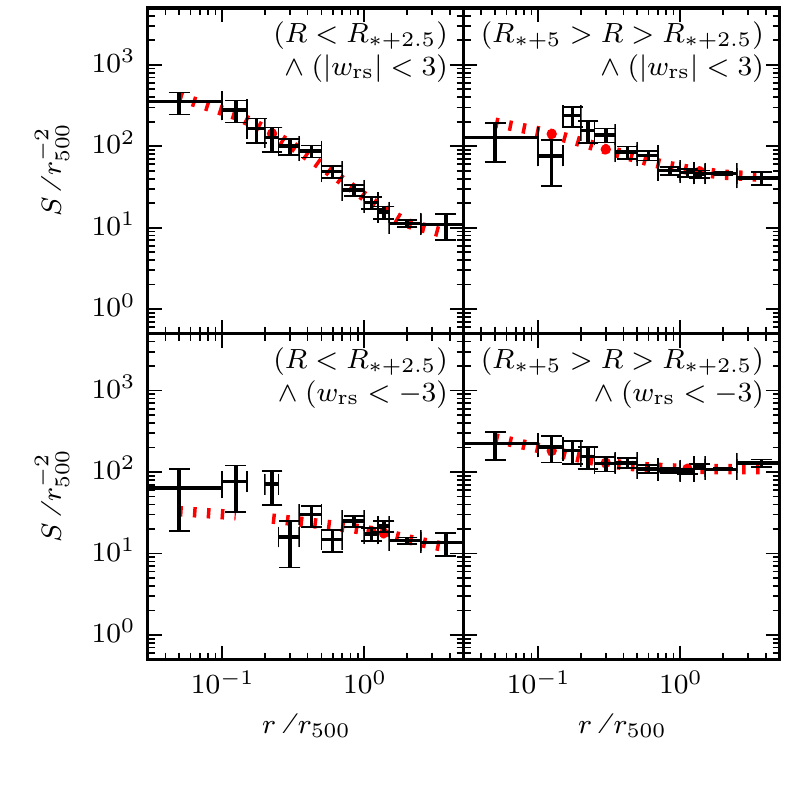}\protect\caption{Radial density profiles for RXC J2234.5--3744. The lines are the same
as described in \figref{Population-radial-density-1}. }
\end{figure}

\begin{figure}
\includegraphics[width=8.2cm]{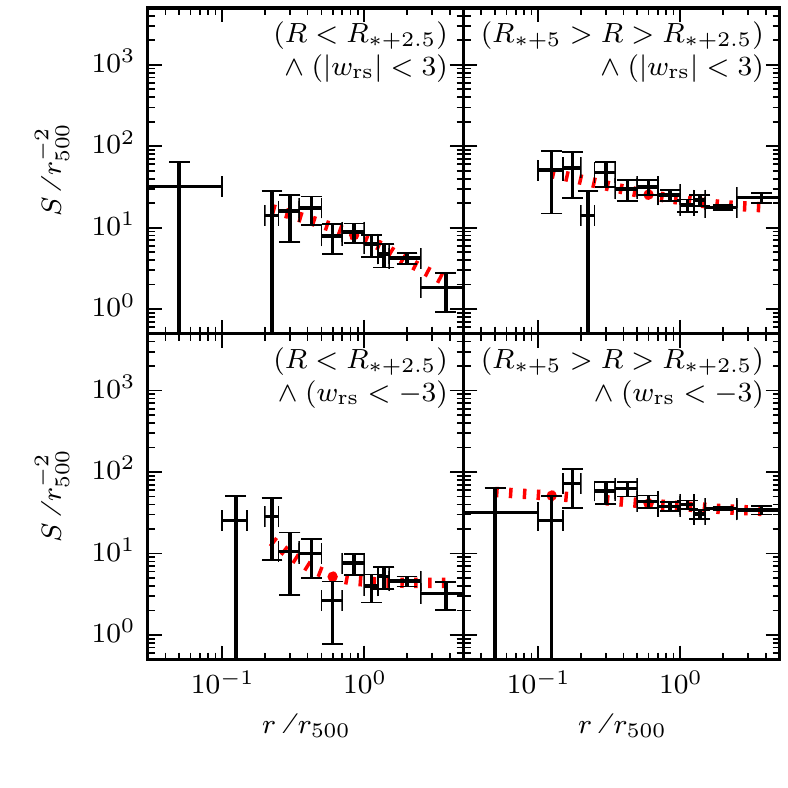}\protect\caption{\label{fig:Population-radial-density-14}Radial density profiles for
RXC J2319.6--7313. The lines are the same as described in \figref{Population-radial-density-1}. }
\end{figure}

\clearpage{}

\begin{table*}
\protect\caption{\label{tab:Individual-NFW-fitting-results.}NFW fitting results.}

\makeatletter{}
\begin{tabular}{lllll}
\hline
Object & Galaxy filter & \ensuremath{\beta} & \ensuremath{\surfacedensity_0} & \NFWareaconstant \\
 &  &  & \ensuremath{\si{\rfhsi\tothe{-3}}} & \ensuremath{\si{\rfhsi\tothe{-2}}} \\
\hline
RXC\,J0006.0--3443 & Bright, red & \ensuremath{2.7 \pm 0.5} & \ensuremath{49.7 \pm 9.0} & \ensuremath{7.2 \pm 2.1} \\
RXC\,J0049.4--2931 & Bright, red & \ensuremath{3.4 \pm 0.5} & \ensuremath{59.0 \pm 10.9} & \ensuremath{1.9 \pm 0.5} \\
RXC\,J0345.7--4112 & Bright, red & \ensuremath{3.2 \pm 1.0} & \ensuremath{12.2 \pm 4.5} & \ensuremath{2.3 \pm 0.6} \\
RXC\,J0547.6--3152 & Bright, red & \ensuremath{2.67 \pm 0.35} & \ensuremath{50.1 \pm 6.8} & \ensuremath{5.4 \pm 1.4} \\
RXC\,J0605.8--3518 & Bright, red & \ensuremath{3.2 \pm 0.4} & \ensuremath{55.5 \pm 8.8} & \ensuremath{7.2 \pm 0.8} \\
RXC\,J0616.8--4748 & Bright, red & \ensuremath{2.58 \pm 0.26} & \ensuremath{30.1 \pm 3.4} & \ensuremath{2.7 \pm 0.5} \\
RXC\,J0645.4--5413 & Bright, red & \ensuremath{2.74 \pm 0.29} & \ensuremath{96.2 \pm 13.0} & \ensuremath{9.8 \pm 1.5} \\
RXC\,J0821.8+0112 & Bright, red & \ensuremath{2.5 \pm 0.4} & \ensuremath{25.8 \pm 4.0} & \ensuremath{1.8 \pm 0.9} \\
RXC\,J2023.0--2056 & Bright, red & \ensuremath{2.7 \pm 0.7} & \ensuremath{23.6 \pm 5.9} & \ensuremath{0.2 \pm 1.5} \\
RXC\,J2048.1--1750 & Bright, red & \ensuremath{2.3 \pm 0.5} & \ensuremath{37.3 \pm 7.7} & \ensuremath{5.8 \pm 2.2} \\
RXC\,J2129.8--5048 & Bright, red & \ensuremath{3.3 \pm 0.7} & \ensuremath{46.1 \pm 10.7} & \ensuremath{3.3 \pm 1.4} \\
RXC\,J2218.6--3853 & Bright, red & \ensuremath{2.9 \pm 1.0} & \ensuremath{20.1 \pm 6.4} & \ensuremath{5.1 \pm 1.1} \\
RXC\,J2234.5--3744 & Bright, red & \ensuremath{3.18 \pm 0.15} & \ensuremath{129.4 \pm 6.9} & \ensuremath{8.4 \pm 0.8} \\
RXC\,J2319.6--7313 & Bright, red & \ensuremath{1.6 \pm 0.4} & \ensuremath{4.5 \pm 0.9} & \ensuremath{2.2 \pm 0.5} \\
RXC\,J0006.0--3443 & Faint, red & \ensuremath{2.0 \pm 0.5} & \ensuremath{20.5 \pm 3.6} & \ensuremath{36.1 \pm 3.4} \\
RXC\,J0049.4--2931 & Faint, red & \ensuremath{3.21 \pm 0.30} & \ensuremath{59.7 \pm 7.0} & \ensuremath{15.5 \pm 0.7} \\
RXC\,J0345.7--4112 & Faint, red & \ensuremath{2.8 \pm 0.9} & \ensuremath{39.7 \pm 12.5} & \ensuremath{5.4 \pm 3.1} \\
RXC\,J0547.6--3152 & Faint, red & \ensuremath{2.4 \pm 0.6} & \ensuremath{33.5 \pm 7.6} & \ensuremath{25.5 \pm 3.3} \\
RXC\,J0605.8--3518 & Faint, red & \ensuremath{2.8 \pm 0.6} & \ensuremath{45.6 \pm 10.3} & \ensuremath{33.7 \pm 2.1} \\
RXC\,J0616.8--4748 & Faint, red & \ensuremath{2.3 \pm 0.7} & \ensuremath{22.3 \pm 6.6} & \ensuremath{26.8 \pm 2.6} \\
RXC\,J0645.4--5413 & Faint, red & \ensuremath{1.6 \pm 0.5} & \ensuremath{27.6 \pm 6.5} & \ensuremath{44.0 \pm 4.6} \\
RXC\,J0821.8+0112 & Faint, red & \ensuremath{1.7 \pm 0.5} & \ensuremath{15.1 \pm 3.2} & \ensuremath{13.3 \pm 3.1} \\
RXC\,J2023.0--2056 & Faint, red & \ensuremath{2.1 \pm 0.7} & \ensuremath{16.7 \pm 3.9} & \ensuremath{12.1 \pm 3.9} \\
RXC\,J2048.1--1750 & Faint, red & \ensuremath{2.3 \pm 0.5} & \ensuremath{51.4 \pm 10.0} & \ensuremath{45.6 \pm 4.2} \\
RXC\,J2129.8--5048 & Faint, red & \ensuremath{2.8 \pm 0.5} & \ensuremath{54.2 \pm 8.9} & \ensuremath{21.2 \pm 2.7} \\
RXC\,J2218.6--3853 & Faint, red & \ensuremath{3.9 \pm 1.5} & \ensuremath{36.0 \pm 17.8} & \ensuremath{29.9 \pm 1.6} \\
RXC\,J2234.5--3744 & Faint, red & \ensuremath{2.6 \pm 0.7} & \ensuremath{40.4 \pm 10.2} & \ensuremath{42.1 \pm 4.0} \\
RXC\,J2319.6--7313 & Faint, red & \ensuremath{2.3 \pm 1.3} & \ensuremath{11.2 \pm 4.8} & \ensuremath{17.8 \pm 2.1} \\
RXC\,J0006.0--3443 & Bright, blue & \ensuremath{1.7 \pm 0.9} & \ensuremath{4.9 \pm 1.4} & \ensuremath{5.9 \pm 1.7} \\
RXC\,J0049.4--2931 & Bright, blue & \ensuremath{1.2 \pm 1.2} & \ensuremath{1.4 \pm 0.9} & \ensuremath{2.0 \pm 0.8} \\
RXC\,J0345.7--4112 & Bright, blue & \ensuremath{6.6 \pm 2.1} & \ensuremath{\left(1.8 \pm 1.0\right) \times 10^{2}} & \ensuremath{0.96 \pm 0.20} \\
RXC\,J0547.6--3152 & Bright, blue & \ensuremath{4.4 \pm 6.1} & \ensuremath{\left(2.6 \pm 3.8\right) \times 10^{1}} & \ensuremath{6.3 \pm 1.2} \\
RXC\,J0605.8--3518 & Bright, blue & \ensuremath{9.4 \pm 13.0} & \ensuremath{\left(1.4 \pm 2.5\right) \times 10^{2}} & \ensuremath{6.8 \pm 0.5} \\
RXC\,J0616.8--4748 & Bright, blue & \ensuremath{2.9 \pm 1.3} & \ensuremath{7.3 \pm 3.5} & \ensuremath{3.8 \pm 0.6} \\
RXC\,J0645.4--5413 & Bright, blue & \ensuremath{\left(0.1 \pm 2.4\right) \times 10^{2}} & \ensuremath{\left(0.5 \pm 6.5\right) \times 10^{2}} & \ensuremath{9.7 \pm 0.7} \\
RXC\,J0821.8+0112 & Bright, blue & \ensuremath{2.0 \pm 0.9} & \ensuremath{5.2 \pm 1.7} & \ensuremath{0.8 \pm 1.0} \\
RXC\,J2023.0--2056 & Bright, blue & \ensuremath{4.1 \pm 2.5} & \ensuremath{33.9 \pm 21.7} & \ensuremath{0.4 \pm 1.2} \\
RXC\,J2048.1--1750 & Bright, blue & \ensuremath{1.1 \pm 0.5} & \ensuremath{3.4 \pm 1.0} & \ensuremath{5.9 \pm 1.1} \\
RXC\,J2129.8--5048 & Bright, blue & \ensuremath{7.5 \pm 3.3} & \ensuremath{\left(1.6 \pm 1.4\right) \times 10^{2}} & \ensuremath{2.47 \pm 0.26} \\
RXC\,J2218.6--3853 & Bright, blue & \ensuremath{2.7 \pm 0.9} & \ensuremath{7.3 \pm 3.1} & \ensuremath{3.6 \pm 0.5} \\
RXC\,J2234.5--3744 & Bright, blue & \ensuremath{1.0 \pm 1.4} & \ensuremath{2.4 \pm 1.2} & \ensuremath{11.9 \pm 3.7} \\
RXC\,J2319.6--7313 & Bright, blue & \ensuremath{5.3 \pm 2.6} & \ensuremath{55.2 \pm 34.1} & \ensuremath{4.3 \pm 0.5} \\
RXC\,J0006.0--3443 & Faint, blue & \ensuremath{-0.6 \pm 0.6} & \ensuremath{0.99 \pm 0.28} & \ensuremath{38.2 \pm 10.6} \\
RXC\,J0049.4--2931 & Faint, blue & \ensuremath{3.0 \pm 0.7} & \ensuremath{20.4 \pm 5.8} & \ensuremath{29.9 \pm 0.9} \\
RXC\,J0345.7--4112 & Faint, blue & \ensuremath{\left(2.0 \pm 5.0\right) \times 10^{1}} & \ensuremath{\left(0.0 \pm 1.7\right) \times 10^{6}} & \ensuremath{7.7 \pm 0.4} \\
RXC\,J0547.6--3152 & Faint, blue & \ensuremath{0.4 \pm 0.7} & \ensuremath{2.6 \pm 1.0} & \ensuremath{44.5 \pm 4.7} \\
RXC\,J0605.8--3518 & Faint, blue & \ensuremath{1.8 \pm 3.2} & \ensuremath{3.6 \pm 5.1} & \ensuremath{70.4 \pm 4.4} \\
RXC\,J0616.8--4748 & Faint, blue & \ensuremath{-1.1 \pm 1.1} & \ensuremath{0.10 \pm 0.08} & \ensuremath{37.3 \pm 2.7} \\
RXC\,J0645.4--5413 & Faint, blue & \ensuremath{0.6 \pm 1.0} & \ensuremath{1.5 \pm 0.8} & \ensuremath{76.2 \pm 2.0} \\
RXC\,J0821.8+0112 & Faint, blue & \ensuremath{0.9 \pm 0.7} & \ensuremath{3.6 \pm 1.0} & \ensuremath{7.9 \pm 3.6} \\
RXC\,J2023.0--2056 & Faint, blue & \ensuremath{4.4 \pm 3.9} & \ensuremath{14.8 \pm 21.4} & \ensuremath{6.9 \pm 1.3} \\
RXC\,J2048.1--1750 & Faint, blue & \ensuremath{1.4 \pm 0.4} & \ensuremath{11.7 \pm 2.0} & \ensuremath{56.9 \pm 2.6} \\
RXC\,J2129.8--5048 & Faint, blue & \ensuremath{-1.5 \pm 0.5} & \ensuremath{0.078 \pm 0.029} & \ensuremath{0.7 \pm 10.5} \\
RXC\,J2218.6--3853 & Faint, blue & \ensuremath{2.6 \pm 13.3} & \ensuremath{1.5 \pm 8.1} & \ensuremath{47.6 \pm 3.2} \\
RXC\,J2234.5--3744 & Faint, blue & \ensuremath{4.4 \pm 1.4} & \ensuremath{70.2 \pm 32.7} & \ensuremath{108.4 \pm 2.5} \\
RXC\,J2319.6--7313 & Faint, blue & \ensuremath{1.5 \pm 1.3} & \ensuremath{4.0 \pm 2.4} & \ensuremath{33.3 \pm 2.9} \\
\hline
\end{tabular}
 
\end{table*}


\clearpage{}

\section{Background count density analysis}

The background count densities for all of the objects, using the four
main galaxy population filters are shown in \tabref{Background-count-density-table-long}.

\makeatletter{}
\begin{table*}
\protect\caption{\label{tab:Background-count-density-table-long}Background count density
measurements. In the case of the RXC~J2023.0-2056 bright blue filter,
no objects are detected in the region used for measuring $\simplebackground$.}

\makeatletter{}
\begin{tabular}{llllll}
\hline
Object & Galaxy filter & \simplebackground & \sectorbackgroundalpha & \NFWareaconstant & \ensuremath{\simplebackground/\NFWareaconstant} \\
 &  & \ensuremath{\si{\arcminuteword\tothe{-2}}} & \ensuremath{\si{\arcminuteword\tothe{-2}}} & \ensuremath{\si{\arcminuteword\tothe{-2}}} &  \\
\hline
RXC\,J0006.0--3443 & Bright, red & $0.166 \pm 0.020$ & $0.02$ & $0.12 \pm 0.04$ & $1.34$ \\
RXC\,J0006.0--3443 & Faint, red & $0.70 \pm 0.04$ & $0.04$ & $0.62 \pm 0.06$ & $1.12$ \\
RXC\,J0006.0--3443 & Bright, blue & $0.128 \pm 0.017$ & $0.02$ & $0.102 \pm 0.029$ & $1.25$ \\
RXC\,J0006.0--3443 & Faint, blue & $1.14 \pm 0.05$ & $0.07$ & $0.66 \pm 0.18$ & $1.73$ \\
RXC\,J0049.4--2931 & Bright, red & $0.064 \pm 0.011$ & $0.01$ & $0.051 \pm 0.013$ & $1.25$ \\
RXC\,J0049.4--2931 & Faint, red & $0.427 \pm 0.027$ & $0.03$ & $0.411 \pm 0.019$ & $1.04$ \\
RXC\,J0049.4--2931 & Bright, blue & $0.074 \pm 0.011$ & $0.01$ & $0.054 \pm 0.022$ & $1.38$ \\
RXC\,J0049.4--2931 & Faint, blue & $0.81 \pm 0.04$ & $0.04$ & $0.795 \pm 0.024$ & $1.02$ \\
RXC\,J0345.7--4112 & Bright, red & $0.029 \pm 0.010$ & $0.01$ & $0.027 \pm 0.007$ & $1.08$ \\
RXC\,J0345.7--4112 & Faint, red & $0.095 \pm 0.019$ & $0.02$ & $0.06 \pm 0.04$ & $1.50$ \\
RXC\,J0345.7--4112 & Bright, blue & $0.011 \pm 0.006$ & $0.00$ & $0.0111 \pm 0.0024$ & $0.98$ \\
RXC\,J0345.7--4112 & Faint, blue & $0.098 \pm 0.019$ & $0.02$ & $0.089 \pm 0.004$ & $1.11$ \\
RXC\,J0547.6--3152 & Bright, red & $0.182 \pm 0.024$ & $0.05$ & $0.134 \pm 0.034$ & $1.35$ \\
RXC\,J0547.6--3152 & Faint, red & $0.71 \pm 0.05$ & $0.06$ & $0.63 \pm 0.08$ & $1.13$ \\
RXC\,J0547.6--3152 & Bright, blue & $0.179 \pm 0.024$ & $0.05$ & $0.155 \pm 0.030$ & $1.15$ \\
RXC\,J0547.6--3152 & Faint, blue & $1.26 \pm 0.06$ & $0.22$ & $1.10 \pm 0.12$ & $1.14$ \\
RXC\,J0605.8--3518 & Bright, red & $0.209 \pm 0.019$ & $0.03$ & $0.184 \pm 0.020$ & $1.13$ \\
RXC\,J0605.8--3518 & Faint, red & $0.90 \pm 0.04$ & $0.06$ & $0.87 \pm 0.05$ & $1.03$ \\
RXC\,J0605.8--3518 & Bright, blue & $0.180 \pm 0.017$ & $0.02$ & $0.175 \pm 0.014$ & $1.03$ \\
RXC\,J0605.8--3518 & Faint, blue & $1.84 \pm 0.06$ & $0.06$ & $1.81 \pm 0.11$ & $1.01$ \\
RXC\,J0616.8--4748 & Bright, red & $0.087 \pm 0.012$ & $0.01$ & $0.060 \pm 0.012$ & $1.44$ \\
RXC\,J0616.8--4748 & Faint, red & $0.644 \pm 0.033$ & $0.04$ & $0.60 \pm 0.06$ & $1.07$ \\
RXC\,J0616.8--4748 & Bright, blue & $0.088 \pm 0.012$ & $0.03$ & $0.086 \pm 0.013$ & $1.03$ \\
RXC\,J0616.8--4748 & Faint, blue & $0.99 \pm 0.04$ & $0.05$ & $0.84 \pm 0.06$ & $1.18$ \\
RXC\,J0645.4--5413 & Bright, red & $0.302 \pm 0.022$ & $0.08$ & $0.231 \pm 0.035$ & $1.31$ \\
RXC\,J0645.4--5413 & Faint, red & $1.19 \pm 0.04$ & $0.07$ & $1.04 \pm 0.11$ & $1.14$ \\
RXC\,J0645.4--5413 & Bright, blue & $0.225 \pm 0.019$ & $0.08$ & $0.230 \pm 0.016$ & $0.98$ \\
RXC\,J0645.4--5413 & Faint, blue & $1.86 \pm 0.06$ & $0.07$ & $1.81 \pm 0.05$ & $1.03$ \\
RXC\,J0821.8+0112 & Bright, red & $0.055 \pm 0.011$ & $0.02$ & $0.031 \pm 0.016$ & $1.77$ \\
RXC\,J0821.8+0112 & Faint, red & $0.301 \pm 0.027$ & $0.03$ & $0.23 \pm 0.05$ & $1.28$ \\
RXC\,J0821.8+0112 & Bright, blue & $0.029 \pm 0.008$ & $0.01$ & $0.013 \pm 0.017$ & $2.13$ \\
RXC\,J0821.8+0112 & Faint, blue & $0.224 \pm 0.023$ & $0.02$ & $0.14 \pm 0.06$ & $1.60$ \\
RXC\,J2023.0--2056 & Bright, red & $0.011 \pm 0.011$ & $0.01$ & $0.002 \pm 0.014$ & $5.66$ \\
RXC\,J2023.0--2056 & Faint, red & $0.12 \pm 0.04$ & $0.04$ & $0.106 \pm 0.034$ & $1.15$ \\
RXC\,J2023.0--2056 & Bright, blue & $0.0 \pm 0$ & $0.00$ & $0.004 \pm 0.010$ & $0.00$ \\
RXC\,J2023.0--2056 & Faint, blue & $0.066 \pm 0.027$ & $0.02$ & $0.061 \pm 0.011$ & $1.09$ \\
RXC\,J2048.1--1750 & Bright, red & $0.245 \pm 0.022$ & $0.05$ & $0.16 \pm 0.06$ & $1.55$ \\
RXC\,J2048.1--1750 & Faint, red & $1.37 \pm 0.05$ & $0.08$ & $1.24 \pm 0.11$ & $1.11$ \\
RXC\,J2048.1--1750 & Bright, blue & $0.238 \pm 0.021$ & $0.03$ & $0.161 \pm 0.029$ & $1.48$ \\
RXC\,J2048.1--1750 & Faint, blue & $1.70 \pm 0.06$ & $0.06$ & $1.54 \pm 0.07$ & $1.10$ \\
RXC\,J2129.8--5048 & Bright, red & $0.051 \pm 0.014$ & $0.01$ & $0.038 \pm 0.017$ & $1.33$ \\
RXC\,J2129.8--5048 & Faint, red & $0.290 \pm 0.034$ & $0.07$ & $0.248 \pm 0.031$ & $1.17$ \\
RXC\,J2129.8--5048 & Bright, blue & $0.027 \pm 0.010$ & $0.01$ & $0.0289 \pm 0.0031$ & $0.95$ \\
RXC\,J2129.8--5048 & Faint, blue & $0.204 \pm 0.028$ & $0.04$ & $0.01 \pm 0.12$ & $24.40$ \\
RXC\,J2218.6--3853 & Bright, red & $0.130 \pm 0.015$ & $0.02$ & $0.115 \pm 0.025$ & $1.13$ \\
RXC\,J2218.6--3853 & Faint, red & $0.673 \pm 0.034$ & $0.05$ & $0.678 \pm 0.035$ & $0.99$ \\
RXC\,J2218.6--3853 & Bright, blue & $0.095 \pm 0.013$ & $0.02$ & $0.081 \pm 0.011$ & $1.18$ \\
RXC\,J2218.6--3853 & Faint, blue & $1.11 \pm 0.04$ & $0.04$ & $1.08 \pm 0.07$ & $1.03$ \\
RXC\,J2234.5--3744 & Bright, red & $0.226 \pm 0.022$ & $0.02$ & $0.169 \pm 0.017$ & $1.34$ \\
RXC\,J2234.5--3744 & Faint, red & $0.92 \pm 0.04$ & $0.04$ & $0.84 \pm 0.08$ & $1.10$ \\
RXC\,J2234.5--3744 & Bright, blue & $0.288 \pm 0.024$ & $0.02$ & $0.24 \pm 0.07$ & $1.21$ \\
RXC\,J2234.5--3744 & Faint, blue & $2.19 \pm 0.07$ & $0.06$ & $2.17 \pm 0.05$ & $1.01$ \\
RXC\,J2319.6--7313 & Bright, red & $0.088 \pm 0.012$ & $0.01$ & $0.051 \pm 0.010$ & $1.74$ \\
RXC\,J2319.6--7313 & Faint, red & $0.431 \pm 0.028$ & $0.02$ & $0.41 \pm 0.05$ & $1.05$ \\
RXC\,J2319.6--7313 & Bright, blue & $0.101 \pm 0.013$ & $0.01$ & $0.100 \pm 0.011$ & $1.00$ \\
RXC\,J2319.6--7313 & Faint, blue & $0.81 \pm 0.04$ & $0.06$ & $0.77 \pm 0.07$ & $1.06$ \\
\hline
\end{tabular}
 
\end{table*}


\clearpage{}

 
\end{document}